\pdfoutput=1

\documentclass[11pt]{article}

\usepackage{acl}

\usepackage{times}
\usepackage{latexsym}
\usepackage{booktabs}
\usepackage[T1]{fontenc}

\usepackage[utf8]{inputenc}

\usepackage{microtype}

\usepackage{inconsolata}

\usepackage{graphicx}
\usepackage{amsmath}
\usepackage{amssymb}
\usepackage{bm}
\usepackage{graphicx}
\usepackage{subcaption}
\usepackage{multirow}
\usepackage{fontawesome}

\usepackage[most, breakable, skins]{tcolorbox}
\usepackage[dvipsnames]{xcolor}
\usepackage{pifont}

\NewDocumentCommand{\ganqu}
{ mO{} }{\textcolor{blue}{\textsuperscript{\textit{ganqu}}\textsf{\textbf{\small[#1]}}}}
\definecolor{ababcol}{HTML}{F14738}
\definecolor{myhailuo2}{HTML}{F97669}
\definecolor{querycol}{HTML}{7964E8}
\definecolor{goldanswercol}{HTML}{FFB43B}
\definecolor{otherscol}{HTML}{FC5BCF}
\definecolor{myhailuo3light}{HTML}{FFA9FA}
\definecolor{myhailuo1dark}{HTML}{FC8900}
\definecolor{myhailuo2dark}{HTML}{F14738}
\definecolor{myhailuo3dark}{HTML}{D12AAA}
\definecolor{myhailuo4dark}{HTML}{4C4DC2}

\definecolor{myhailuo1}{HTML}{FFB43B}
\definecolor{myhailuo2}{HTML}{F97669}
\definecolor{myhailuo3}{HTML}{FC5BCF}
\definecolor{myhailuo4}{HTML}{7964E8}

\definecolor{myhailuo1light}{HTML}{FFD085}
\definecolor{myhailuo2light}{HTML}{FFA19F}
\definecolor{myhailuo3light}{HTML}{FFA9FA}
\definecolor{myhailuo4light}{HTML}{BDACFB}

\colorlet{myorange}{Orange!20}
\colorlet{mygreen}{LimeGreen!25}
\colorlet{myyellow}{Yellow!30}
\colorlet{myblue}{CornflowerBlue!25}
\colorlet{mybrown}{RawSienna!25}
\colorlet{mypurple}{Orchid!25}
\colorlet{myred}{Red!60}
\colorlet{myorangefull}{YellowOrange!60}
\colorlet{mybrownfull}{RawSienna!60}

\colorlet{myorangethick}{Orange!40}
\colorlet{mygreenthick}{LimeGreen!50}
\colorlet{myyellowthick}{Yellow!60}
\colorlet{mybluethick}{CornflowerBlue!50}

\tcbset{
    showcase/.style={
        fonttitle=\large,
        colback=white!20,  
        colframe=black,   
        coltitle=white,   
        boxrule=0.5mm,    
        arc=2mm,          
        outer arc=2mm,    
        left=1mm,         
        right=1mm,        
        top=1mm,          
        bottom=1mm,       
        width=\textwidth, 
        before skip=0.1pt,
        after skip=0.1pt,
    },
    context/.style={
        fontupper=\scriptsize,
        fonttitle=\large,
        colframe=querycol,     
        coltitle=white,   
        colback=white,    
        boxrule=0.3mm,    
        arc=2mm,          
        outer arc=2mm,    
        left=1mm,         
        right=1mm,        
        top=1mm,          
        bottom=1mm,       
        before skip=1pt,
        after skip=0.1pt, 
    },
    query/.style={
        fontupper=\scriptsize,
        fontlower=\scriptsize,
        colframe=querycol,     
        coltitle=white,   
        colback=white,    
        boxrule=0.1mm,    
        arc=2mm,          
        outer arc=2mm,    
        left=1mm,         
        right=1mm,        
        top=1mm,          
        bottom=1mm,       
        before skip=1pt,
        after skip=0.1pt,
    },
    abab/.style={
        fontupper=\scriptsize,
        fonttitle=,
        colframe=ababcol, 
        coltitle=white,   
        boxrule=0.5mm,    
        arc=2mm,          
        outer arc=2mm,    
        left=1mm,         
        right=1mm,        
        top=1mm,          
        bottom=1mm,       
        width=0.33\textwidth, 
        before skip=0.1pt,
        after skip=0.1pt, 
    },
    others/.style={
        fontupper=\scriptsize,
        colframe=myhailuo3light, 
        coltitle=white,
        boxrule=0.5mm,    
        arc=2mm,          
        outer arc=2mm,    
        left=1mm,         
        right=1mm,        
        top=1mm,          
        bottom=1mm,       
        width=0.33\textwidth, 
        before skip=0.1pt,
        after skip=0.1pt, 
    },
    goldanswer/.style={
        fontupper=\scriptsize,
        colframe=goldanswercol,     
        coltitle=white,   
        boxrule=0.5mm,    
        arc=2mm,          
        outer arc=2mm,    
        left=1mm,         
        right=1mm,        
        top=1mm,          
        bottom=1mm,       
        width=0.33\textwidth, 
        before skip=0.1pt,
        after skip=0.1pt, 
    },
}

\title{\benchmark{}: Visual Game Generation for Code Large Language Models}

\author{
  {\bf Wei Zhang}\textsuperscript{\rm 1},
  {\bf Jack Yang}, 
  {\bf Renshuai Tao}\textsuperscript{\rm 3},
  {\bf Lingzheng Chai},
  {\bf Shawn Guo},
  {\bf Jiajun Wu}, \\
  {\bf Xiaoming Chen}\textsuperscript{\rm 4},  
  {\bf Ganqu Cui}\textsuperscript{\rm 1*},
  {\bf Ning Ding}\textsuperscript{\rm 1}, 
  {\bf Xander Xu}\textsuperscript{\rm 2}, 
  {\bf Hu Wei}\textsuperscript{\rm 2*},
  {\bf Bowen Zhou}\textsuperscript{\rm 1}\thanks{\ \ Corresponding Author.} \\
   \textsuperscript{\rm 1}Shanghai AI Lab;  
   \textsuperscript{\rm 2}Alibaba Group;
   \textsuperscript{\rm 3}Beijing Jiaotong University;
   \textsuperscript{\rm 4}AIStrong; \\
}

\definecolor{ClosedColor}{HTML}{D8EEF2}  
\definecolor{OpenColor}{HTML}{FDEBDD}    
\definecolor{HeaderColor}{gray}{.85}
\usepackage{colortbl}
\arrayrulecolor{black}    

\newcommand{\benchmark}{V-GameGym}

\begin{document}
\maketitle
\begin{abstract}
Code large language models have demonstrated remarkable capabilities in programming tasks, yet current benchmarks primarily focus on single modality rather than visual game development. Most existing code-related benchmarks evaluate syntax correctness and execution accuracy, overlooking critical game-specific metrics such as playability, visual aesthetics, and user engagement that are essential for real-world deployment. To address the gap between current LLM capabilities in algorithmic problem-solving and competitive programming versus the comprehensive requirements of practical game development, we present \textbf{\benchmark{}}, a comprehensive benchmark comprising 2,219 high-quality samples across 100 thematic clusters derived from real-world repositories, adopting a novel clustering-based curation methodology to ensure both diversity and structural completeness. Further, we introduce a multimodal evaluation framework with an automated LLM-driven pipeline for visual code synthesis using complete UI sandbox environments. Our extensive analysis reveals that \benchmark{} effectively bridges the gap between code generation accuracy and practical game development workflows, providing quantifiable quality metrics for visual programming and interactive element generation.
\end{abstract}

\section{Introduction}

Recent advances in code large language models (code LLMs) have demonstrated remarkable capabilities in programming tasks, building upon foundational models such as Qwen-Coder~\cite{qwen25coder}, StarCoder~\citep{starcoder,starcoder2}, and DeepSeek-Coder~\cite{deepseek_coder}, establishing strong baselines for code generation~\citep{humaneval,bigcodebench,fullstack}, completion~\cite{execrepobench}, and understanding tasks~\cite{CodeXGLUE}. These LLMs adopt specialized training strategies combining pre-training on large code corpora from repositories like GitHub, followed by post-training to align outputs with programming best practices.

\begin{figure}[t!]
  \centering
  \includegraphics[width=1.0\columnwidth]{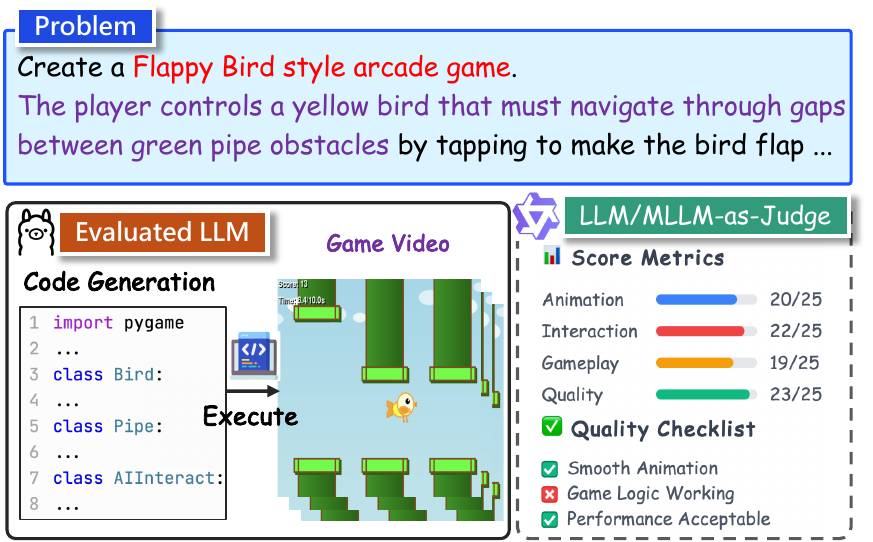}   
  \caption{A visual programming about the flappy bird style arcade game.}
  \label{fig:figures/intro.pdf}
  \vspace{-22pt}
\end{figure}

The recent LLMs like Claude 4 and GLM-4.5~\citep{claude4,glm45} exhibit enhanced reasoning capabilities for complex programming scenarios. Further, Kimi-K2~\cite{kimi_k2} focuses on long-context code comprehension and generation. \textit{The focus of these advanced LLMs is not on solving algorithmic problems, but rather on visual programming to provide more intuitive demonstrations of model performance.} The open-source community~\cite{gamegpt} has begun developing specialized evaluations for game generation tasks. Visual game synthesis~\cite{game_rl} further advances this domain by incorporating multi-modal understanding to generate games with coherent visual and interactive elements. However, these approaches primarily focus on code generation accuracy and syntax correctness, overlooking critical game-specific evaluation metrics such as playability, visual aesthetics, user engagement, and performance optimization. The absence of comprehensive evaluation frameworks and targeted improvement methodologies limits the practical deployment of code LLMs in professional game development workflows.

\begin{figure*}[h!]
  \centering
  \includegraphics[width=0.85\textwidth]{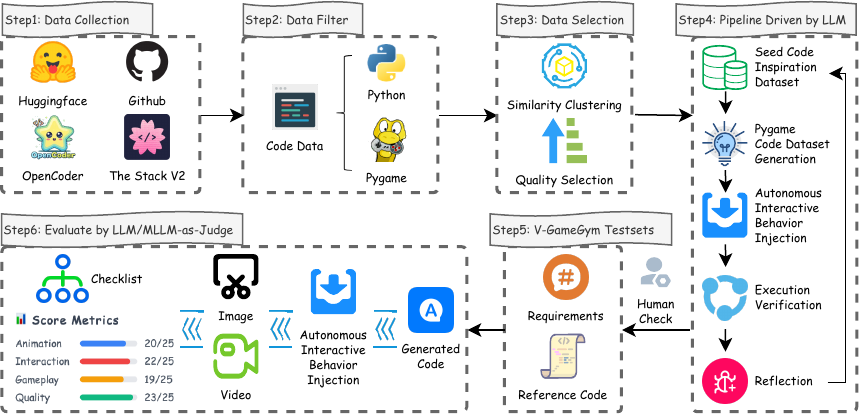}   
  \caption{Overview of the \benchmark{} framework from data collection to evaluation.}
  \label{fig:figures/framework.pdf}
  \vspace{-20pt}
\end{figure*}

In this work, we first introduce a comprehensive benchmark, \textbf{\benchmark{}}, comprising 2,219 high-quality samples across 100 thematic clusters derived from real-world Pygame repositories. The process begins by filtering Python source files from large open-source repositories (OpenCoder and The Stack v2) to identify Pygame-related projects, then applies a clustering-based curation strategy that partitions the code corpus using high-dimensional feature vectors and selects the highest-quality program from each cluster based on structural completeness metrics. The curated seed dataset is then processed through an automated LLM-driven pipeline that analyzes code intent, transforms interactive programs into self-contained demonstrations, verifies execution in sandboxed environments with automated error correction, and generates natural language requirement specifications. Finally, the dataset undergoes human validation by 8 graduate students who manually check approximately 2,219 Pygame programs using a complete UI sandbox environment to ensure code integrity and quality.

The contributions are summarized as follows: 
\textbf{(1)} We propose \benchmark{} comprised of 2,219 manually verified samples sourced from 2,190 distinct repositories, a comprehensive code generation benchmark for evaluating multimodal game development capabilities, encompassing 100 clusters with diverse functional characteristics. 
\textbf{(2)} We introduce a novel clustering-based curation methodology that combines high-dimensional feature extraction with quality-based selection, ensuring both diversity and structural completeness in the dataset.
\textbf{(3)} We systematically construct a multimodal evaluation framework with an automated LLM-driven pipeline for code transformation and requirement synthesis, validated through comprehensive human annotation involving 8 graduate students. Notably, extensive analysis reveals that \benchmark{} effectively captures the complexity spectrum of real-world game development tasks with quantifiable quality metrics.

\begin{figure*}[t!]
  \centering
  \includegraphics[width=0.95\textwidth]{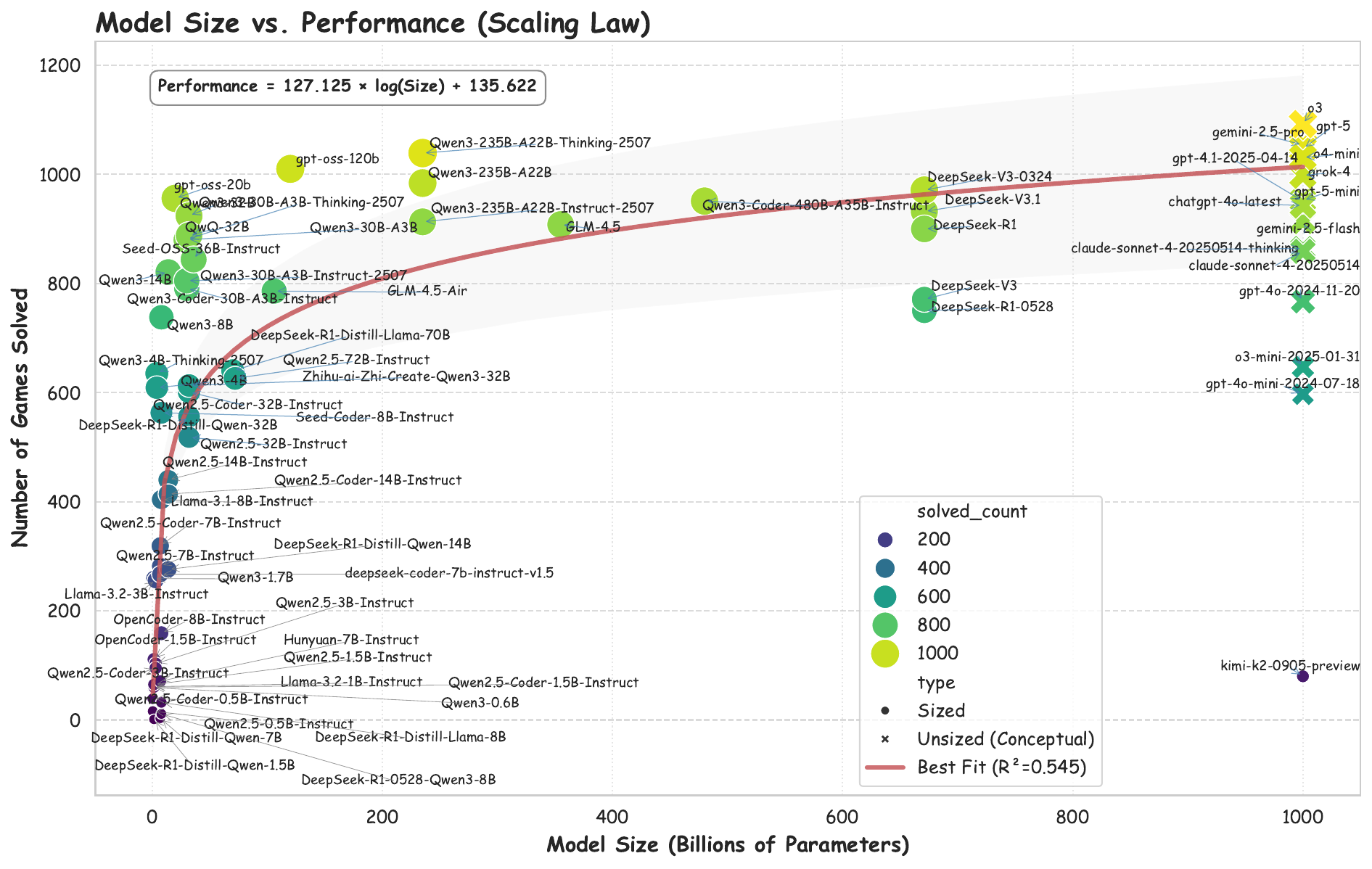}\vspace{-10pt}
  \caption{Correlation between model size and games solved.}
  \label{fig:figures/figures_evals/12_model_size_vs_performance.pdf}
  \vspace{-15pt}
\end{figure*}

\section{\benchmark{}}

\subsection{\benchmark{} Task Definition}

Let $\mathbb{I}$ and $\mathbb{C}$ denote the spaces of natural language instructions and program source codes, respectively. The model under evaluation, $\mathcal{M}_\theta$, is a generative model parameterized by $\theta$ that approximates the conditional probability distribution $P(\mathcal{C}|\mathcal{I})$ where $\mathcal{I} \in \mathbb{I}$ and $\mathcal{C} \in \mathbb{C}$. The comprehensive process of generation and evaluation for a given instruction $\mathcal{I}$ is defined by the following sequence.

\paragraph{Code Generation}
A code instance $\mathcal{C}$ is sampled from the model's output distribution: $\mathcal{C} \sim P_\theta(\cdot|\mathcal{I})$.

\paragraph{Execution \& Artifact Synthesis}
The generated code $\mathcal{C}$ is executed by a deterministic environment function $\mathcal{E}: \mathbb{C} \to \mathbb{A}$, which synthesizes a set of multimedia artifacts $(\mathcal{V}, \mathcal{S}) \in \mathbb{A}$. Here, $\mathbb{A} = \mathbb{V} \times \mathbb{S}$ represents the artifact space, composed of the video space $\mathbb{V}$ and the image space $\mathbb{S}$: $(\mathcal{V}, \mathcal{S}) = \mathcal{E}(\mathcal{C})$.

\begin{figure}[t!]
  \centering
  \includegraphics[width=0.6\columnwidth]{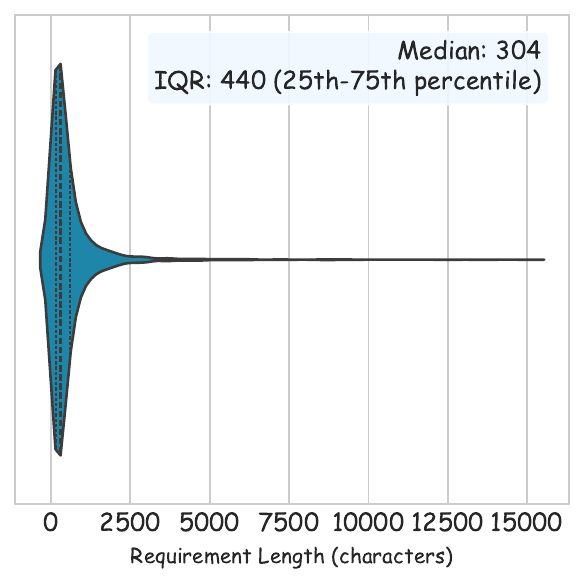}\vspace{-5pt}
  \caption{Overall Requirement Length Distribution}
  \label{fig:figures/figures_testset/figures_requirements/001_requirement_length_distribution_density.pdf}
  \vspace{-5pt}
\end{figure}

\begin{figure}[t!]
  \centering
  \includegraphics[width=0.6\columnwidth]{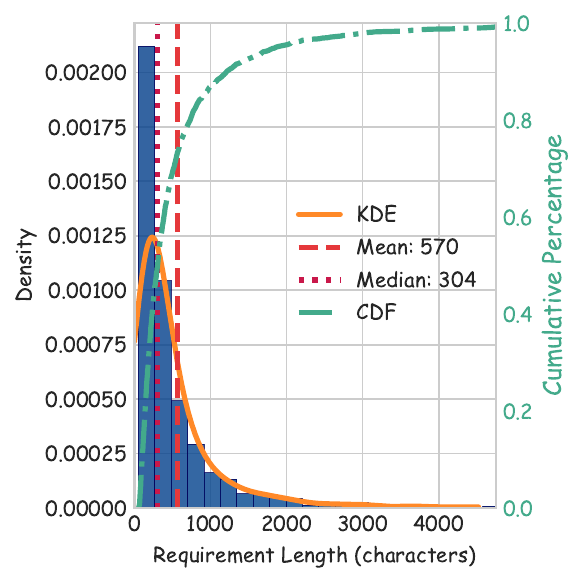}
  \caption{Linear-Scale Requirement Length Histogram}
  \label{fig:figures/figures_testset/figures_requirements/002_requirement_length_distribution_linear_scale.pdf}
  \vspace{-15pt}
\end{figure}

\begin{figure}[h!]
  \centering
  \includegraphics[width=0.6\columnwidth]{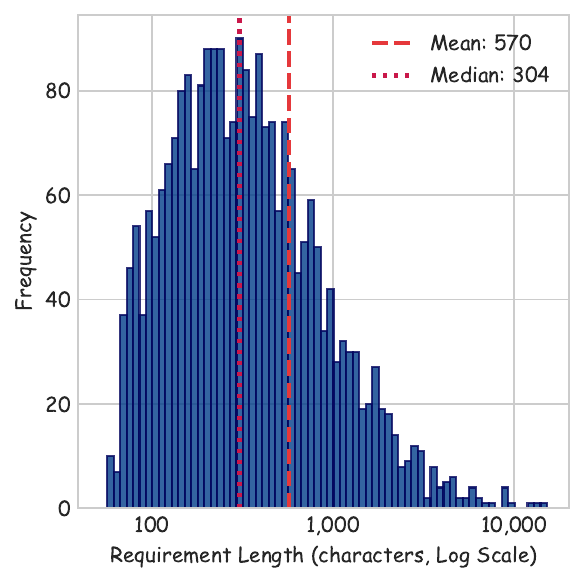}
  \caption{Log-Scale Requirement Length Histogram}\vspace{-5pt}
  \label{fig:figures/figures_testset/figures_requirements/003_requirement_length_distribution_log_scale.pdf}
  \vspace{-5pt}
\end{figure}
\begin{figure}[t!]
  \centering
  \includegraphics[width=0.6\columnwidth]{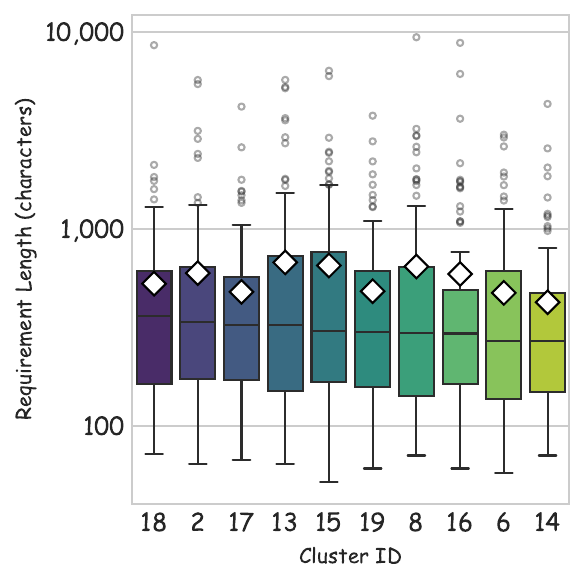}
  \caption{Requirement Length Comparison by Cluster}\vspace{-5pt}
  \label{fig:figures/figures_testset/figures_requirements/004_top10_clusters_requirement_length_distribution.pdf}
  \vspace{-25pt}
\end{figure}

\paragraph{Multimodal Scoring}
The quality of the generation is quantified by a comprehensive scoring function by aggregating scores from multiple assessment modalities:
\begin{MiddleEquation}\begin{equation}
    \text{Score}(\mathcal{I}, \mathcal{C}, \mathcal{V}, \mathcal{S}) = \sum_{1 \le k \le 3} w_k \cdot \phi_k
\end{equation}\end{MiddleEquation}where $\phi_{\text{1}}(\mathcal{I}, \mathcal{C})$, $\phi_{\text{2}}(\mathcal{I}, \mathcal{S})$, and $\phi_{\text{3}}(\mathcal{I}, \mathcal{V})$ represent modality-specific assessment functions for code, static visuals, and dynamic gameplay, respectively, and $w_k$ are their corresponding weights satisfying $\sum_k w_k = 1$.

\paragraph{Score Distribution Metrics}
To provide granular insights into model performance patterns, we categorize each game's final score into four quality bands. \textbf{Excellent (80-100)}: Games demonstrating superior implementation quality with minimal issues. \textbf{Good (60-80)}: Games with solid functionality and minor deficiencies. \textbf{Fair (40-60)}: Games with basic functionality but notable limitations. \textbf{Poor (0-40)}: Games with significant implementation failures or non-functional code.


\definecolor{HeaderColor}{HTML}{4A5568}        
\definecolor{ProprietaryColor}{HTML}{E3F2FD}   
\definecolor{Size400PlusColor}{HTML}{FFEBEE}   
\definecolor{Size100To400Color}{HTML}{FFF3E0}  
\definecolor{Size30To100Color}{HTML}{F9FBE7}   
\definecolor{Size10To30Color}{HTML}{E8F5E8}    
\definecolor{SizeBelow10Color}{HTML}{F3E5F5}   

\subsection{\benchmark{} Construction}

\paragraph{Data Collection}
Our raw data is sourced from two extensive, publicly available code corpora: OpenCoder~\cite{huang2025opencoder} and The Stack v2~\cite{lozhkov2024starcoder2stackv2}. To construct a domain-specific dataset, we engineered a high-throughput filtering pipeline. This pipeline leverages parallel processing to stream and analyze all Python source files, systematically isolating code that explicitly contains the ``pygame'' keyword. This procedure allowed us to efficiently distill a targeted corpus of Pygame-related projects from the broader, more generalized repositories, forming the foundation for all subsequent curation steps.

\paragraph{Clustering-based Curation}
To ensure the resulting dataset exhibits both high quality and functional diversity, we implemented a rigorous curation strategy that can be described as a formalized selection principle. The process first partitions the entire corpus based on a high-dimensional feature representation and then selects the most structurally complete program from each partition.

Let \( D = \{c_1, c_2, \dots, c_n\} \) be the initial corpus of code samples. Let \( \bm{v}(c) \in \mathbb{R}^d \) be the high-dimensional feature vector extraction function described previously, which maps a code sample \( c \) to its quantitative fingerprint (encompassing size, structure, API usage, and semantics). Let \( S_{\text{quality}}(c) \) be the scalar heuristic score that evaluates the structural completeness of a program.

The corpus \( D \) is first partitioned into \( k \) clusters, \( C = \{C_1, C_2, \dots, C_k\} \), using the \( \texttt{MiniBatchKMeans} \) algorithm on the feature vectors \( \bm{v}(c) \). The final curated seed dataset, \( D_{\text{seed}} \), is then constructed by selecting the single element from each cluster that maximizes the quality score \( S_{\text{quality}} \):
\begin{MiddleEquation}\begin{equation}
D_{\text{seed}} = \bigcup_{i=1}^{k} \left\{ \underset{c \in C_i}{\arg\max} \, S_{\text{quality}}(c) \right\}
\label{eq:selection_principle}
\end{equation}\end{MiddleEquation}where the clusters \(C_i\) are the result of \(\texttt{MiniBatchKMeans}(D, \bm{v}, k)\).

\begin{table}[t!]
\centering
\resizebox{0.85\columnwidth}{!}{%
\begin{tabular}{@{}lll@{}}
\toprule
\textbf{Category} & \textbf{Metric} & \textbf{Value} \\ 
\midrule
\multicolumn{3}{l}{\textit{\textbf{General Statistics}}} \\
& Total Samples (Unique Games) & 2,219 \\
& Unique Source Repositories & 2,190 \\
& Unique Clusters & 100 \\
\midrule
\multicolumn{3}{l}{\textit{\textbf{Requirement Metrics (Average per game)}}} \\
& Requirement Length (characters) & 1,210 \\
& Word Count & 178 \\
& Number of Sentences & 9.6 \\
\midrule

\multicolumn{3}{l}{\textit{\textbf{Reference Code Metrics (Average per game)}}} \\
& Lines of Code & 257 \\
& Code Length (characters) & 8,533 \\
& Number of Functions & 2.8 \\
& Number of Classes & 2.4 \\
\midrule

\multicolumn{3}{l}{\textit{\textbf{Execution \& Quality Metrics}}} \\
& Execution Success Rate & 100.0\% \\
& Video Coverage & 100.0\% \\
& Average Images per Game & 9.9 \\
& Average Recording Duration & 10.0 s \\
\bottomrule
\end{tabular}%
}
\caption{Overall Statistics of the \benchmark{} Dataset.}
\label{tab:dataset_overview}
\vspace{-20pt}
\end{table}

This selection principle, articulated in Equation \ref{eq:selection_principle}, formally captures our two-stage methodology. The clustering operation partitions the dataset based on functional and structural similarity (as defined by \( \bm{v} \)), thereby ensuring diversity. The subsequent \( \arg\max \) operation within each disjoint set \( C_i \) guarantees that the selected program is the most complete and runnable exemplar of that particular functional group, based on the quality heuristic:
\begin{MiddleEquation}\begin{equation}
S_{\text{quality}}(c) = \sum_{f \in \mathcal{C}_{\text{struct}}} w_f \cdot \mathbb{I}(f \in c) + S_{\text{len}}(L(c))
\nonumber
\end{equation}\end{MiddleEquation}
This decoupling of the clustering metric from the selection metric is intentional. It allows us to group programs by a rich definition of functional behavior (\( \bm{v} \)) while applying a simpler, targeted heuristic (\( S_{\text{quality}} \)) to ensure each chosen sample meets a minimum standard of structural integrity.

\begin{table*}[h!]
  \centering
  \small
  \renewcommand{\arraystretch}{1.0}
  \vspace{-30pt}
  \resizebox{0.85\textwidth}{!}{%
    \begin{tabular}{lr|c|ccc|cccc}
      \toprule
      Model & Size & Final & Code & Image & Video & Excellent & Good & Fair & Poor \\
      \midrule
      \rowcolor{HeaderColor}\multicolumn{10}{c}{\textcolor{white}{\textbf{Proprietary LLMs}}} \\
      \midrule
      \rowcolor{ProprietaryColor} \small gpt-5 & \faLock{} & \textbf{45.0} & \underline{96.6} & 17.6 & 20.7 & \textbf{83} & 288 & 676 & 1172 \\
      \rowcolor{ProprietaryColor} \small o3 & \faLock{} & \underline{44.8} & 92.3 & \textbf{20.2} & 21.9 & \underline{65} & \textbf{341} & \textbf{686} & 1127\\
      \rowcolor{ProprietaryColor} \small gpt-5-mini & \faLock{} & 43.5 & \textbf{96.7} & 15.7 & 18.0 & 61 & 236 & 655 & 1267 \\
      \rowcolor{ProprietaryColor} \small gemini-2.5-pro & \faLock{} & 43.5 & 89.1 & 19.1 & \underline{22.2} & 45 & \underline{337} & 672 & 1165  \\
      \rowcolor{ProprietaryColor} \small o4-mini & \faLock{} & 43.0 & 87.8 & 19.8 & 21.4 & 36 & 313 & \underline{682} & 1188 \\
      \rowcolor{ProprietaryColor} \small gpt-4.1-2025-04-14 & \faLock{} & 42.5 & 91.8 & 17.6 & 18.1 & 47 & 263 & 641 & 1268 \\
      \rowcolor{ProprietaryColor} \small grok-4 & \faLock{} & 42.0 & 83.9 & 19.8 & \textbf{22.4} & 21 & 327 & 650 & 1221 \\
      \rowcolor{ProprietaryColor} \small gemini-2.5-flash & \faLock{} & 42.0 & 92.8 & 16.5 & 16.7 & 28 & 252 & 634 & 1304 \\
      \rowcolor{ProprietaryColor} \small chatgpt-4o-latest & \faLock{} & 41.2 & 82.5 & \underline{19.9} & 21.3 & 25 & 305 & 613 & 1276 \\
      \rowcolor{ProprietaryColor} \small claude-sonnet-4-20250514-thinking & \faLock{} & 40.5 & 90.3 & 14.4 & 16.9 & 36 & 204 & 624 & 1355 \\
      \rowcolor{ProprietaryColor} \small claude-sonnet-4-20250514 & \faLock{} & 40.2 & 87.7 & 15.7 & 17.4 & 36 & 207 & 616 & 1360 \\
      \rowcolor{ProprietaryColor} \small o3-mini-2025-01-31 & \faLock{} & 38.2 & 89.3 & 11.9 & 13.3 & 26 & 204 & 417 & \underline{1572} \\
      \rowcolor{ProprietaryColor} \small gpt-4o-mini-2024-07-18 & \faLock{} & 33.9 & 70.4 & 15.5 & 15.8 & 4 & 134 & 459 & \textbf{1622} \\
      \midrule
      \rowcolor{HeaderColor}\multicolumn{10}{c}{\textcolor{white}{\textbf{400B+ Open-Weight LLMs}}} \\
      \midrule
      \rowcolor{Size400PlusColor} \small Qwen3-Coder-480B-A35B-Instruct & 32B/480B & \textbf{41.3} & \underline{85.3} & 18.3 & \underline{20.5} & 20 & 287 & \textbf{644} & 1268 \\
      \rowcolor{Size400PlusColor} \small DeepSeek-V3-0324 & 37B/671B & \underline{41.1} & 83.6 & \underline{19.3} & \textbf{20.5} & 22 & \textbf{311} & \underline{638} & 1248 \\
      \rowcolor{Size400PlusColor} \small DeepSeek-V3.1 & 37B/671B & 40.9 & 83.1 & \textbf{19.3} & 20.2 & \underline{25} & \underline{296} & 611 & 1287 \\
      \rowcolor{Size400PlusColor} \small DeepSeek-R1-0528 & 37B/671B & 38.7 & \textbf{88.1} & 13.4 & 14.6 & \textbf{32} & 174 & 544 & \underline{1469} \\
      \rowcolor{Size400PlusColor} \small kimi-k2-0905-preview & 32B/1000B & 23.5 & 66.3 & 2.0 & 2.2 & 0 & 18 & 62 & \textbf{2135} \\
      \midrule
      \rowcolor{HeaderColor}\multicolumn{10}{c}{\textcolor{white}{\textbf{100B-400B Open-Weight LLMs}}} \\
      \midrule
      \rowcolor{Size100To400Color} \small Qwen3-235B-A22B-Thinking-2507 & 22B/235B & \underline{42.3} & 84.5 & \textbf{20.0} & \underline{22.4} & 22 & \underline{322} & \textbf{695} & 1180 \\
      \rowcolor{Size100To400Color} \small Qwen3-235B-A22B & 235B & 41.2 & 81.3 & \underline{19.8} & \textbf{22.6} & 14 & 302 & \underline{668} & 1235 \\
      \rowcolor{Size100To400Color} \small Qwen3-235B-A22B-Instruct-2507 & 22B/235B & 41.1 & 85.3 & 18.2 & 19.7 & 16 & 308 & 589 & 1306 \\
      \rowcolor{Size100To400Color} \small GLM-4.5 & 32B/355B & 40.0 & 84.7 & 17.0 & 18.3 & \underline{31} & 216 & 661 & \underline{1311} \\
      \rowcolor{Size100To400Color} \small GLM-4.5-Air & 12B/106B & 39.4 & \underline{85.4} & 16.3 & 16.5 & 23 & 230 & 533 & \textbf{1433} \\
      \rowcolor{Size100To400Color} \small gpt-oss-120b & 5.1B/117B & \textbf{43.4} & \textbf{90.1} & 19.7 & 20.3 & \textbf{52} & \textbf{324} & 634 & 1209  \\
      \midrule
      \rowcolor{HeaderColor}\multicolumn{10}{c}{\textcolor{white}{\textbf{30B-100B Open-Weight LLMs}}} \\
      \midrule
      \rowcolor{Size30To100Color} \small Qwen3-32B & 32B & \textbf{40.4} & 81.6 & 18.9 & \underline{20.6} & 8 & \underline{274} & \textbf{642} & 1295 \\
      \rowcolor{Size30To100Color} \small Seed-OSS-36B-Instruct & 36B & \underline{40.3} & \textbf{88.3} & 16.4 & 16.2 & \textbf{25} & 234 & 585 & 1375 \\
      \rowcolor{Size30To100Color} \small Qwen3-30B-A3B-Thinking-2507 & 3B/30B & 40.0 & 80.7 & \underline{18.9} & 20.4 & 13 & \textbf{279} & 589 & 1338 \\
      \rowcolor{Size30To100Color} \small QwQ-32B & 32B & 39.6 & 79.7 & 18.5 & 20.6 & 10 & 268 & \underline{610} & 1331 \\
      \rowcolor{Size30To100Color} \small Qwen3-30B-A3B & 3B/30B & 39.6 & 78.4 & \textbf{19.7} & \textbf{20.7} & 9 & 274 & 597 & 1339 \\
      \rowcolor{Size30To100Color} \small Qwen3-Coder-30B-A3B-Instruct & 3B/30B & 39.0 & \underline{83.8} & 16.6 & 16.7 & \underline{22} & 226 & 543 & 1428 \\
      \rowcolor{Size30To100Color} \small Qwen3-30B-A3B-Instruct-2507 & 30B & 38.6 & 81.4 & 16.5 & 17.8 & 11 & 223 & 571 & 1414 \\
      \rowcolor{Size30To100Color} \small DeepSeek-R1-Distill-Llama-70B & 70B & 35.3 & 74.1 & 15.8 & 16.0 & 4 & 188 & 448 & 1579 \\
      \rowcolor{Size30To100Color} \small Qwen2.5-72B-Instruct & 72B & 34.6 & 73.2 & 14.7 & 15.9 & 3 & 174 & 449 & 1593 \\
      \rowcolor{Size30To100Color} \small Qwen2.5-Coder-32B-Instruct & 32B & 34.4 & 74.5 & 13.8 & 14.9 & 9 & 167 & 425 & 1618 \\
      \rowcolor{Size30To100Color} \small DeepSeek-R1-Distill-Qwen-32B & 32B & 33.4 & 71.9 & 14.4 & 13.9 & 0 & 145 & 411 & \underline{1663} \\
      \rowcolor{Size30To100Color} \small Qwen2.5-32B-Instruct & 32B & 31.8 & 66.4 & 14.0 & 15.1 & 2 & 127 & 389 & \textbf{1701} \\
      \midrule
      \rowcolor{HeaderColor}\multicolumn{10}{c}{\textcolor{white}{\textbf{10B-30B Open-Weight LLMs}}} \\
      \midrule
      \rowcolor{Size10To30Color} \small gpt-oss-20b & 3.6B/21B & \textbf{42.2} & \textbf{88.8} & \textbf{18.6} & \textbf{19.2} & \textbf{31} & \textbf{299} & \textbf{626} & 1263 \\
      \rowcolor{Size10To30Color} \small Qwen3-14B & 14B & \underline{38.8} & \underline{79.1} & \underline{18.4} & \underline{18.8} & \underline{9} & \underline{245} & \underline{567} & 1398 \\
      \rowcolor{Size10To30Color} \small Qwen2.5-Coder-14B-Instruct & 14B & 30.2 & 68.5 & 10.9 & 11.2 & 0 & 87 & 327 & \underline{1804} \\
      \rowcolor{Size10To30Color} \small DeepSeek-R1-Distill-Qwen-14B & 14B & 27.4 & 65.3 & 8.7 & 8.3 & 1 & 77 & 198 & \textbf{1943} \\
      \midrule
      \rowcolor{HeaderColor}\multicolumn{10}{c}{\textcolor{white}{\textbf{Below 10B Open-Weight LLMs}}} \\
      \midrule
      \rowcolor{SizeBelow10Color} \small Qwen3-8B & 8B & \textbf{36.9} & \textbf{76.2} & \textbf{17.2} & \textbf{17.3} & \textbf{5} & \textbf{187} & \textbf{546} & 1480 \\
      \rowcolor{SizeBelow10Color} \small Qwen3-4B & 4B & \underline{34.4} & 72.7 & 15.1 & 15.5 & 1 & 144 & 464 & 1610 \\
      \rowcolor{SizeBelow10Color} \small Qwen3-4B-Thinking-2507 & 4B & 34.3 & 70.0 & \underline{16.1} & \underline{16.8} & 2 & \underline{168} & \underline{465} & 1584 \\
      \rowcolor{SizeBelow10Color} \small Seed-Coder-8B-Instruct & 8B & 33.9 & \underline{73.2} & 14.0 & 14.4 & \underline{4} & 137 & 422 & 1656 \\
      \rowcolor{SizeBelow10Color} \small Llama-3.1-8B-Instruct & 8B & 29.4 & 62.9 & 13.0 & 12.3 & 1 & 84 & 319 & 1815 \\
      \rowcolor{SizeBelow10Color} \small Qwen2.5-Coder-7B-Instruct & 7B & 27.6 & 63.9 & 9.1 & 9.7 & 0 & 69 & 250 & 1899 \\
      \rowcolor{SizeBelow10Color} \small Qwen3-1.7B & 1.7B & 25.0 & 57.3 & 9.1 & 8.7 & 0 & 43 & 216 & 1959 \\
      \rowcolor{SizeBelow10Color} \small Llama-3.2-3B-Instruct & 3B & 24.5 & 55.5 & 9.5 & 8.5 & 0 & 46 & 210 & 1963 \\
      \rowcolor{SizeBelow10Color} \small deepseek-coder-7b-instruct-v1.5 & 7B & 24.0 & 53.8 & 9.0 & 9.2 & 0 & 38 & 229 & 1952 \\
      \rowcolor{SizeBelow10Color} \small Hunyuan-7B-Instruct & 7B & 21.0 & 57.8 & 2.8 & 2.2 & 0 & 12 & 59 & 2148 \\
      \rowcolor{SizeBelow10Color} \small Qwen2.5-Coder-3B-Instruct & 3B & 20.4 & 53.8 & 4.4 & 2.9 & 0 & 10 & 85 & 2124 \\
      \rowcolor{SizeBelow10Color} \small OpenCoder-8B-Instruct & 8B & 20.1 & 49.0 & 6.5 & 4.7 & 0 & 7 & 152 & 2060 \\
      \rowcolor{SizeBelow10Color} \small Llama-3.2-1B-Instruct & 1B & 15.5 & 40.3 & 4.3 & 1.9 & 0 & 4 & 57 & 2158 \\
      \rowcolor{SizeBelow10Color} \small DeepSeek-R1-Distill-Llama-8B & 8B & 15.1 & 42.5 & 1.7 & 1.0 & 0 & 6 & 26 & 2187 \\
      \rowcolor{SizeBelow10Color} \small Qwen3-0.6B & 0.6B & 13.7 & 35.1 & 4.8 & 1.1 & 0 & 3 & 56 & 2160 \\
      \rowcolor{SizeBelow10Color} \small Qwen2.5-Coder-0.5B-Instruct & 0.5B & 12.8 & 34.6 & 3.6 & 0.0 & 0 & 0 & 39 & 2179 \\
      \rowcolor{SizeBelow10Color} \small DeepSeek-R1-Distill-Qwen-7B & 7B & 12.1 & 35.8 & 0.3 & 0.1 & 0 & 0 & 3 & \underline{2216} \\
      \rowcolor{SizeBelow10Color} \small DeepSeek-R1-Distill-Qwen-1.5B & 1.5B & 8.5 & 25.4 & 0.0 & 0.0 & 0 & 0 & 1 & \textbf{2218} \\
      \bottomrule
    \end{tabular}%
  }%
  \caption{Comprehensive Performance Evaluation, showing Final Score, Code Score, Image Score, Video Score, and score distribution. \textbf{Bold} indicate highest performance; \underline{underlined} indicate second-highest performance.}
  \label{tab:combined_results}
  \vspace{-20pt}
\end{table*}

\paragraph{Test Set Construction}
The seed dataset is then processed through an automated Language Model (Claude-Sonnet-4)-driven pipeline to construct the final test set, composed of (requirement, code) instruction pairs. This pipeline operationalizes a closed-loop ``analyze-inject-validate-generate'' workflow. The process commences with an \textbf{Intent Analysis} stage, where the LLM parses the seed code to infer its core game mechanics and objectives. This is followed by a \textbf{Autonomous Interactive Behavior Injection} stage, which refactors the original, often interactive, code into a self-contained, autonomous demonstration that executes for a fixed duration. The transformed artifact then undergoes \textbf{Execution Verification} within a sandboxed environment. Any execution failures initiate a \textbf{Self-Correction} loop, wherein the error logs are fed back to the LLM for automated debugging and regeneration. Upon successful validation, \textbf{Requirement Generation} module prompts the LLM to synthesize a high-level, natural language requirement specification for the program, emulating the perspective of a product manager. This rigorous process ensures that every entry in the final test set is correct, executable, and paired with a corresponding high-level description.

\paragraph{Human Check and Annotation}
To ensure the quality of \benchmark{}, 8 graduate students used a UI sandbox and LLM assistance to verify nearly 2,219 Pygame code cases and their visual outputs.

\subsection{Data Statistics Overview}

\paragraph{Data Statistics} Table~\ref{tab:dataset_overview} presents a comprehensive statistical overview of the \benchmark{} dataset. The benchmark is substantial in scale, comprising 2,219 unique games sourced from 2,190 distinct repositories and organized into 100 thematic clusters. The complexity of the tasks is reflected in the metrics for both the natural language requirements and the reference code; on average, each game's requirements consist of 178 words, while the corresponding reference code implementation spans 257 lines. Critically, the dataset's high quality is underscored by a 100\% execution success rate and complete video coverage for all samples, ensuring its reliability for evaluation purposes.

\paragraph{Analysis of \benchmark{} Requirements} Analysis of the requirement texts reveals a highly right-skewed length distribution, as visualized in the violin plot and histogram (Figures~\ref{fig:figures/figures_testset/figures_requirements/001_requirement_length_distribution_density.pdf},~\ref{fig:figures/figures_testset/figures_requirements/002_requirement_length_distribution_linear_scale.pdf}). This distribution is characterized by a preponderance of concise specifications, evidenced by a mean length (570) substantially exceeding the median (297), with 80\% of texts falling under 1000 characters. On a logarithmic scale, the distribution approximates a log-normal form (Figure~\ref{fig:figures/figures_testset/figures_requirements/003_requirement_length_distribution_log_scale.pdf}). Crucially, this length variation correlates with task type. A box plot comparison across the top 10 requirement clusters (Figure~\ref{fig:figures/figures_testset/figures_requirements/004_top10_clusters_requirement_length_distribution.pdf}) demonstrates significant heterogeneity, suggesting that the clustering effectively segments tasks by their underlying complexity or documentation style.

\paragraph{Scaling Law of Visual Game Generation}
Figure~\ref{fig:figures/figures_evals/12_model_size_vs_performance.pdf}, the results reveal a statistically observable positive correlation between the count of model parameters and the performance of the task. The models with smaller parameters (0.5B-3B) exhibit consistently lower performance metrics, typically achieving fewer than 400 solved games, while intermediate-scale models (7B-32B parameters) demonstrate moderate performance ranges of 200-600 solved games. Large-scale models (70B+ parameters) achieve superior performance outcomes, with solved game counts reaching 600-1000+. However, the data suggests that the relationship exhibits logarithmic characteristics rather than linear scaling, indicating diminishing marginal returns as parameter count increases. Significantly, the LLMs (e.g. gpt-oss-20b, Seed-OSS-36B-Instruct, and Qwen3-Coder-30B-A3B) with similar parameters suggest that architectural innovations, training methodologies, and algorithmic optimizations may constitute equally critical factors in achieving state-of-the-art performance. We can obtain the formula for model size and performance as: $M = A * \log (N) + B$, where $M$ is the number of the resolved problems and $N$ is the number of model parameters ($A = 127.2, B=135.6$).

\section{Experiment Setup}

\paragraph{Experiment Code LLMs} We evaluate all LLMs on Ubuntu 22.04, equipped with an Intel Xeon (R) Gold 6348 CPU @2.60GHz, eight NVIDIA H800 GPUs (80 GB), and 528 GB of memory. The software setup includes NVIDIA-SMI version 535.104.05 and CUDA 12.3. We set temperature to $0.0$ for LLMs when inferring by sglang v0.5.1~\cite{sglang}.


\paragraph{Evaluated Models} For a comprehensive and thorough evaluation, we assess 70 widely used models, including both proprietary and open-source ones. For proprietary models, we evaluate series from leading labs such as OpenAI's GPT (e.g., gpt-5)~\cite{gpt4, gpt45} and reasoning models (o3, o4-mini)~\cite{o3-and-o4-mini}, Anthropic's claude-sonnet-4~\cite{claude4}, Google's gemini-2.5 series~\cite{gemini2.5-pro, gemini2.5-flash}, and xAI's grok-4~\cite{grok4}. For open-source models, our testing spans a diverse range from major tech companies. This includes the extensive Qwen family~\cite{qwen25coder, qwq-32b, qwen3}, ByteDance's Seed series~\cite{seed-oss, seedcoder}, Moonshot AI's Kimi-K2~\cite{kimi_k2}, various DeepSeek models~\cite{deepseekv3, deepseekr1}, Meta's Llama series (Llama-3.1, Llama-3.2)~\cite{llama3} and Zhipu AI's GLM models~\cite{glm45}. The evaluation also incorporates community and research-driven models like OpenAI's gpt-oss~\cite{gpt-oss} and OpenCoder~\cite{huang2025opencoder}.

\paragraph{Judge Models} We employ an LLM-as-Judge using \textbf{\texttt{Qwen3-Coder-480B-A35B-Instruct}} to evaluate code scores and \textbf{\texttt{Qwen2.5-VL-72B}} for image/video scores.

\section{Analysis}

\paragraph{Main Result} Note that the sum of distribution counts may not equal 2,219 for some models due to execution failures that prevent complete end-to-end evaluation pipeline completion. In Table~\ref{tab:combined_results}, several important trends can be observed. \textbf{Clear Performance Hierarchy} Proprietary models generally lead, with GPT-5 topping the list at 45.0 points. Among open-source models, large-parameter models (400B+) perform best, such as Qwen3-Coder-480B and the DeepSeek-V3 series, both exceeding 40 points. \textbf{Imbalanced Capability Dimensions} All models perform strongly in code generation (most over 70 points) but are generally weaker in image and video evaluation (most under 25 points), indicating that current models still have significant room for improvement in visual representation and dynamic effect generation. \textbf{Pronounced Scale Effect} Open-source models exhibit a clear positive correlation between scale and performance, improving from an average of around 20 points for models under 10B to over 40 points for 400B+ models. \textbf{Long-Tail Distribution of Quality} Most generated games fall into the ``Poor'' and ``Fair'' categories, with limited samples reaching the ``Excellent'' standard, reflecting that high-quality game code generation remains challenging.

\begin{figure}[t!]
\centering
\includegraphics[width=0.9\columnwidth]{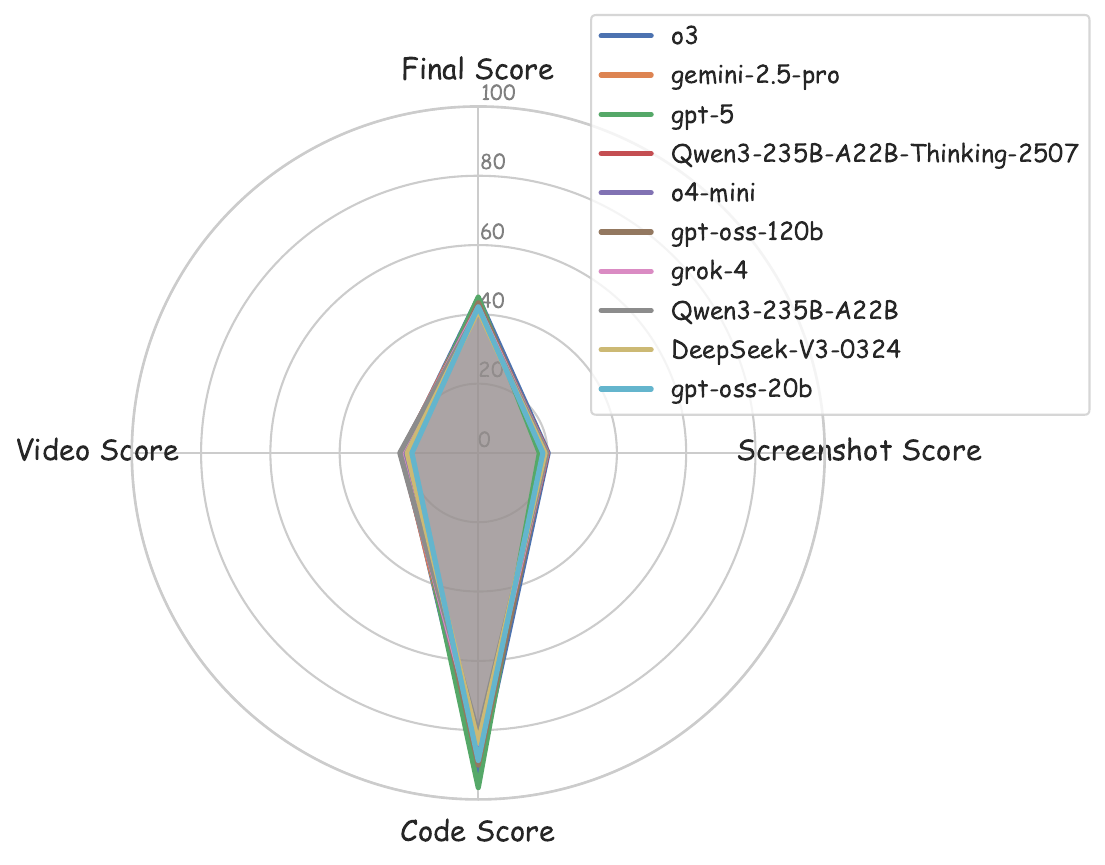}
\caption{Radar chart comparing the top 10 models across four key performance dimensions.}
\label{fig:figures/figures_evals/03_top_model_radar_chart.pdf}
\vspace{-15pt}
\end{figure}

\paragraph{Multi-Dimensional Capability Analysis of Top-Performing Models} Figure~\ref{fig:figures/figures_evals/03_top_model_radar_chart.pdf} highlights distinct model capabilities. A clear code-visual trade-off exists: GPT-5 excels at code (96.6) but is weak in vision (17.6/20.7), while o3 is more balanced and leads in image score (20.2). Notably, open-source models like gpt-oss-120b now rival proprietary systems in game development, narrowing the capability gap.

\paragraph{Evaluation Score Distribution Across Task Dimensions}

\begin{figure}[t!]
  \centering
  \includegraphics[width=0.9\columnwidth]{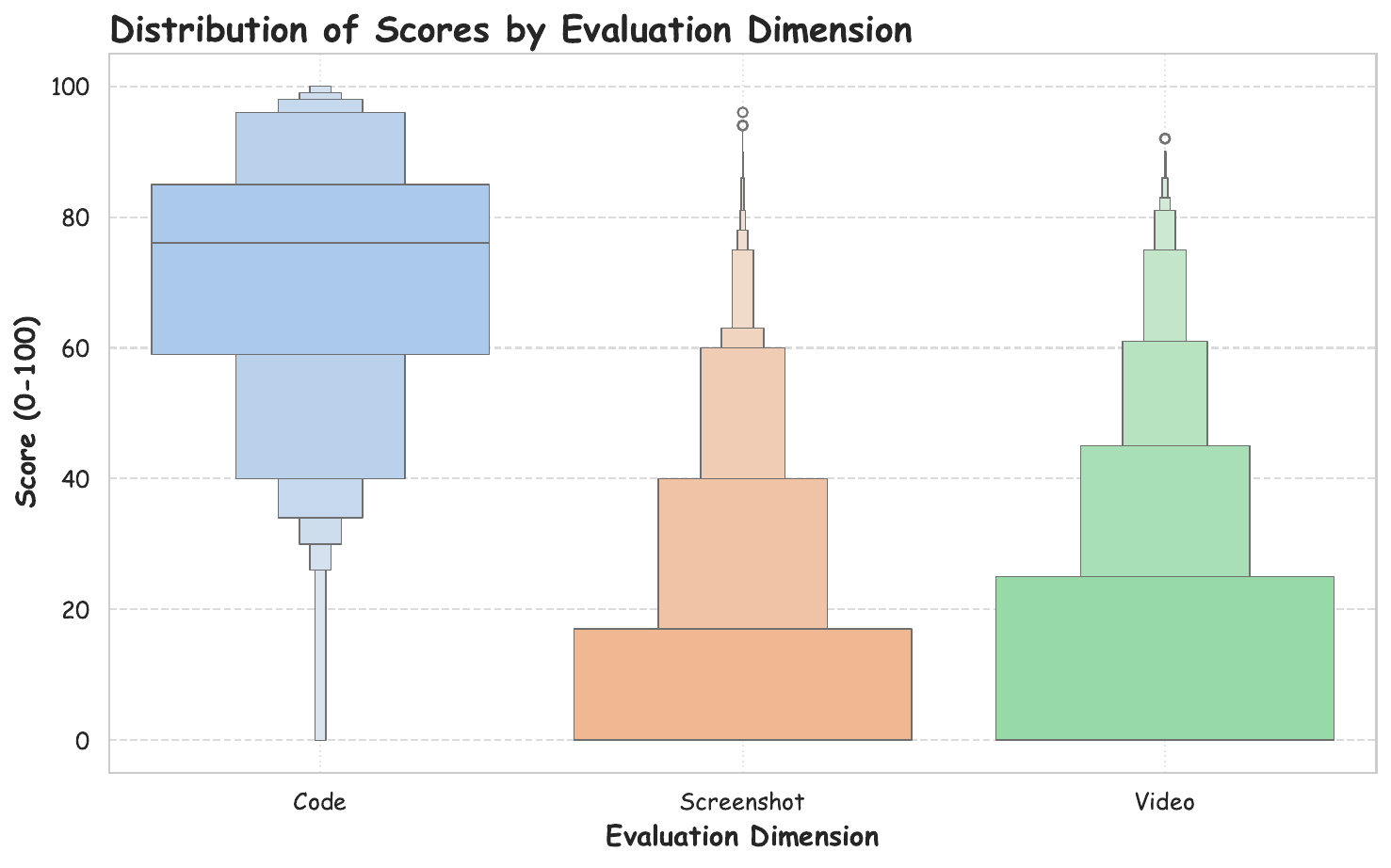}
  \caption{Distribution of evaluation scores across three key dimensions: Code, Screenshot, and Video.}
  \label{fig:figures/figures_evals/13_eval_score_distribution.pdf}
  \vspace{-15pt}
\end{figure}

Figure~\ref{fig:figures/figures_evals/13_eval_score_distribution.pdf} shows a clear hierarchy in AI capabilities for game development. Models excel at code evaluation, achieving high and varied scores, but struggle significantly with visual tasks like screenshot and video evaluations. The consistently low scores in these visual areas highlight a major weakness in the models' visual understanding and generation abilities.


\paragraph{Game Difficulty Distribution}

\begin{figure}[t!]
  \centering
  \includegraphics[width=1.0\columnwidth]{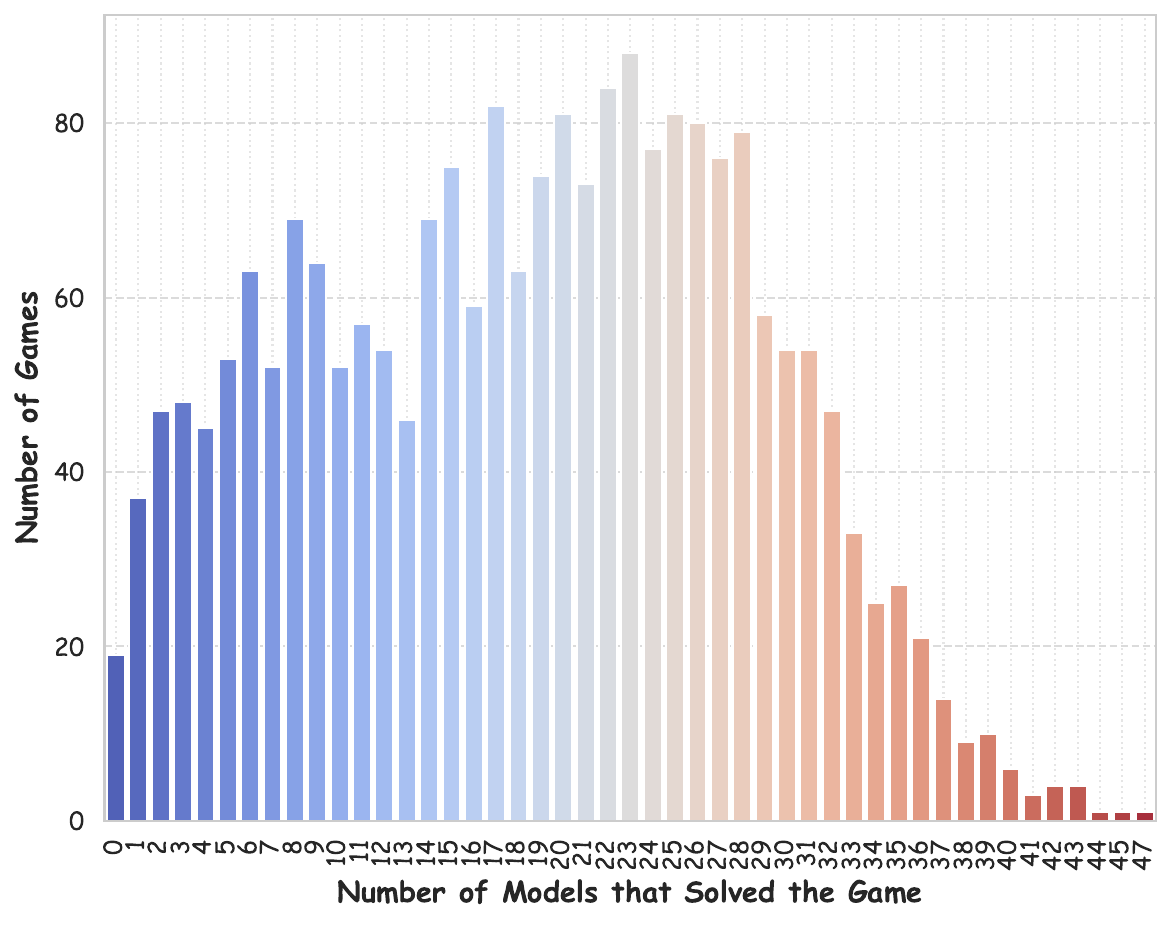}
  \caption{Game difficulty distribution by number of solving models.}\vspace{-10pt}
  \label{fig:figures/figures_evals/04_game_difficulty_distribution.pdf}
  \vspace{-15pt}
\end{figure}

Figure~\ref{fig:figures/figures_evals/04_game_difficulty_distribution.pdf} presents a typical right-skewed distribution, with most games concentrated in the low solution rate range (where only a few models can solve them), indicating that the games in the test set are generally difficult. The peak on the left shows a considerable number of games that no model could solve, while the number of simple games that could be solved by most models is small. This distribution is beneficial for distinguishing the capability differences among various models.

\paragraph{Performance Analysis Across Game Difficulty Tiers}

\begin{figure}[t!]
  \centering
  \includegraphics[width=1.0\columnwidth]{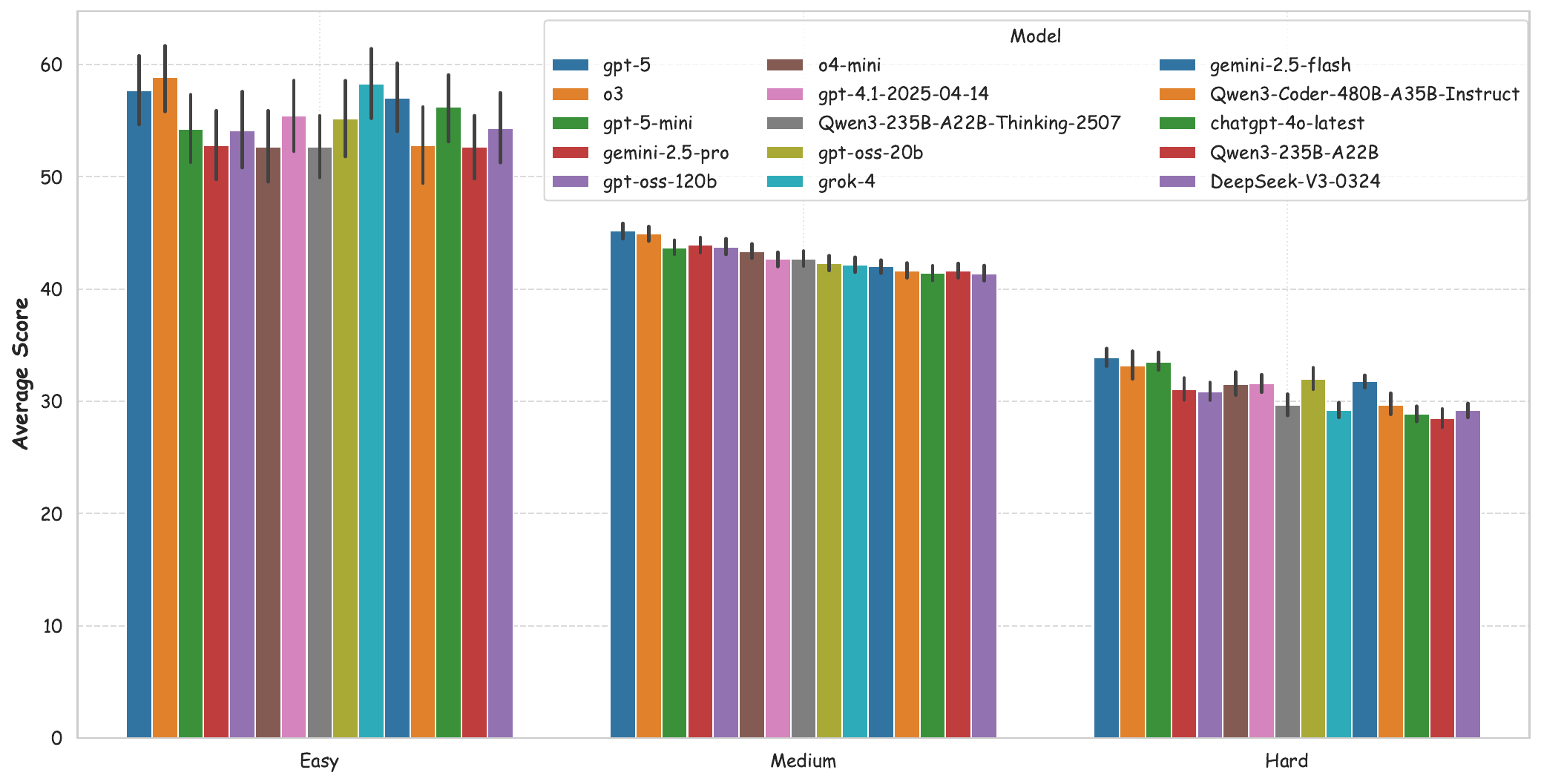}
  \caption{Performance comparison of top 15 models across Easy, Medium, and Hard difficulty tiers, showing consistent ranking patterns and scaling challenges.}
  \label{fig:figures/figures_evals/07_performance_by_difficulty.pdf}
  \vspace{-15pt}
\end{figure}

In Figure~\ref{fig:figures/figures_evals/07_performance_by_difficulty.pdf}, the difficulty tier analysis reveals that while all models experience performance degradation on harder games, the relative ranking between top-tier models remains remarkably stable across difficulty levels. This consistency validates the benchmark's discriminative power and suggests that superior models maintain their advantages regardless of task complexity.

\paragraph{Evaluation Dimension Correlations}

\begin{figure}[t!]
  \centering
  \includegraphics[width=0.9\columnwidth]{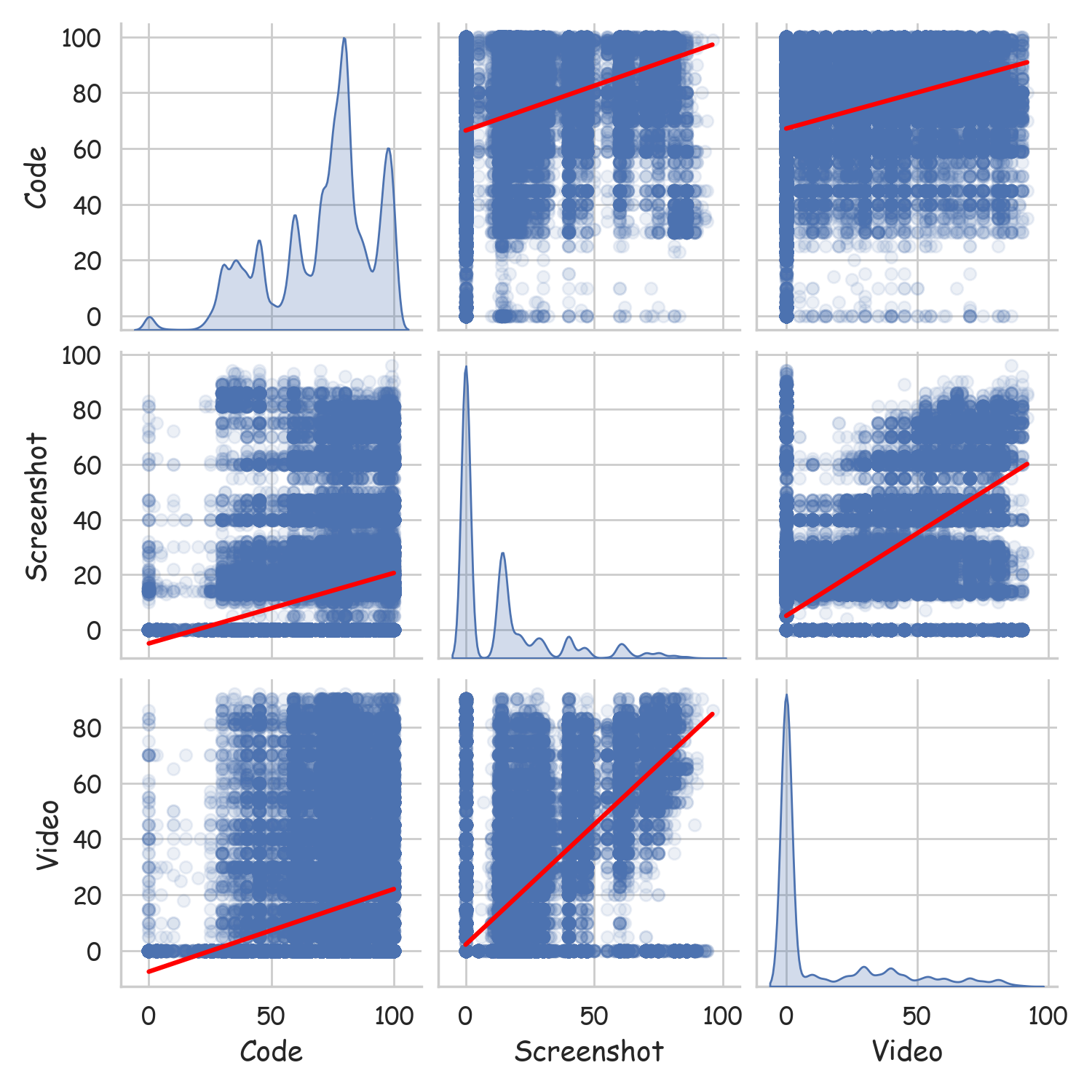}
  \caption{Correlation matrix between Code, Screenshot, and Video evaluation dimensions, demonstrating the interdependence of multimodal capabilities in game development.}
  \label{fig:figures/figures_evals/09_capability_correlation_matrix.png}
  \vspace{-15pt}
\end{figure}

In Figure~\ref{fig:figures/figures_evals/09_capability_correlation_matrix.png}, the correlation analysis shows moderate to strong positive correlations between all evaluation dimensions, indicating that models with superior code generation capabilities tend to also excel in visual assessment tasks. This suggests that game development requires integrated multimodal understanding rather than isolated technical skills.

\paragraph{Overall Correlation vs. Top-Tier Specialization} While this positive correlation holds true across the entire model population, a more nuanced picture emerges when examining the elite models, as highlighted in Figure~\ref{fig:figures/figures_evals/03_top_model_radar_chart.pdf}. For instance, models like GPT-5 demonstrate a specialization, achieving near-perfect code scores at the expense of comparatively lower visual scores, suggesting a potential ``capability trade-off'' at the frontier of performance. This indicates that while foundational capabilities are interconnected, advanced models may adopt different strategies to allocate their ``reasoning budget'', prioritizing either logical code structure or visual aesthetics.

\section{Related Work}
\paragraph{Code Large Language Models} Code-specific large language models (Code LLMs) \cite{starcoder,code_llama,guo2024deepseekcoder,codearena,execrepobench} demonstrate remarkable performance in software engineering and agentic tasks, with foundational models like Qwen2.5/3-Coder~\citep{qwen25coder}, Seed-Coder~\cite{code_llama}, GLM-4.5~\cite{glm45}, and Kimi-K2~\cite{kimi_k2} excelling in general code generation and understanding. The success of multi-agent collaboration \cite{multi_agents_survey,autonomous_agents_survey} inspires the use of a language-specific agent to formulate a multilingual instruction dataset. Subsequently, instruction tuning \cite{instructGPT,llama_adapter,self_instructions} enhances the ability of the LLMs to generalize and follow instructions~\cite{self_instructions,codealpaca,wizardcoder,magicoder,wavecoder}. A series of code benchmarks is proposed to evaluate different aspects of the code LLMs, including realistic~\citep{fullstack,bigcodebench,adc} and multilingual scenarios~\cite{multipl_e,mceval,mdeval,LiveRepoReflection}.

\paragraph{Game for Large Language Models}
The intersection of games and large language models (LLMs) has emerged as a rich area of research encompassing multiple paradigms and applications. Early works established the potential of using game environments as training grounds for LLMs, and then they extended to more complex games (e.g. minecraft~\citep{mindagent}, social deduction games~\citep{avalonbench,werewolf}, text-based adventure games~\citep{textarena}). Subsequent research~\cite{spin_bench} has explored LLMs as players in various game contexts, from traditional board games requiring strategic reasoning to complex multiplayer online environments that demand natural language communication and coordination. The recent work KORGym~\cite{korgym} offers over fifty games in either textual or visual formats. But these benchmarks focus on text reasoning, ignoring the evaluation for the code large language model. In this work, we introduce \benchmark{} to evaluate the coding capability of LLMs to create the visual games.

\section{Conclusion}

We introduce \benchmark{}, a multimodal benchmark for evaluating code LLMs in visual game generation. Built by curating 2,219 high-quality Pygame samples, our framework assesses both code generation and visual capabilities. Our evaluation of 70 models reveals a significant performance gap between proprietary and open-source systems, with top models succeeding only 45\%. The benchmark highlights critical limitations in visual understanding and dynamic gameplay generation, providing a foundation for advancing AI-assisted game development.

\bibliography{custom}

\appendix

\onecolumn

\clearpage


\section{Comprehensive Leaderboard Ranking Models}

\begin{figure*}[h!]
  \centering
  \includegraphics[width=0.8\textwidth]{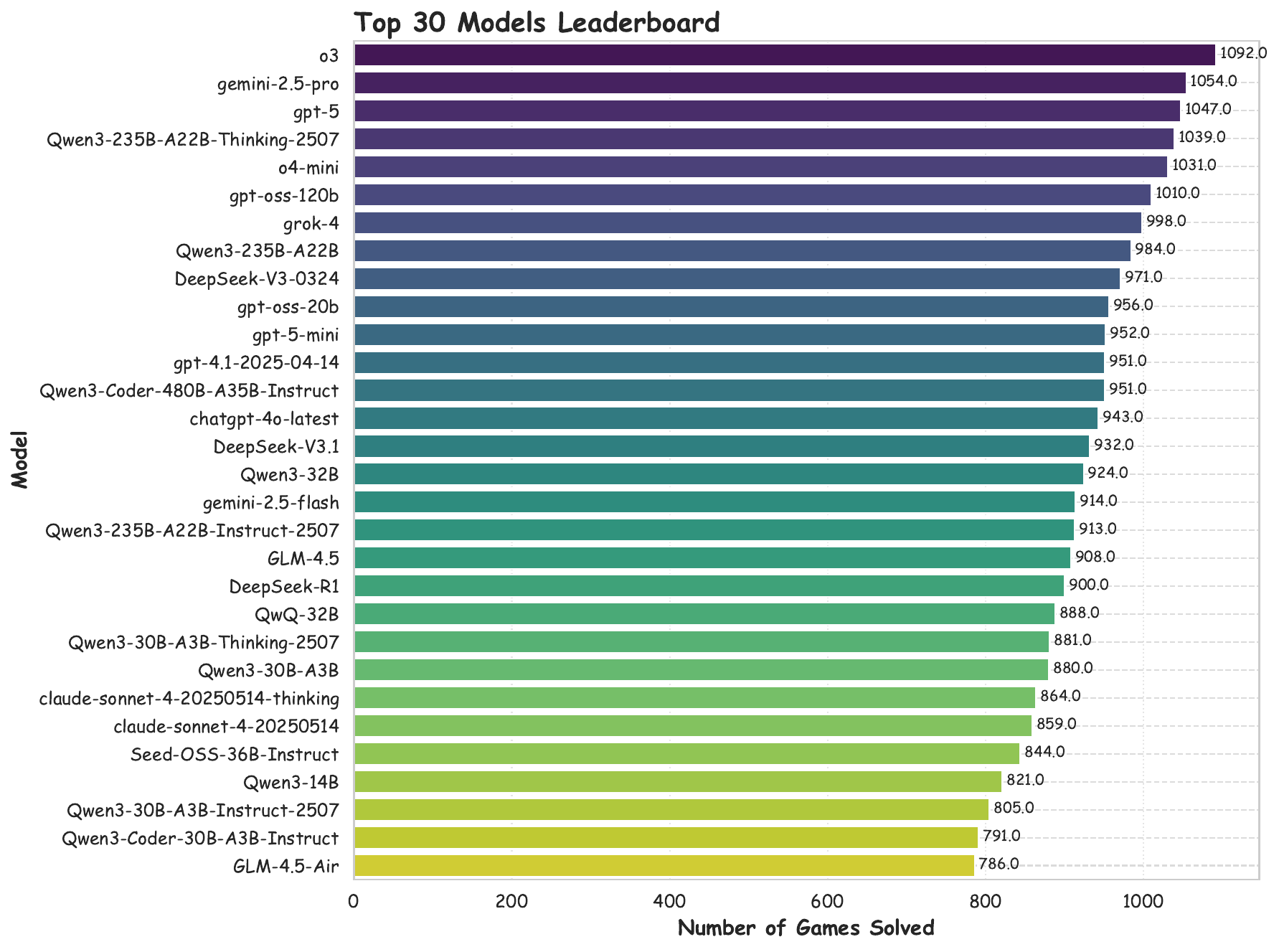}
  \caption{Comprehensive Leaderboard Ranking Models by the Number of Games Successfully Solved Out.}
  \label{fig:figures/figures_evals/01_model_leaderboard.pdf}
\end{figure*}

Figure~\ref{fig:figures/figures_evals/01_model_leaderboard.pdf} reveals a clear performance hierarchy with o3 achieving the highest success rate by solving 1,092 games, followed closely by Gemini-2.5-Pro (1,054 games) and GPT-5 (1,047 games). Notably, proprietary models dominate the top positions, with 5 out of the top 6 performers being closed-source systems. Among open-source models, the Qwen3 series demonstrates exceptional performance, with multiple variants appearing in the top rankings. The Qwen3-235B-A22B-Thinking-2507 model achieves the highest open-source performance at 1,039 games solved, ranking 4th overall. The strong showing of thinking-enhanced variants (e.g., Qwen3-235B-A22B-Thinking-2507) suggests that reasoning-augmented architectures provide substantial benefits for complex code generation tasks. The performance gap between the leading models and lower-ranked ones is substantial, with success rates ranging from approximately 49\% (1,092/2,219) at the top to 35\% (786/2,219) for the 30th-ranked model. This distribution indicates that while current state-of-the-art models can successfully generate functional game code for roughly half of the benchmark tasks, there remains significant room for improvement in achieving consistent, high-quality game development capabilities across diverse requirements.

\section{Complete Leaderboard}

Now, we show the complete leaderboard in Table~\ref{tab:combined_results_all}.

\begin{table*}[h!]
  \centering
  \small
  \renewcommand{\arraystretch}{1.0}
  \vspace{-30pt}
  \resizebox{0.85\textwidth}{!}{%
    \begin{tabular}{lr|c|ccc|cccc}
      \toprule
      Model & Size & Final & Code & Image & Video & Excellent & Good & Fair & Poor \\
      \midrule
      \rowcolor{HeaderColor}\multicolumn{10}{c}{\textcolor{white}{\textbf{Proprietary LLMs}}} \\
      \midrule
      \rowcolor{ProprietaryColor} \small gpt-5 & \faLock{} & \textbf{45.0} & \underline{96.6} & 17.6 & 20.7 & \textbf{83} & 288 & 676 & 1172 \\
      \rowcolor{ProprietaryColor} \small o3 & \faLock{} & \underline{44.8} & 92.3 & \textbf{20.2} & 21.9 & \underline{65} & \textbf{341} & \textbf{686} & 1127\\
      \rowcolor{ProprietaryColor} \small gpt-5-mini & \faLock{} & 43.5 & \textbf{96.7} & 15.7 & 18.0 & 61 & 236 & 655 & 1267 \\
      \rowcolor{ProprietaryColor} \small gemini-2.5-pro & \faLock{} & 43.5 & 89.1 & 19.1 & \underline{22.2} & 45 & \underline{337} & 672 & 1165  \\
      \rowcolor{ProprietaryColor} \small o4-mini & \faLock{} & 43.0 & 87.8 & 19.8 & 21.4 & 36 & 313 & \underline{682} & 1188 \\
      \rowcolor{ProprietaryColor} \small gpt-4.1-2025-04-14 & \faLock{} & 42.5 & 91.8 & 17.6 & 18.1 & 47 & 263 & 641 & 1268 \\
      \rowcolor{ProprietaryColor} \small grok-4 & \faLock{} & 42.0 & 83.9 & 19.8 & \textbf{22.4} & 21 & 327 & 650 & 1221 \\
      \rowcolor{ProprietaryColor} \small gemini-2.5-flash & \faLock{} & 42.0 & 92.8 & 16.5 & 16.7 & 28 & 252 & 634 & 1304 \\
      \rowcolor{ProprietaryColor} \small chatgpt-4o-latest & \faLock{} & 41.2 & 82.5 & \underline{19.9} & 21.3 & 25 & 305 & 613 & 1276 \\
      \rowcolor{ProprietaryColor} \small claude-sonnet-4-20250514-thinking & \faLock{} & 40.5 & 90.3 & 14.4 & 16.9 & 36 & 204 & 624 & 1355 \\
      \rowcolor{ProprietaryColor} \small claude-sonnet-4-20250514 & \faLock{} & 40.2 & 87.7 & 15.7 & 17.4 & 36 & 207 & 616 & 1360 \\
      \rowcolor{ProprietaryColor} \small o3-mini-2025-01-31 & \faLock{} & 38.2 & 89.3 & 11.9 & 13.3 & 26 & 204 & 417 & \underline{1572} \\
      \rowcolor{ProprietaryColor} \small gpt-4o-2024-11-20 & \faLock{} & 37.6 & 76.6 & 17.5 & 18.6 & 12 & 224 & 531 & 1452 \\
      \rowcolor{ProprietaryColor} \small gpt-4o-mini-2024-07-18 & \faLock{} & 33.9 & 70.4 & 15.5 & 15.8 & 4 & 134 & 459 & \textbf{1622} \\
      \midrule
      \rowcolor{HeaderColor}\multicolumn{10}{c}{\textcolor{white}{\textbf{400B+ Open-Weight LLMs}}} \\
      \midrule
      \rowcolor{Size400PlusColor} \small Qwen3-Coder-480B-A35B-Instruct & 32B/480B & \textbf{41.3} & \underline{85.3} & 18.3 & \underline{20.5} & 20 & 287 & \textbf{644} & 1268 \\
      \rowcolor{Size400PlusColor} \small DeepSeek-V3-0324 & 37B/671B & \underline{41.1} & 83.6 & \underline{19.3} & \textbf{20.5} & 22 & \textbf{311} & \underline{638} & 1248 \\
      \rowcolor{Size400PlusColor} \small DeepSeek-V3.1 & 37B/671B & 40.9 & 83.1 & \textbf{19.3} & 20.2 & \underline{25} & \underline{296} & 611 & 1287 \\
      \rowcolor{Size400PlusColor} \small DeepSeek-R1 & 37B/671B & 40.1 & 81.0 & 19.2 & 20.1 & 15 & 278 & 607 & 1319 \\
      \rowcolor{Size400PlusColor} \small DeepSeek-R1-0528 & 37B/671B & 38.7 & \textbf{88.1} & 13.4 & 14.6 & \textbf{32} & 174 & 544 & \underline{1469} \\
      \rowcolor{Size400PlusColor} \small DeepSeek-V3 & 37B/671B & 36.7 & 73.4 & 17.7 & 18.9 & 3 & 204 & 564 & 1447 \\
      \rowcolor{Size400PlusColor} \small kimi-k2-0905-preview & 32B/1000B & 23.5 & 66.3 & 2.0 & 2.2 & 0 & 18 & 62 & \textbf{2135} \\
      \midrule
      \rowcolor{HeaderColor}\multicolumn{10}{c}{\textcolor{white}{\textbf{100B-400B Open-Weight LLMs}}} \\
      \midrule
      \rowcolor{Size100To400Color} \small Qwen3-235B-A22B-Thinking-2507 & 22B/235B & \underline{42.3} & 84.5 & \textbf{20.0} & \underline{22.4} & 22 & \underline{322} & \textbf{695} & 1180 \\
      \rowcolor{Size100To400Color} \small Qwen3-235B-A22B & 235B & 41.2 & 81.3 & \underline{19.8} & \textbf{22.6} & 14 & 302 & \underline{668} & 1235 \\
      \rowcolor{Size100To400Color} \small Qwen3-235B-A22B-Instruct-2507 & 22B/235B & 41.1 & 85.3 & 18.2 & 19.7 & 16 & 308 & 589 & 1306 \\
      \rowcolor{Size100To400Color} \small GLM-4.5 & 32B/355B & 40.0 & 84.7 & 17.0 & 18.3 & \underline{31} & 216 & 661 & \underline{1311} \\
      \rowcolor{Size100To400Color} \small GLM-4.5-Air & 12B/106B & 39.4 & \underline{85.4} & 16.3 & 16.5 & 23 & 230 & 533 & \textbf{1433} \\
      \rowcolor{Size100To400Color} \small gpt-oss-120b & 5.1B/117B & \textbf{43.4} & \textbf{90.1} & 19.7 & 20.3 & \textbf{52} & \textbf{324} & 634 & 1209  \\
      \midrule
      \rowcolor{HeaderColor}\multicolumn{10}{c}{\textcolor{white}{\textbf{30B-100B Open-Weight LLMs}}} \\
      \midrule
      \rowcolor{Size30To100Color} \small Qwen3-32B & 32B & \textbf{40.4} & 81.6 & 18.9 & \underline{20.6} & 8 & \underline{274} & \textbf{642} & 1295 \\
      \rowcolor{Size30To100Color} \small Seed-OSS-36B-Instruct & 36B & \underline{40.3} & \textbf{88.3} & 16.4 & 16.2 & \textbf{25} & 234 & 585 & 1375 \\
      \rowcolor{Size30To100Color} \small Qwen3-30B-A3B-Thinking-2507 & 3B/30B & 40.0 & 80.7 & \underline{18.9} & 20.4 & 13 & \textbf{279} & 589 & 1338 \\
      \rowcolor{Size30To100Color} \small QwQ-32B & 32B & 39.6 & 79.7 & 18.5 & 20.6 & 10 & 268 & \underline{610} & 1331 \\
      \rowcolor{Size30To100Color} \small Qwen3-30B-A3B & 3B/30B & 39.6 & 78.4 & \textbf{19.7} & \textbf{20.7} & 9 & 274 & 597 & 1339 \\
      \rowcolor{Size30To100Color} \small Qwen3-Coder-30B-A3B-Instruct & 3B/30B & 39.0 & \underline{83.8} & 16.6 & 16.7 & \underline{22} & 226 & 543 & 1428 \\
      \rowcolor{Size30To100Color} \small Qwen3-30B-A3B-Instruct-2507 & 30B & 38.6 & 81.4 & 16.5 & 17.8 & 11 & 223 & 571 & 1414 \\
      \rowcolor{Size30To100Color} \small DeepSeek-R1-Distill-Llama-70B & 70B & 35.3 & 74.1 & 15.8 & 16.0 & 4 & 188 & 448 & 1579 \\
      \rowcolor{Size30To100Color} \small Zhihu-ai-Zhi-Create-Qwen3-32B & 32B & 35.1 & 75.8 & 15.2 & 14.4 & 3 & 184 & 426 & 1606 \\
      \rowcolor{Size30To100Color} \small Qwen2.5-72B-Instruct & 72B & 34.6 & 73.2 & 14.7 & 15.9 & 3 & 174 & 449 & 1593 \\
      \rowcolor{Size30To100Color} \small Qwen2.5-Coder-32B-Instruct & 32B & 34.4 & 74.5 & 13.8 & 14.9 & 9 & 167 & 425 & 1618 \\
      \rowcolor{Size30To100Color} \small DeepSeek-R1-Distill-Qwen-32B & 32B & 33.4 & 71.9 & 14.4 & 13.9 & 0 & 145 & 411 & \underline{1663} \\
      \rowcolor{Size30To100Color} \small Qwen2.5-32B-Instruct & 32B & 31.8 & 66.4 & 14.0 & 15.1 & 2 & 127 & 389 & \textbf{1701} \\
      \midrule
      \rowcolor{HeaderColor}\multicolumn{10}{c}{\textcolor{white}{\textbf{10B-30B Open-Weight LLMs}}} \\
      \midrule
      \rowcolor{Size10To30Color} \small gpt-oss-20b & 3.6B/21B & \textbf{42.2} & \textbf{88.8} & \textbf{18.6} & \textbf{19.2} & \textbf{31} & \textbf{299} & \textbf{626} & 1263 \\
      \rowcolor{Size10To30Color} \small Qwen3-14B & 14B & \underline{38.8} & \underline{79.1} & \underline{18.4} & \underline{18.8} & \underline{9} & \underline{245} & \underline{567} & 1398 \\
      \rowcolor{Size10To30Color} \small Qwen2.5-14B-Instruct & 14B & 30.3 & 66.4 & 11.4 & 13.0 & 0 & 92 & 348 & 1779 \\
      \rowcolor{Size10To30Color} \small Qwen2.5-Coder-14B-Instruct & 14B & 30.2 & 68.5 & 10.9 & 11.2 & 0 & 87 & 327 & \underline{1804} \\
      \rowcolor{Size10To30Color} \small DeepSeek-R1-Distill-Qwen-14B & 14B & 27.4 & 65.3 & 8.7 & 8.3 & 1 & 77 & 198 & \textbf{1943} \\
      \midrule
      \rowcolor{HeaderColor}\multicolumn{10}{c}{\textcolor{white}{\textbf{Below 10B Open-Weight LLMs}}} \\
      \midrule
      \rowcolor{SizeBelow10Color} \small Qwen3-8B & 8B & \textbf{36.9} & \textbf{76.2} & \textbf{17.2} & \textbf{17.3} & \textbf{5} & \textbf{187} & \textbf{546} & 1480 \\
      \rowcolor{SizeBelow10Color} \small Qwen3-4B & 4B & \underline{34.4} & 72.7 & 15.1 & 15.5 & 1 & 144 & 464 & 1610 \\
      \rowcolor{SizeBelow10Color} \small Qwen3-4B-Thinking-2507 & 4B & 34.3 & 70.0 & \underline{16.1} & \underline{16.8} & 2 & \underline{168} & \underline{465} & 1584 \\
      \rowcolor{SizeBelow10Color} \small Seed-Coder-8B-Instruct & 8B & 33.9 & \underline{73.2} & 14.0 & 14.4 & \underline{4} & 137 & 422 & 1656 \\
      \rowcolor{SizeBelow10Color} \small Llama-3.1-8B-Instruct & 8B & 29.4 & 62.9 & 13.0 & 12.3 & 1 & 84 & 319 & 1815 \\
      \rowcolor{SizeBelow10Color} \small Qwen2.5-Coder-7B-Instruct & 7B & 27.6 & 63.9 & 9.1 & 9.7 & 0 & 69 & 250 & 1899 \\
      \rowcolor{SizeBelow10Color} \small Qwen2.5-7B-Instruct & 7B & 26.1 & 59.8 & 9.2 & 9.2 & 0 & 52 & 230 & 1937 \\
      \rowcolor{SizeBelow10Color} \small Qwen3-1.7B & 1.7B & 25.0 & 57.3 & 9.1 & 8.7 & 0 & 43 & 216 & 1959 \\
      \rowcolor{SizeBelow10Color} \small Llama-3.2-3B-Instruct & 3B & 24.5 & 55.5 & 9.5 & 8.5 & 0 & 46 & 210 & 1963 \\
      \rowcolor{SizeBelow10Color} \small deepseek-coder-7b-instruct-v1.5 & 7B & 24.0 & 53.8 & 9.0 & 9.2 & 0 & 38 & 229 & 1952 \\
      \rowcolor{SizeBelow10Color} \small Hunyuan-7B-Instruct & 7B & 21.0 & 57.8 & 2.8 & 2.2 & 0 & 12 & 59 & 2148 \\
      \rowcolor{SizeBelow10Color} \small Qwen2.5-Coder-3B-Instruct & 3B & 20.4 & 53.8 & 4.4 & 2.9 & 0 & 10 & 85 & 2124 \\
      \rowcolor{SizeBelow10Color} \small OpenCoder-8B-Instruct & 8B & 20.1 & 49.0 & 6.5 & 4.7 & 0 & 7 & 152 & 2060 \\
      \rowcolor{SizeBelow10Color} \small Qwen2.5-3B-Instruct & 3B & 18.7 & 46.9 & 5.0 & 4.1 & 0 & 9 & 94 & 2116 \\
      \rowcolor{SizeBelow10Color} \small Qwen2.5-Coder-1.5B-Instruct & 1.5B & 17.3 & 46.6 & 3.9 & 1.2 & 0 & 2 & 58 & 2159 \\
      \rowcolor{SizeBelow10Color} \small OpenCoder-1.5B-Instruct & 1.5B & 16.9 & 43.1 & 5.6 & 1.9 & 0 & 9 & 102 & 2108 \\
      \rowcolor{SizeBelow10Color} \small Llama-3.2-1B-Instruct & 1B & 15.5 & 40.3 & 4.3 & 1.9 & 0 & 4 & 57 & 2158 \\
      \rowcolor{SizeBelow10Color} \small Qwen2.5-1.5B-Instruct & 1.5B & 15.2 & 40.1 & 3.7 & 1.8 & 0 & 2 & 63 & 2154 \\
      \rowcolor{SizeBelow10Color} \small DeepSeek-R1-Distill-Llama-8B & 8B & 15.1 & 42.5 & 1.7 & 1.0 & 0 & 6 & 26 & 2187 \\
      \rowcolor{SizeBelow10Color} \small DeepSeek-R1-0528-Qwen3-8B & 8B & 14.0 & 41.3 & 0.4 & 0.3 & 0 & 3 & 8 & 2207 \\
      \rowcolor{SizeBelow10Color} \small Qwen3-0.6B & 0.6B & 13.7 & 35.1 & 4.8 & 1.1 & 0 & 3 & 56 & 2160 \\
      \rowcolor{SizeBelow10Color} \small Qwen2.5-Coder-0.5B-Instruct & 0.5B & 12.8 & 34.6 & 3.6 & 0.0 & 0 & 0 & 39 & 2179 \\
      \rowcolor{SizeBelow10Color} \small DeepSeek-R1-Distill-Qwen-7B & 7B & 12.1 & 35.8 & 0.3 & 0.1 & 0 & 0 & 3 & \underline{2216} \\
      \rowcolor{SizeBelow10Color} \small Qwen2.5-0.5B-Instruct & 0.5B & 10.9 & 30.8 & 1.7 & 0.0 & 0 & 0 & 16 & 2200 \\
      \rowcolor{SizeBelow10Color} \small DeepSeek-R1-Distill-Qwen-1.5B & 1.5B & 8.5 & 25.4 & 0.0 & 0.0 & 0 & 0 & 1 & \textbf{2218} \\
      \bottomrule
    \end{tabular}%
  }%
  \caption{Comprehensive Performance Evaluation, showing Final Score, Code Score, Image Score, Video Score, and score distribution. \textbf{Bold} indicate highest performance; \underline{underlined} indicate second-highest performance.}
  \label{tab:combined_results_all}
  \vspace{-30pt}
\end{table*}

\section{Comprehensive Performance Comparison Across Different Evaluation Dimensions}

\begin{figure*}[h!]
  \centering
  \vspace{-20pt}
  \begin{subfigure}[b]{1.0\textwidth}
    \centering
    \includegraphics[width=\textwidth, height=4.8cm]{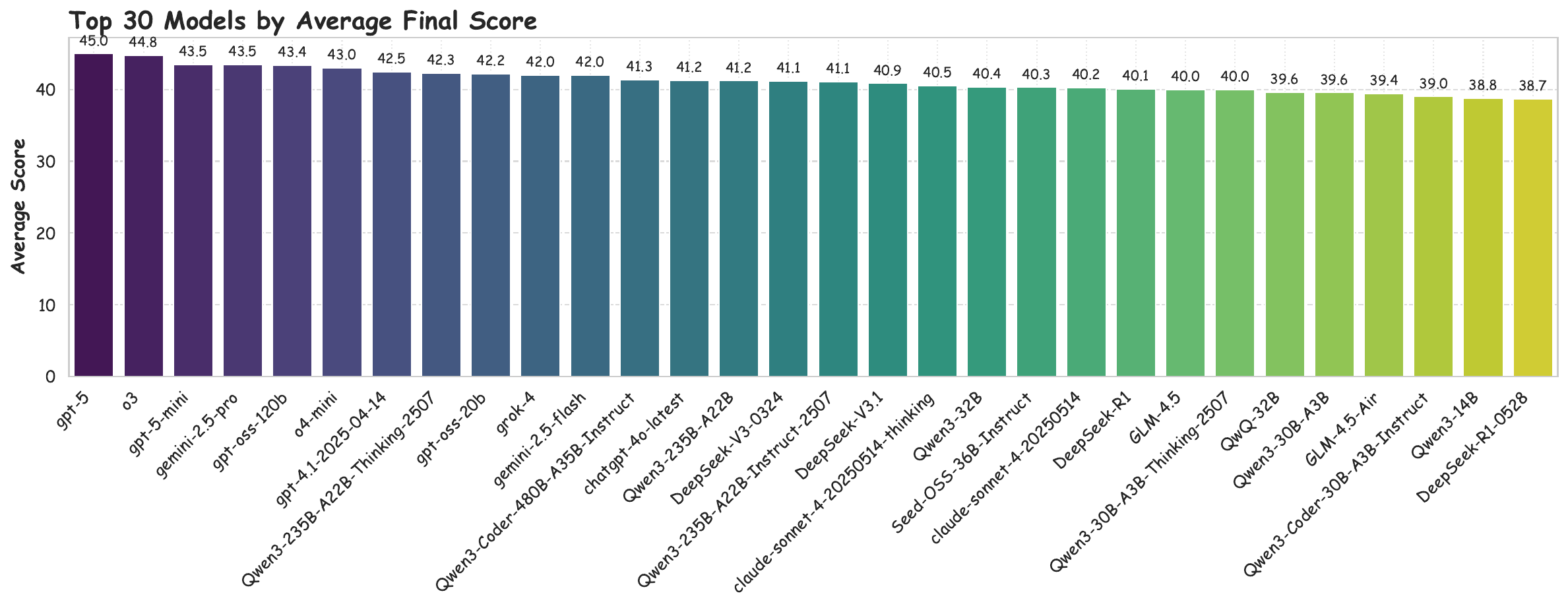}
    \caption{Final Performance}
  \end{subfigure}
  \hfill
  \begin{subfigure}[b]{1.0\textwidth}
    \centering
    \includegraphics[width=\textwidth, height=4.8cm]{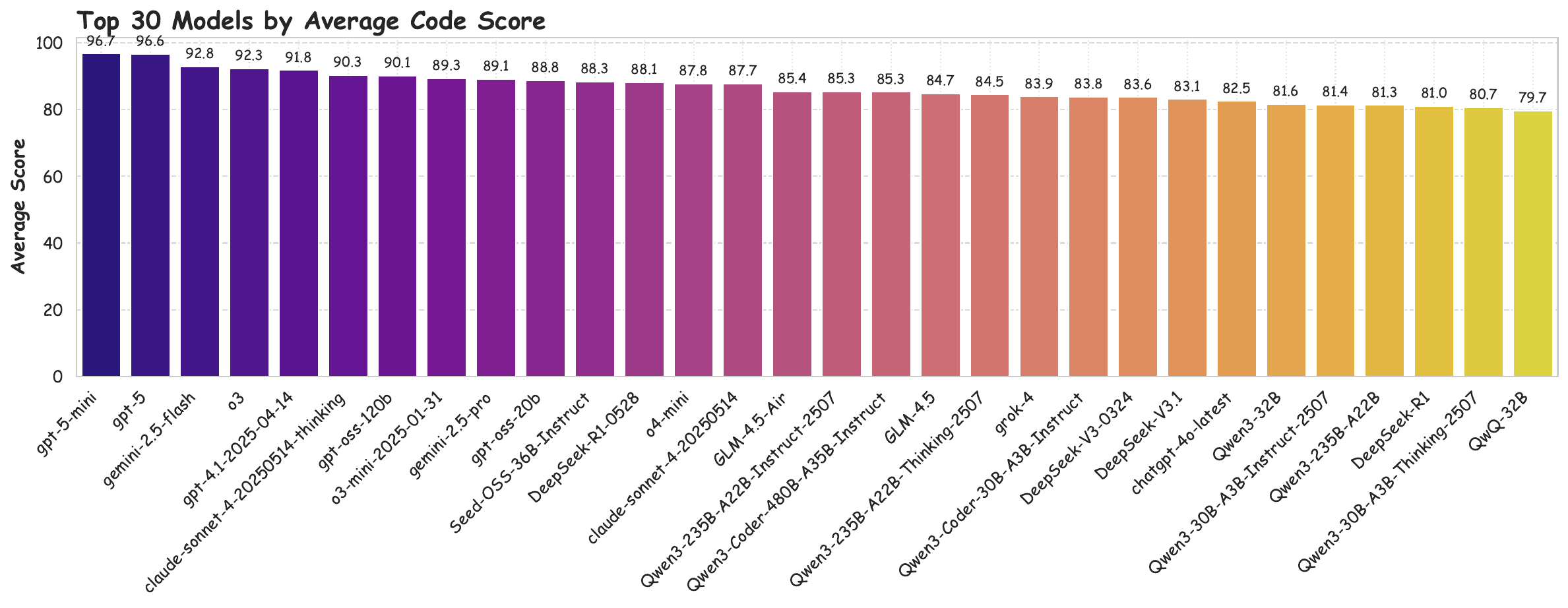}
    \caption{Code Generation Performance}
  \end{subfigure}
  \begin{subfigure}[b]{1.0\textwidth}
    \centering
    \includegraphics[width=\textwidth, height=4.8cm]{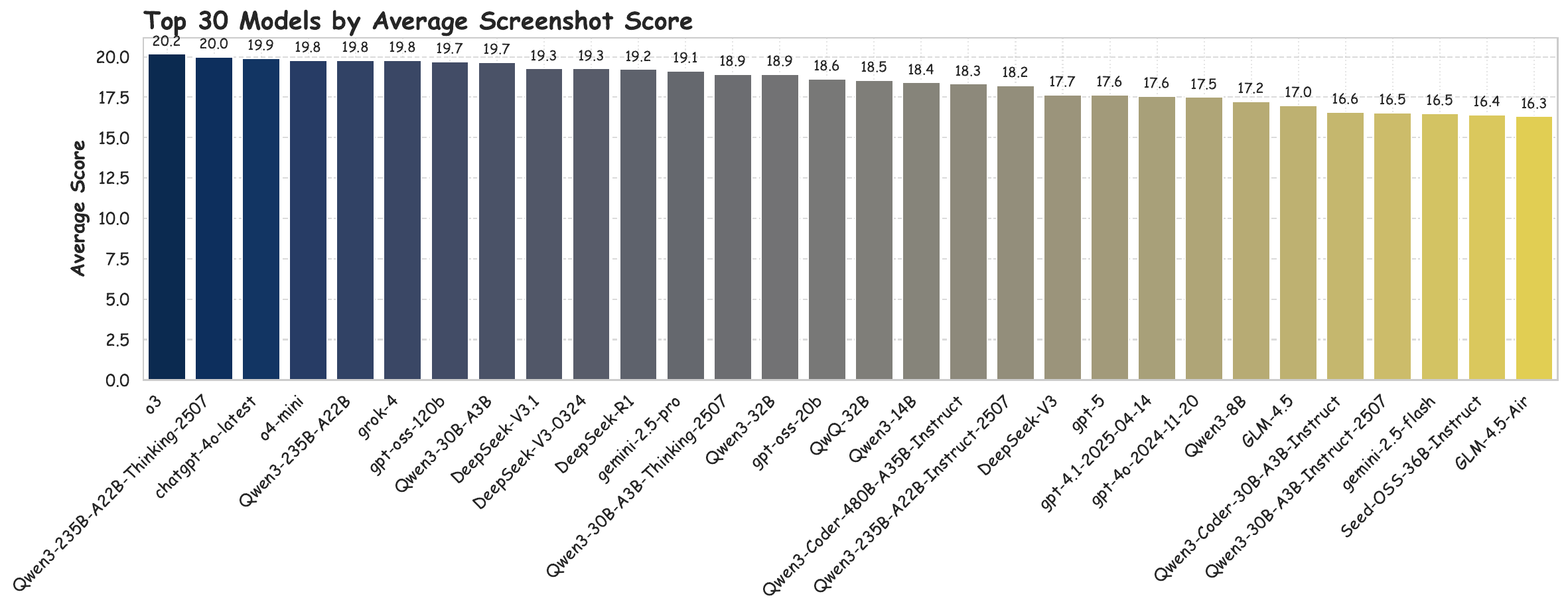}
    \caption{Image Evaluation Performance}
  \end{subfigure}
  \hfill
  \begin{subfigure}[b]{1.0\textwidth}
    \centering
    \includegraphics[width=\textwidth, height=4.8cm]{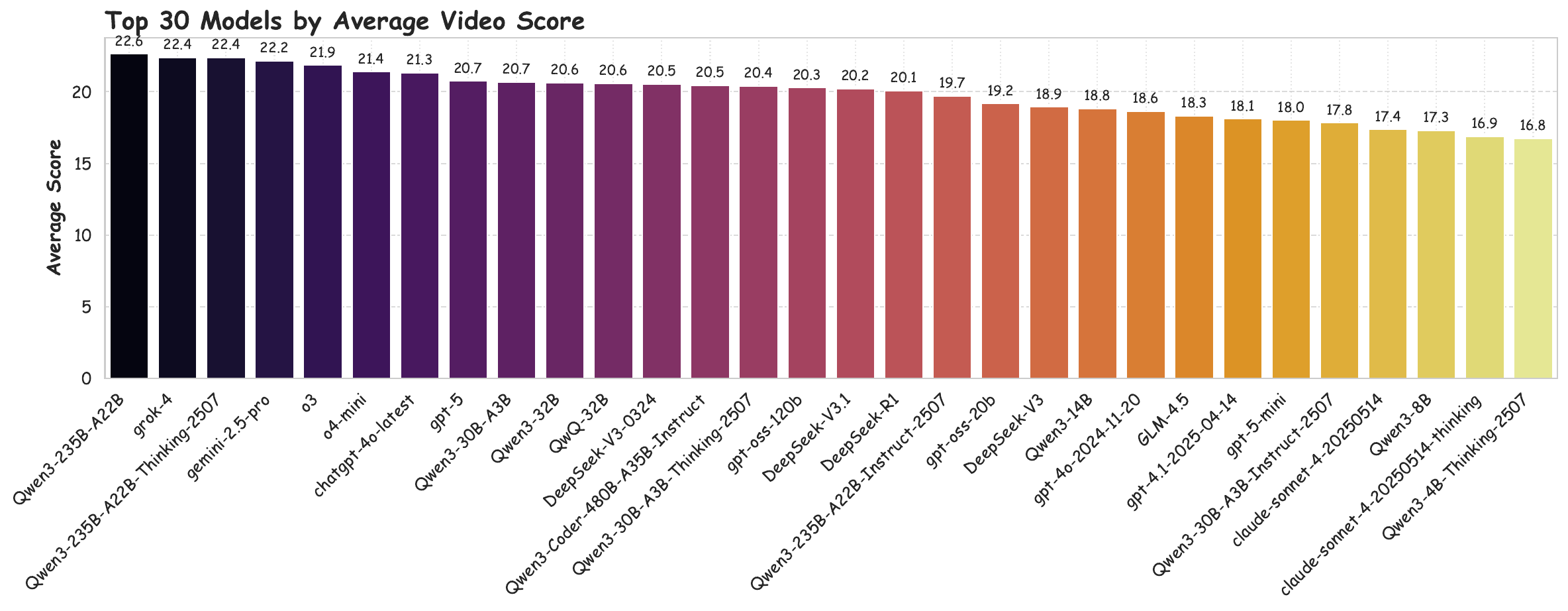}
    \caption{Video Evaluation Performance}
  \end{subfigure}
  \caption{Comprehensive performance comparison across different evaluation dimensions. Performance comparison across four evaluation dimensions: (a) Overall scores with GPT-5 and o3 leading, (b) High code generation scores (70-97), (c-d) Low visual evaluation scores (0-20), revealing models excel at code generation but struggle with visual assessment.}
  \label{fig:comprehensive_performance_comparison}
  \vspace{-20pt}
\end{figure*}

Figure~\ref{fig:comprehensive_performance_comparison} presents a comprehensive performance comparison across four key evaluation dimensions. The final performance ranking (a) shows proprietary models dominating the leaderboard, with GPT-5 achieving the highest score of 45.0, followed closely by O3 at 44.8. Code generation performance (b) reveals the strongest capability across all models, with scores ranging from 70-97 points, indicating mature syntactic and logical programming abilities. However, a significant performance gap emerges in visual assessment tasks: image evaluation (c) shows dramatically lower scores (0-20 points), while video evaluation (d) exhibits similar patterns with scores reaching only up to 22.6. This stark contrast between code generation and visual evaluation performance highlights a fundamental challenge in current language models - while they excel at producing syntactically correct and logically sound code, generating visually appealing and functionally rich interactive games remains substantially more difficult. The consistent ranking patterns across visual modalities suggest that models capable of generating better static visual content also tend to produce superior dynamic gameplay experiences.

\section{\benchmark{} Reference Code Analysis}

\begin{figure*}[h!]
  \centering
  \begin{subfigure}[b]{0.32\textwidth}
    \centering
    \includegraphics[width=\textwidth]{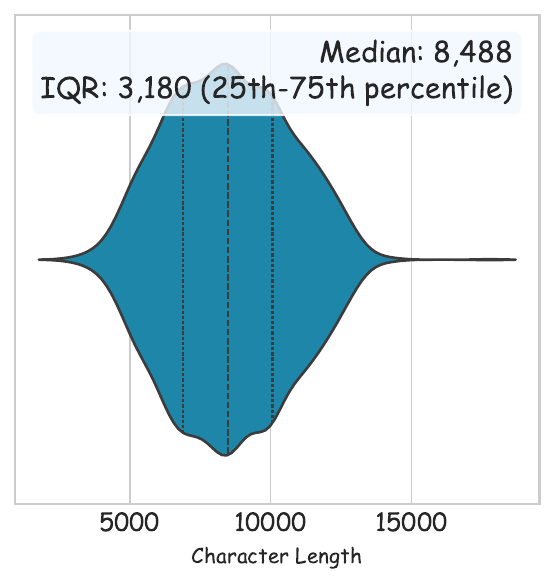}
    \caption{Violin plot showing distribution peaked at ~8,500 characters.}
  \end{subfigure}
  \hfill
  \begin{subfigure}[b]{0.32\textwidth}
    \centering
    \includegraphics[width=\textwidth]{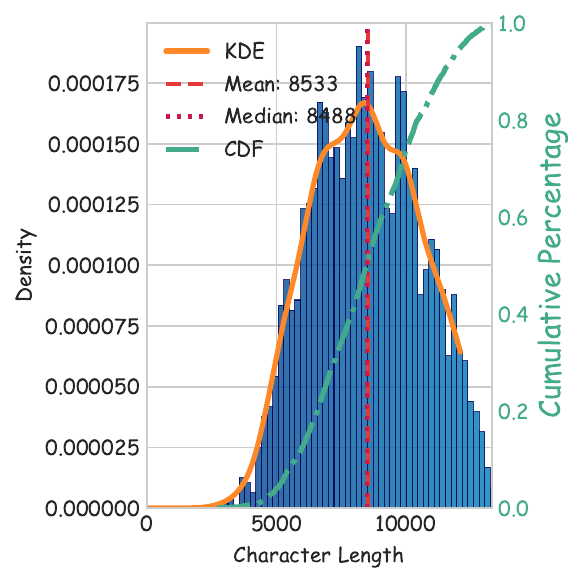}
    \caption{Histogram showing symmetric distribution.}
  \end{subfigure}
  \begin{subfigure}[b]{0.32\textwidth}
    \centering
    \includegraphics[width=\textwidth]{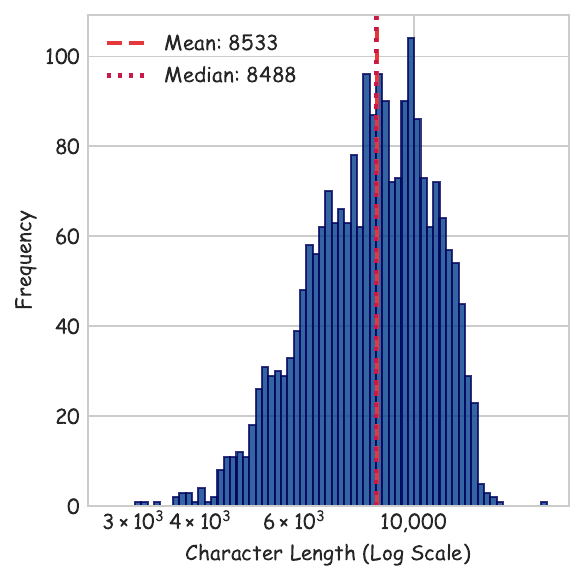}
    \caption{Log-scale view showing full range distribution.}
  \end{subfigure}
  \hfill
  \vspace{-10pt}
  \caption{Comprehensive analysis of reference code character length distribution using three complementary visualization methods: density estimation, linear-scale histogram, and logarithmic-scale representation.}
  \label{fig:figures/figures_testset/figures_refer_code}
  \vspace{-10pt}
\end{figure*}

To comprehensively analyze the character length of reference code within the dataset, we employed three complementary visualization methods. Figure~\ref{fig:figures/figures_testset/figures_refer_code}(a), a Violin Plot, reveals the overall probability density distribution of the data, exhibiting a clear unimodal shape with its peak concentrated around 8,500 characters. This figure intuitively displays the core statistical characteristics of the data: a median of 8,488 characters, with 50\% of the data falling within an interquartile range (IQR) spanning 3,180 characters. Figure~\ref{fig:figures/figures_testset/figures_refer_code}(b) provides a more refined depiction of this distribution through a histogram with a kernel density estimate (KDE) curve under linear coordinates. The calculated mean of 8,533 characters is notably close to the median, suggesting an approximately symmetrical distribution. Concurrently, the cumulative distribution function (CDF) curve on the right offers a quantitative perspective on the data; for instance, approximately 80\% of code samples have a length below 10,000 characters. Finally, to effectively examine the data's full dynamic range, particularly its long-tail portion, Figure~\ref{fig:figures/figures_testset/figures_refer_code}(c) employs a logarithmic axis. This view compresses the larger value ranges, allowing extreme values at both ends of the distribution to be clearly presented, thus completely illustrating the entire distribution from the shortest to the longest code segments. In summary, these three figures collectively provide a detailed and multifaceted representation of the dataset's central tendency, dispersion, and distributional shape.

\section{\benchmark{} Word Cloud Analysis} 
\begin{figure*}[h!]
  \centering
  \includegraphics[width=1.0\textwidth]{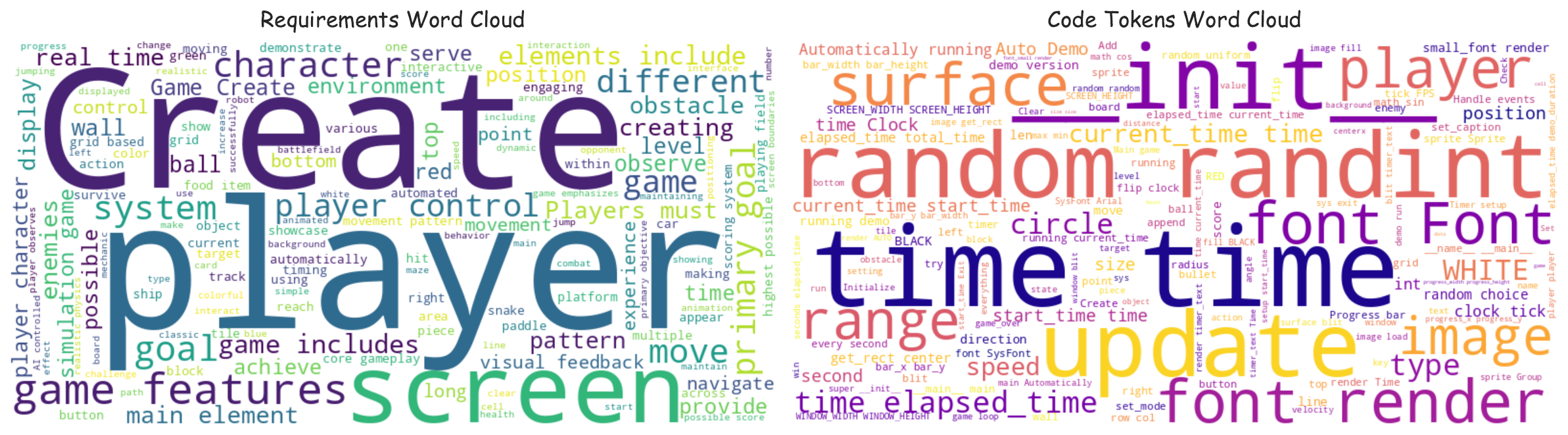}
  \caption{\benchmark{} Comparative Word Cloud Analysis of Requirements and Code Corpora.}
  \label{fig:figures/figures_testset/word_clouds_analysis.pdf}
\end{figure*}

Figure~\ref{fig:figures/figures_testset/word_clouds_analysis.pdf} presents a word cloud analysis comparing the linguistic content of the natural language requirements against the Python code solutions in our dataset. The Requirements Word Cloud (left) highlights a strong focus on core game mechanics, with dominant terms such as \texttt{player}, \texttt{screen}, \texttt{game}, and \texttt{create}. This confirms the prompts are well aligned with the intended domain. The Code Tokens Word Cloud (right) reveals the most frequent Pygame API calls and programming constructs, including \texttt{render}, \texttt{font}, \texttt{random}, and \texttt{time}, outlining the key technical skills required. The clear semantic alignment between the two clouds demonstrates a direct and coherent mapping from the problem descriptions to their programmatic solutions, validating the dataset's suitability for evaluating an LLM's code generation capabilities in this domain.

\section{\benchmark{} Reference Code Patterns Quantitative Analysis}

\begin{figure*}[h!]
  \centering
  \begin{subfigure}[b]{0.32\textwidth}
    \centering
    \includegraphics[width=\textwidth, height=4cm]{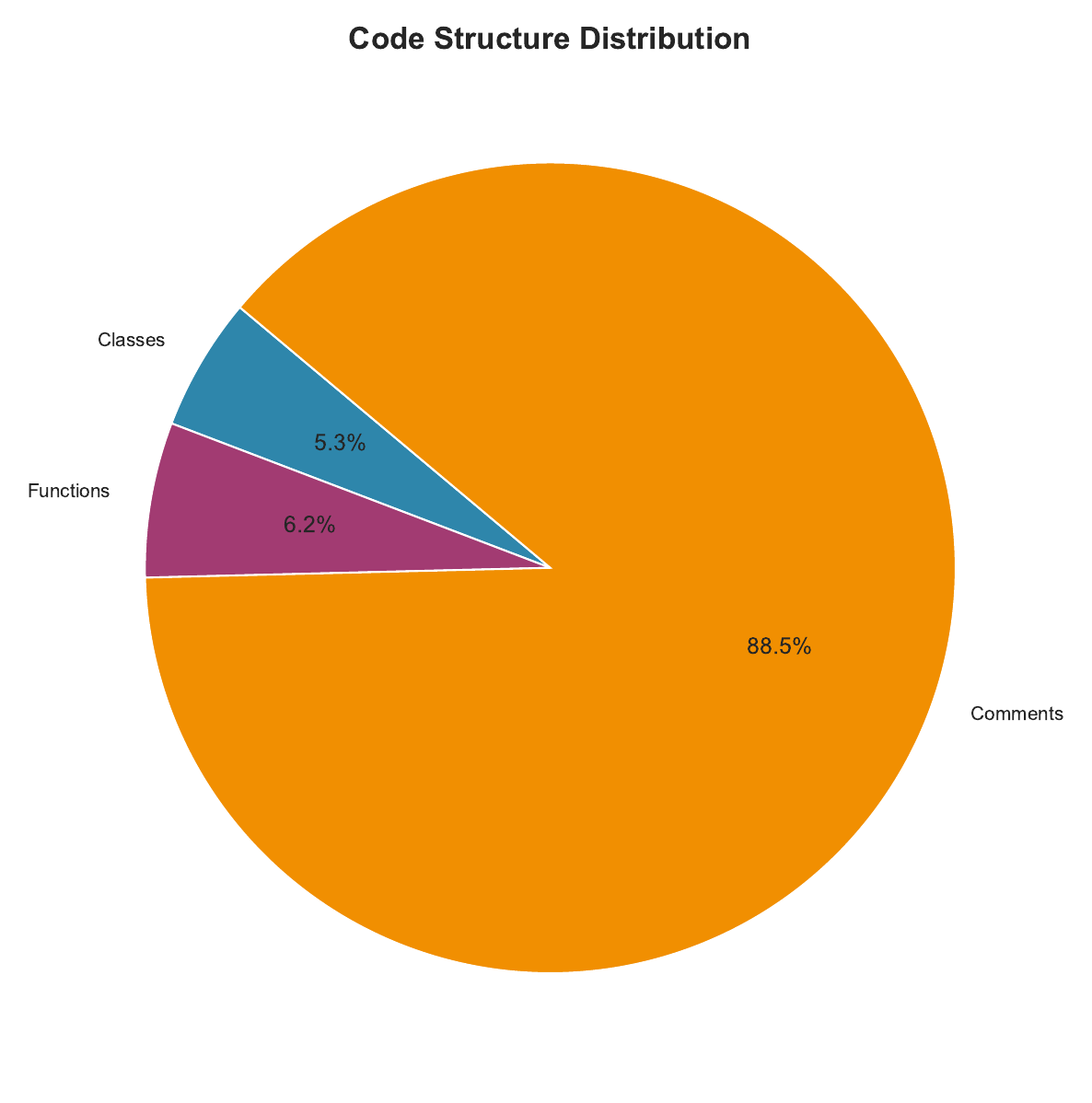}
    \caption{Occurrence frequency of core game loop patterns.}
  \end{subfigure}
  \hfill
  \begin{subfigure}[b]{0.32\textwidth}
    \centering
    \includegraphics[width=\textwidth, height=4cm]{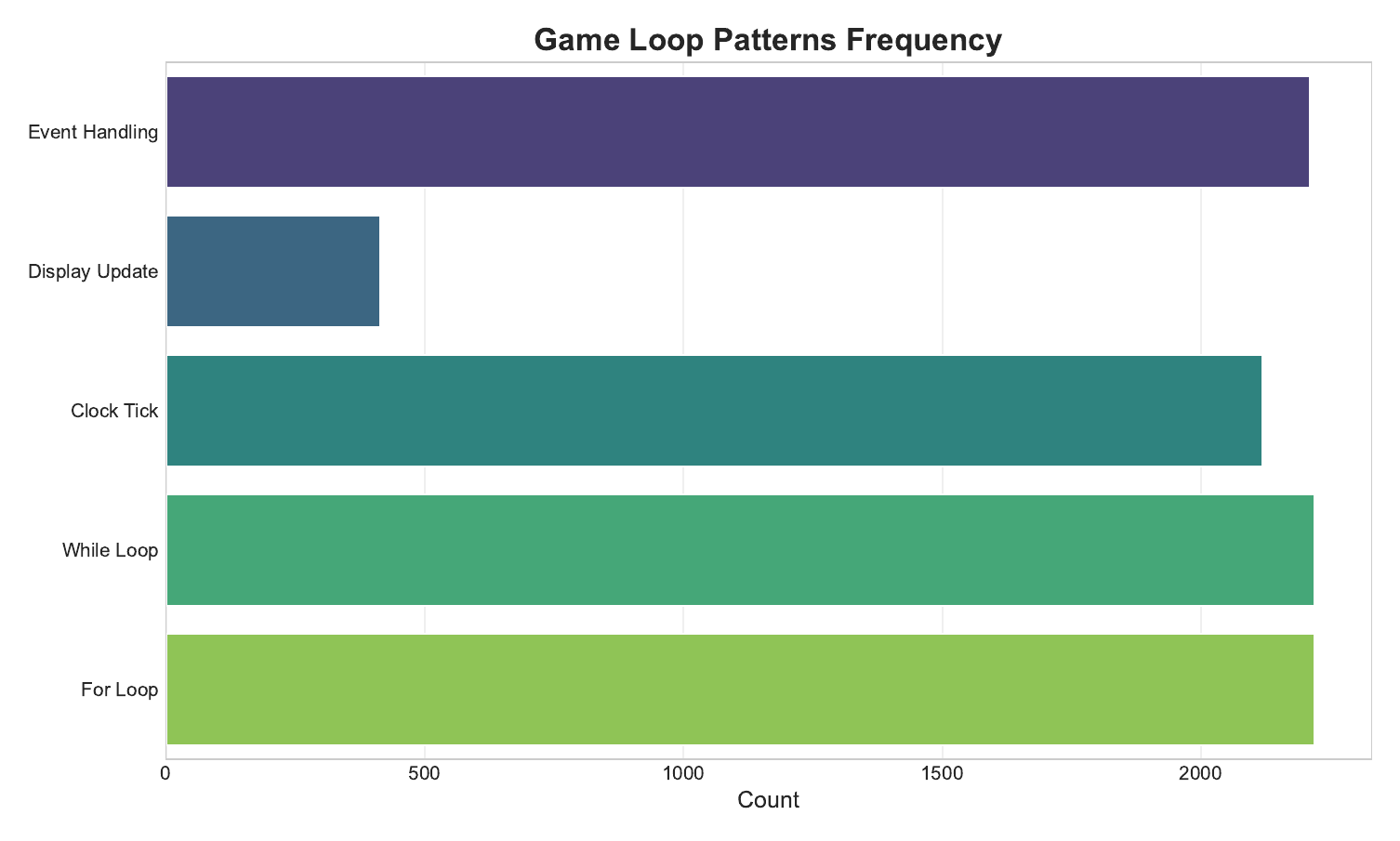}
    \caption{Distribution of code structure elements.}
  \end{subfigure}
  \begin{subfigure}[b]{0.32\textwidth}
    \centering
    \includegraphics[width=\textwidth, height=4cm]{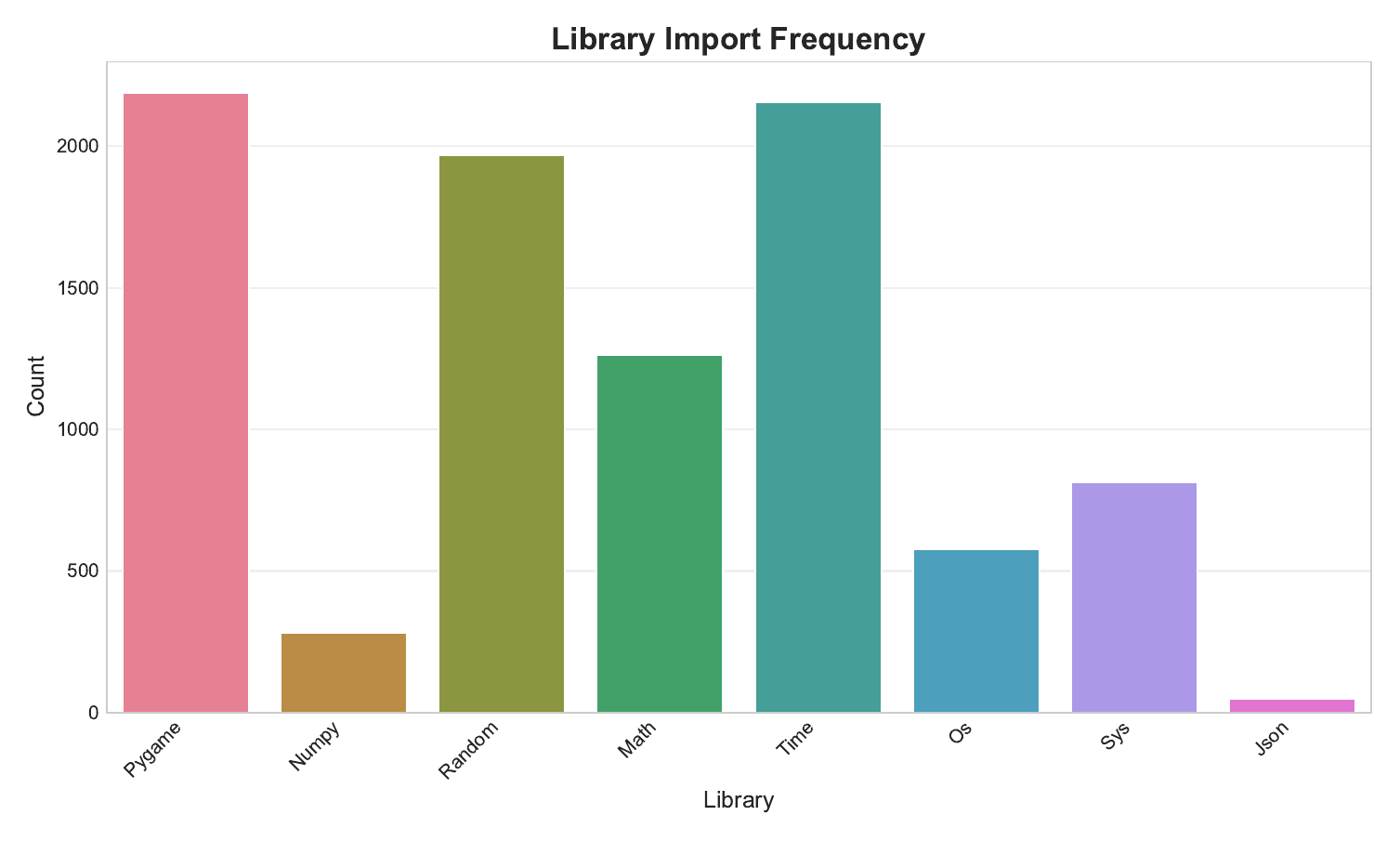}
    \caption{Import frequency of common Python libraries.}
  \end{subfigure}
  \hfill
  \caption{Comprehensive analysis of reference code character length distribution.}
  \label{fig:figures/figures_testset/code_pattern_pdf_outputs}
\end{figure*}

To deeply understand the inherent structure and common practices of code within the \benchmark{} dataset, we conducted a quantitative analysis of code patterns, with results shown in Figure~\ref{fig:figures/figures_testset/code_pattern_pdf_outputs}. The results reveal library usage frequency, core game loop mechanisms, code structure paradigms, and overall complexity distribution, respectively. On one hand, they generally adhere to standard Pygame development paradigms; on the other hand, they exhibit significant diversity in code structure and complexity, making this dataset an ideal resource for training and evaluating code generation models.

\begin{figure*}[h!]
  \centering
  \includegraphics[width=1.0\textwidth, height=8cm]{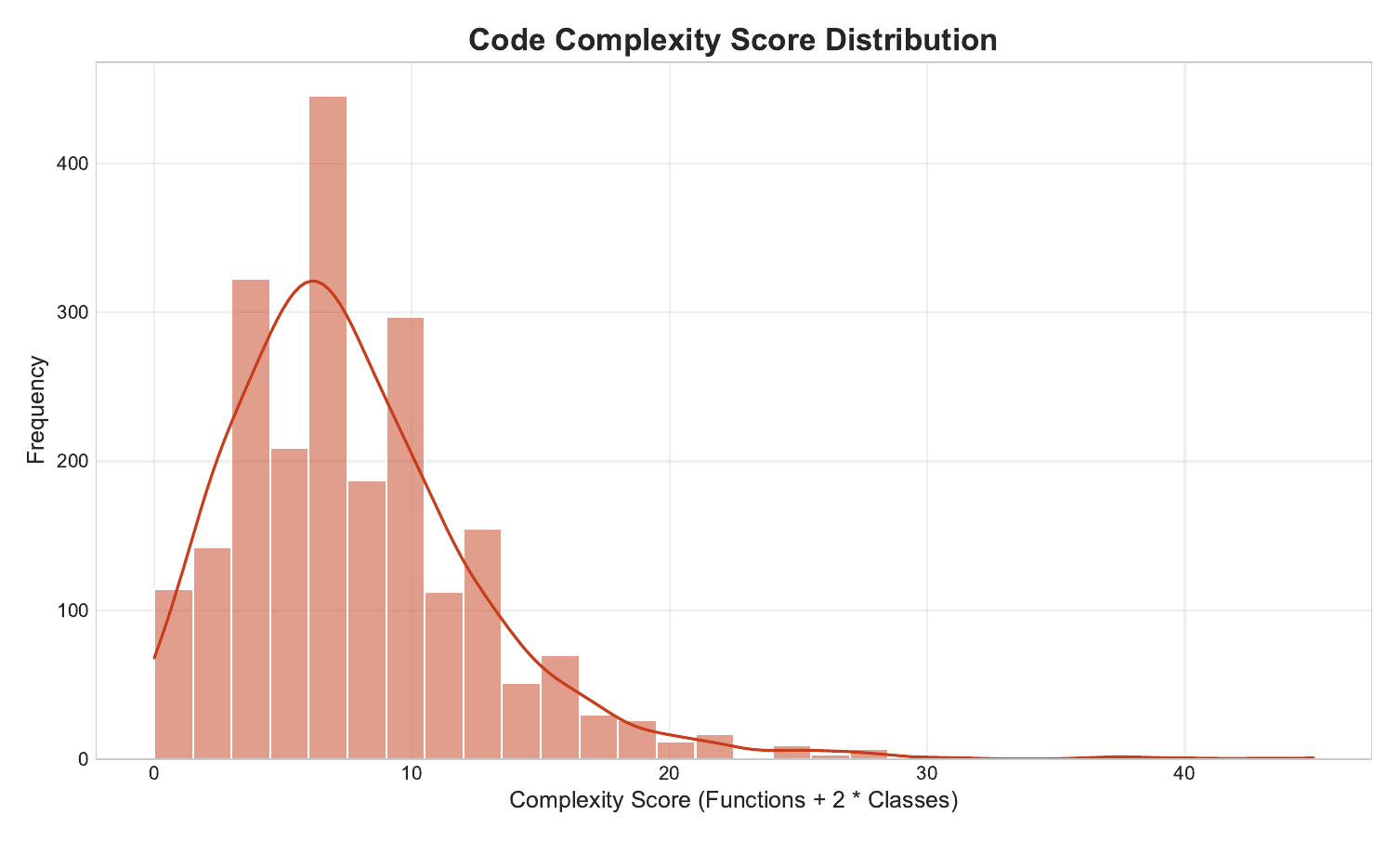}
  \caption{Distribution of code complexity scores.}
  \label{fig:figures/figures_testset/code_pattern_pdf_outputs/code_complexity_distribution.pdf}
\end{figure*}

\paragraph{Library Import Frequency} This plot displays the most frequently imported Python libraries in the \benchmark{} dataset. The results indicate that \texttt{pygame} is widely used as the core framework, while \texttt{random} and \texttt{math} libraries also appear frequently, reflecting the common presence of random elements and mathematical computation requirements in game development. This confirms that the samples in the dataset follow standard \texttt{Pygame} development practices.

\paragraph{Game Loop Patterns} This plot quantifies the occurrences of key code patterns that constitute a typical game loop. The high frequency of calls to \texttt{pygame.event} highlights the central role of event-driven programming in game interaction. Similarly, the frequent use of \texttt{pygame.display.update} and \texttt{clock.tick}, corresponding to screen rendering and frame rate control respectively, is fundamental for building real-time, smooth gaming experiences.

\paragraph{Code Structure Distribution} This pie chart depicts the relative proportions of classes, functions, and comments within the code. Analysis shows that functions are the primary units of code organization, while the use of classes also accounts for a significant proportion, indicating a certain application of object-oriented programming (OOP) principles in the samples. The proportion of comments provides an indirect measure of code readability and maintainability.

\paragraph{Code Complexity Score Distribution} To assess the structural complexity of the code, we defined a complexity score (calculated as number of functions + 2 * number of classes). The histogram in this figure shows that the complexity scores exhibit a right-skewed distribution, indicating that most game codes in the dataset have relatively simple structures, but it also includes a portion of complex projects with highly intricate structures (e.g., a large number of classes and functions).

\section{\benchmark{} Quality Score Prediction Model Results}

\begin{figure*}[h!]
  \centering
  \begin{subfigure}[b]{0.45\textwidth}
    \centering
    \includegraphics[width=\textwidth, height=6cm]{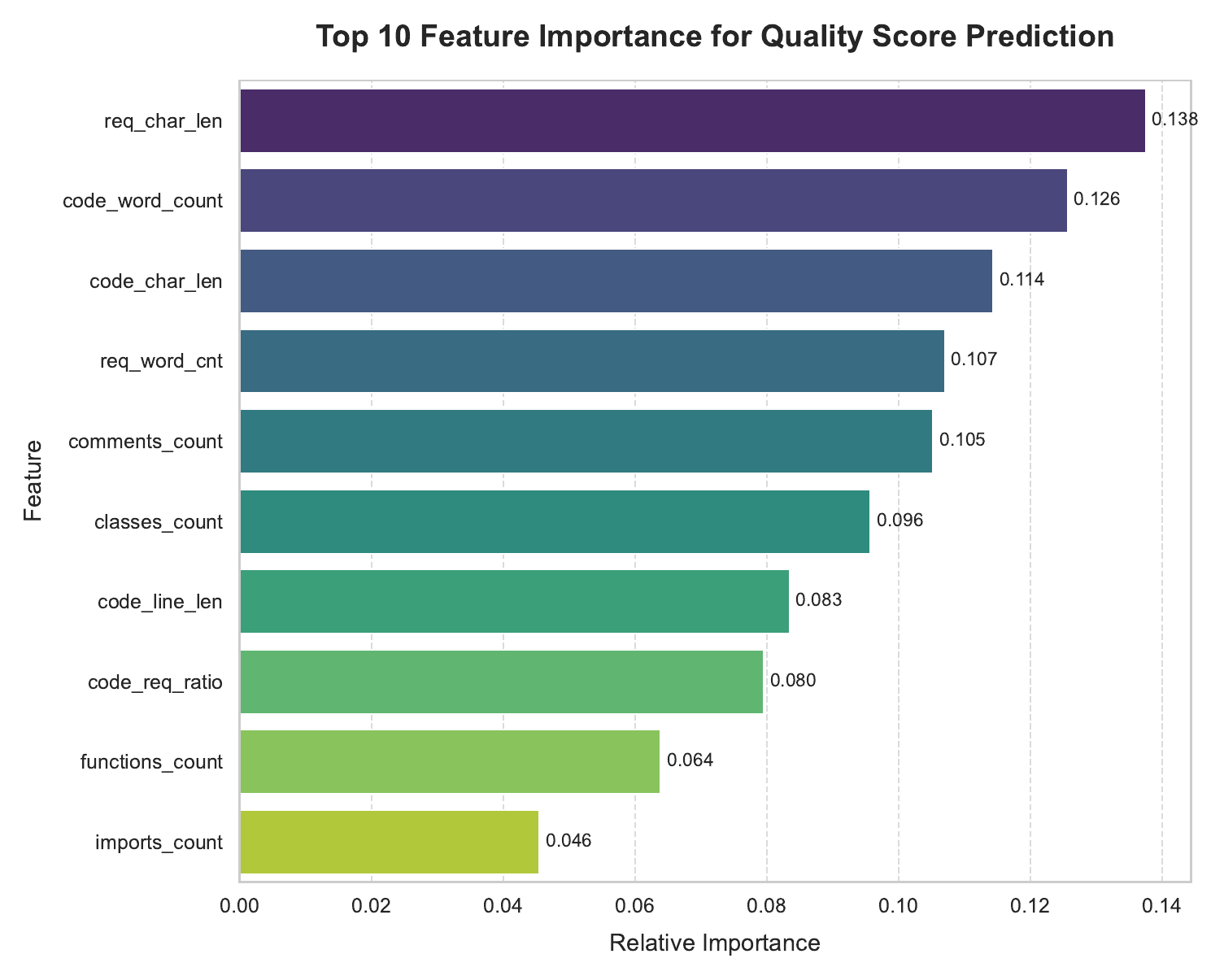}
    \caption{Comparison of actual vs. predicted quality scores.}
  \end{subfigure}
  \begin{subfigure}[b]{0.45\textwidth}
    \centering
    \includegraphics[width=\textwidth, height=6cm]{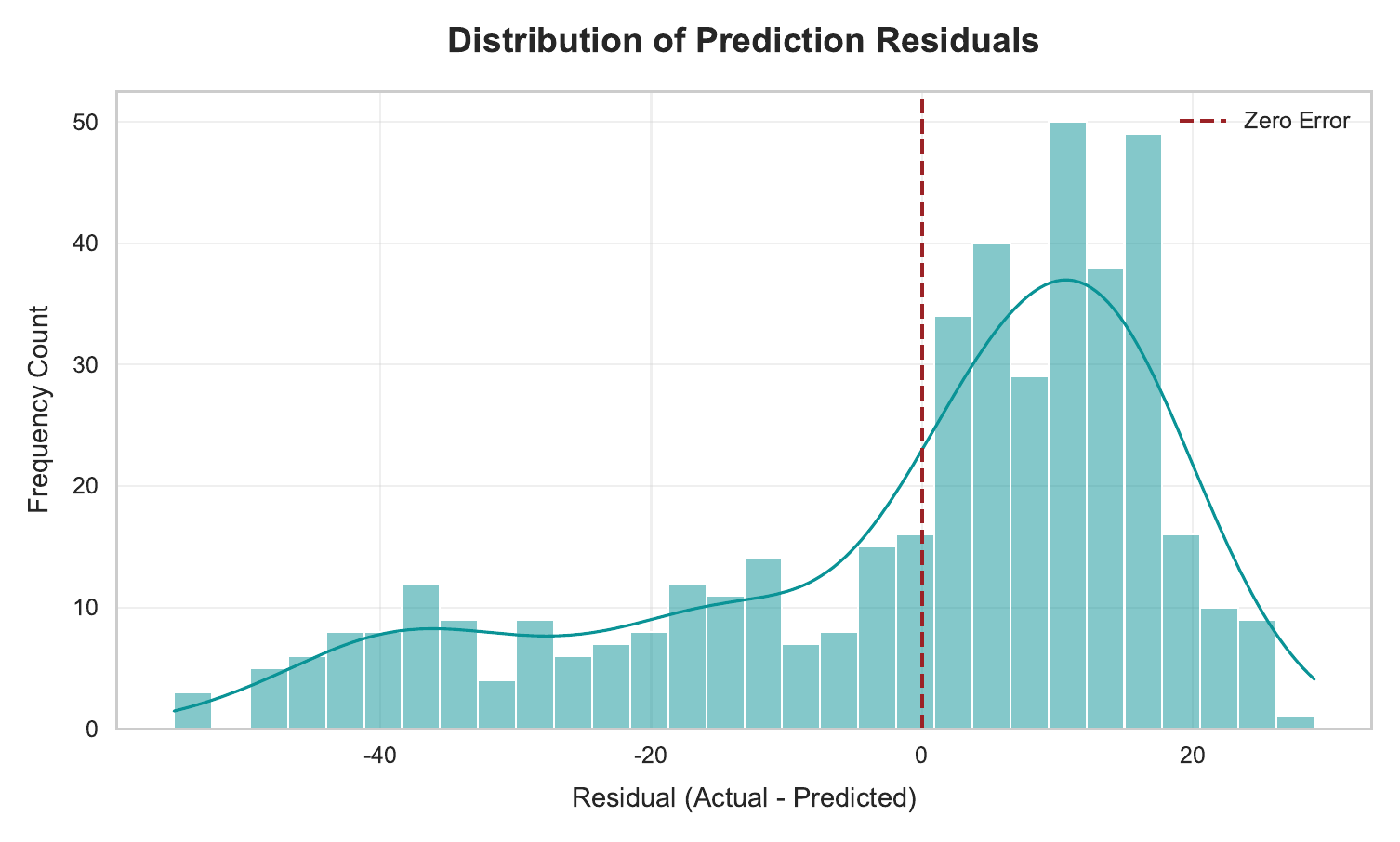}
    \caption{Distribution of prediction residuals.}
  \end{subfigure}
  \hfill
  \caption{Quality Score Prediction Distribution.}
  \label{fig:figures/figures_testset/quality_model}
\end{figure*}

To evaluate the performance of our Random Forest regression model for quality score prediction, a multifaceted analysis was conducted, as illustrated in Figure~\ref{fig:figures/figures_testset/quality_model}.

\paragraph{Feature Importance} This panel presents the top ten most influential features in determining the model's predictions, ranked by their Gini importance. The analysis reveals that metrics related to code volume and complexity, such as \texttt{code\_char\_len} (total characters in the code) and \texttt{code\_word\_count}, are the strongest predictors. This insight underscores the significant relationship between the sheer size of the codebase and its perceived quality score within this dataset.

\paragraph{Residuals Distribution} This histogram displays the distribution of the prediction residuals, calculated as the difference between the actual and predicted scores (Actual - Predicted). The distribution is approximately centered around zero and exhibits a quasi-normal shape, suggesting that the model has no systematic bias (i.e., it does not consistently over- or under-predict). This desirable characteristic indicates that the model's errors are random, which is a key assumption for a well-fitted regression model.

\section{\benchmark{} Distribution of Game Samples Across the Top 30 Source Repositories}

\begin{figure*}[ht]
  \centering
  \includegraphics[width=1.0\textwidth]{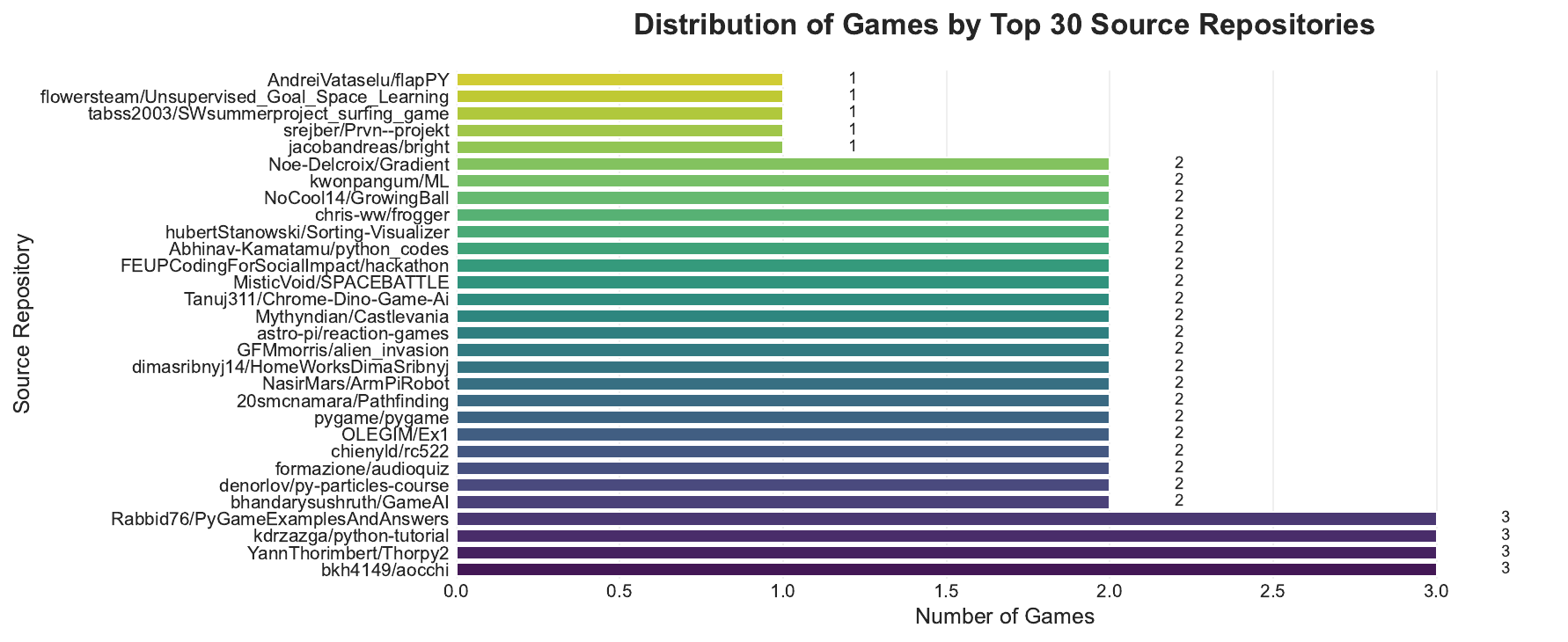}
  \caption{Distribution of game samples across the top 30 source repositories.}
  \label{fig:figures/figures_testset/top_repositories_by_game_count.pdf}
\end{figure*}

Figure~\ref{fig:figures/figures_testset/top_repositories_by_game_count.pdf} provides a quantitative analysis of the contribution frequency from the top 30 source repositories within the curated dataset. The horizontal bar chart illustrates the number of game samples sourced from each unique repository, which are ranked in descending order of their contribution count. A prominent characteristic revealed by the visualization is the highly granular and flat distribution of samples. The data indicates that the contributions are thinly spread across a wide array of sources, with the most frequent repositories supplying a maximum of only three game samples. A substantial cohort of repositories provided two samples each, followed by another group contributing single instances. This flat, long-tail distribution pattern underscores the extensive diversity of the dataset's origins. By sourcing a small number of games from a large pool of independent repositories, we effectively minimize the risk of stylistic and structural bias that could arise from over-representing a few dominant sources. The resulting heterogeneity ensures a broad and more representative collection of programming patterns, architectural designs, and implementation logic. This characteristic is fundamental to the dataset's objective of serving as a robust foundation for training generalizable models in tasks such as automated code generation and program analysis.

\section{Model Similarity Analysis}

\begin{figure*}[ht]
  \centering
  \includegraphics[width=0.85\textwidth]{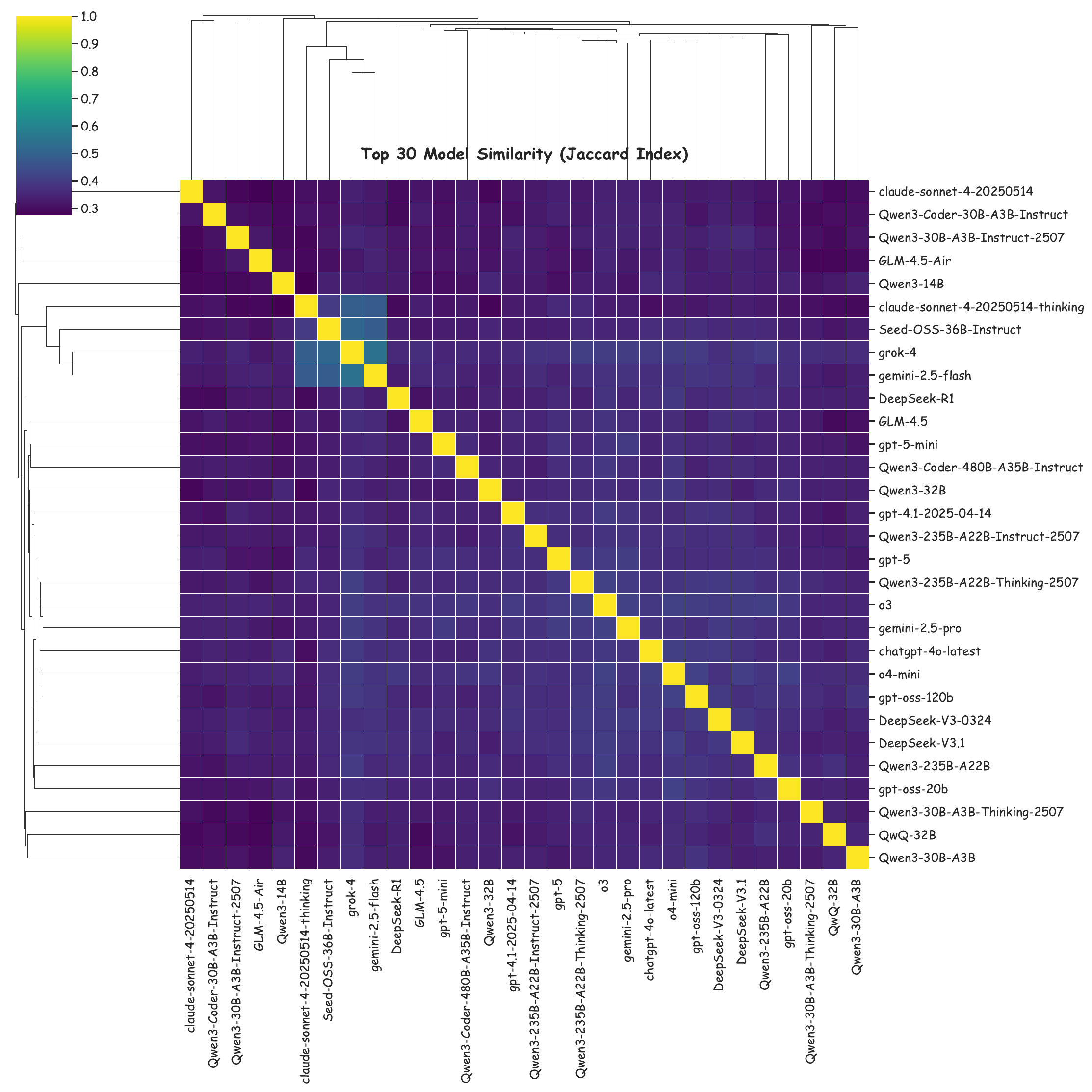}
  \caption{Hierarchical clustering of models based on solved game overlap using Jaccard similarity index.}
  \label{fig:figures/figures_evals/05_model_similarity_clustermap.pdf}
\end{figure*}

In Figure~\ref{fig:figures/figures_evals/05_model_similarity_clustermap.pdf}, the similarity clustering reveals distinct model families with comparable problem-solving patterns. Models from the same architecture family (e.g., Qwen3 variants, DeepSeek series) tend to cluster together, indicating that foundational architecture and training methodologies significantly influence which games models can successfully solve. Interestingly, some cross-family clusters emerge between models of similar scale, suggesting that parameter count plays a crucial role in determining capability overlap beyond architectural differences.

\section{Score Threshold Sensitivity Analysis}

\begin{figure*}[h!]
  \centering
  \includegraphics[width=1.0\textwidth]{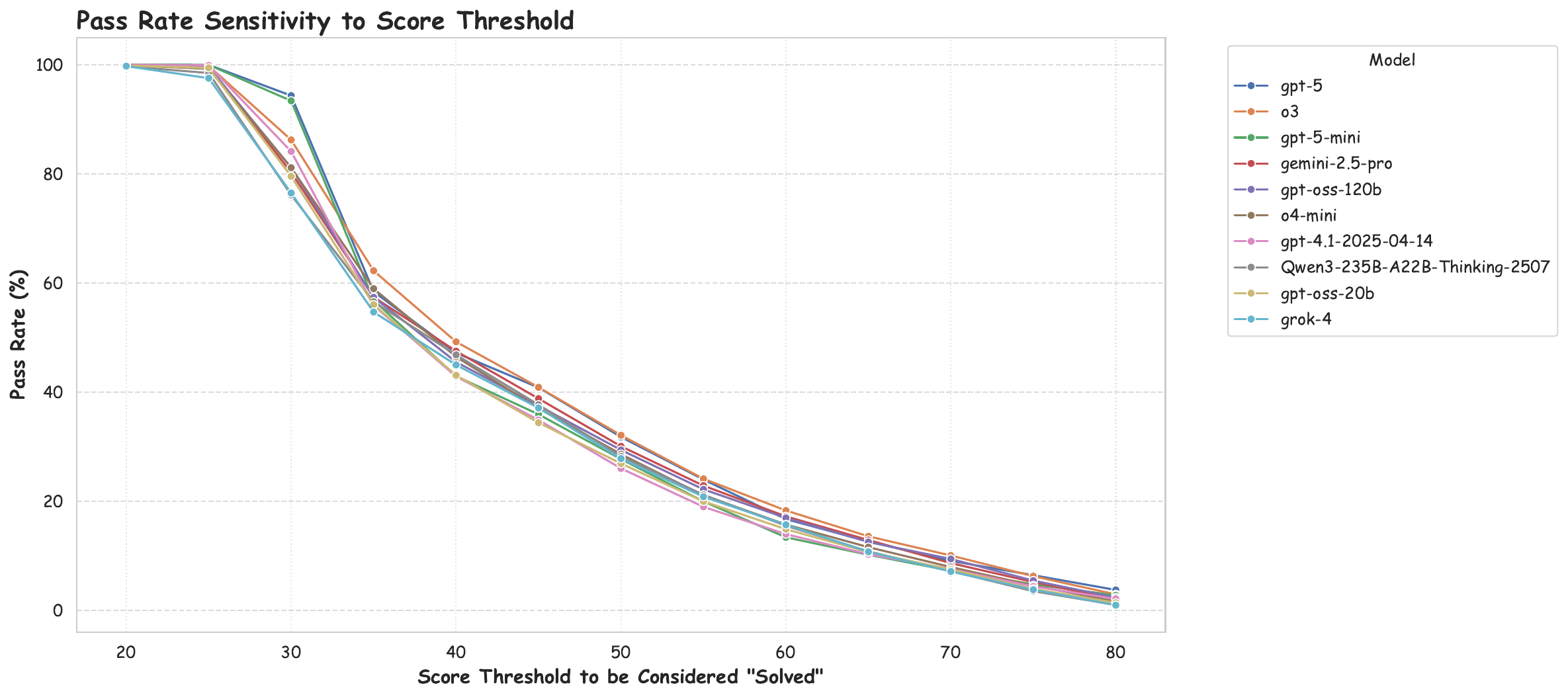}
  \caption{Pass rate variations across different score thresholds, demonstrating ranking stability.}
  \label{fig:figures/figures_evals/06_threshold_sensitivity.pdf}
\end{figure*}

In Figure~\ref{fig:figures/figures_evals/06_threshold_sensitivity.pdf}, the threshold sensitivity analysis demonstrates remarkable ranking stability across different score cutoffs. As the threshold increases from 20 to 80 points, all models show expected performance degradation, but their relative positions remain largely unchanged. This robustness validates our evaluation methodology and suggests that the observed performance differences reflect genuine capability gaps rather than evaluation artifacts. The parallel decline curves indicate that our scoring system maintains discriminative power across the full quality spectrum.

\section{Score Distribution Characteristics}

\begin{figure*}[h!]
  \centering
  \includegraphics[width=0.85\textwidth]{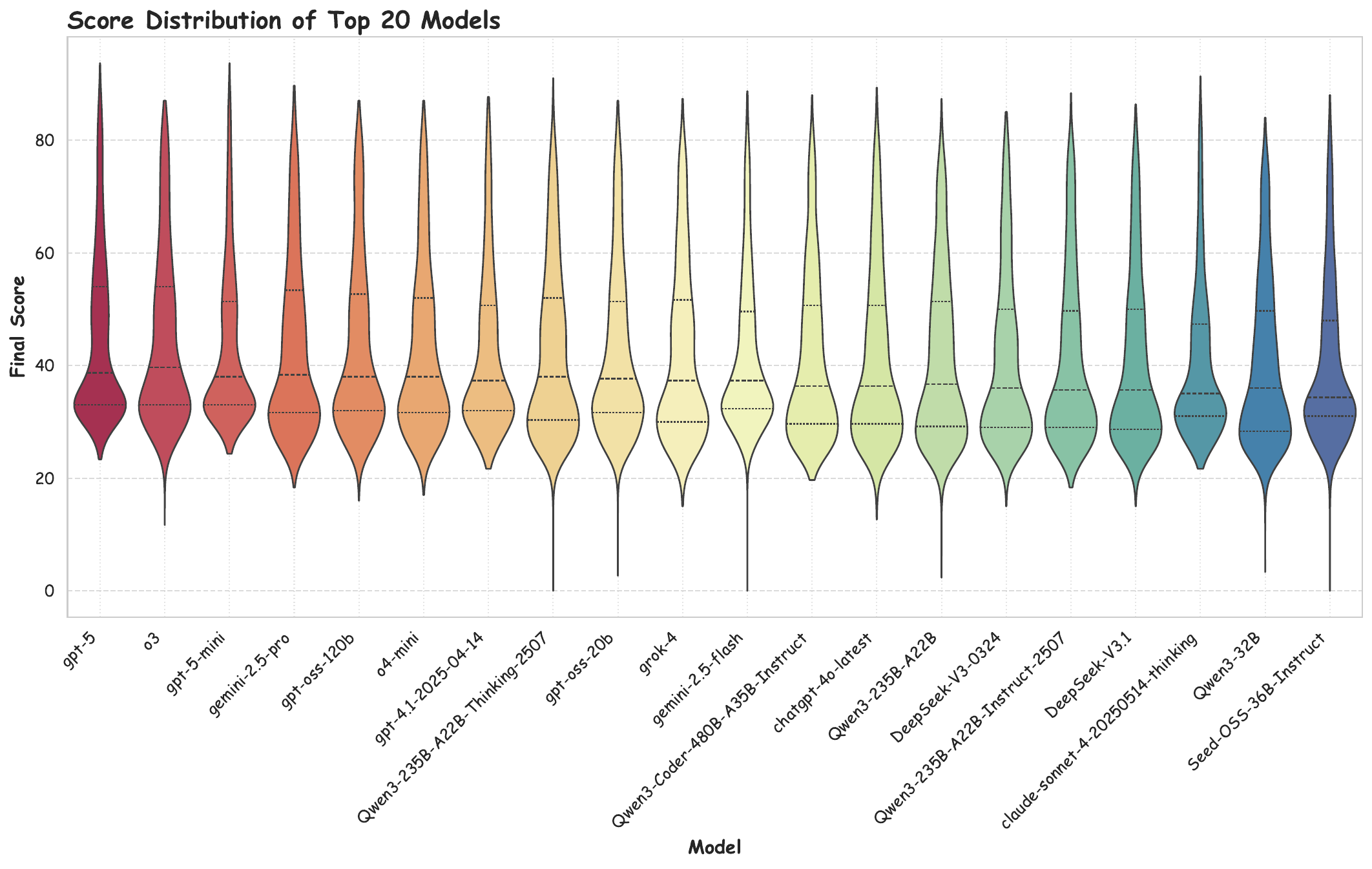}
  \caption{Violin plots showing score distribution patterns for top 20 models.}
  \label{fig:figures/figures_evals/08_score_distribution_violin.pdf}
\end{figure*}

Figure~\ref{fig:figures/figures_evals/08_score_distribution_violin.pdf} reveals diverse score distribution patterns among top-performing models. Some models exhibit narrow, concentrated distributions around their median scores, indicating consistent performance across different game types. Others show broader, multi-modal distributions, suggesting specialized strengths in particular game categories. The distribution shapes provide insights into model reliability - models with tighter distributions may be more predictable for production use, while those with wider distributions might excel in specific domains but struggle with others.

\section{Representative Head-to-Head Comparisons}

\begin{figure*}[h!]
    \centering
    \begin{subfigure}[b]{0.24\textwidth}
        \centering
        \includegraphics[width=\textwidth]{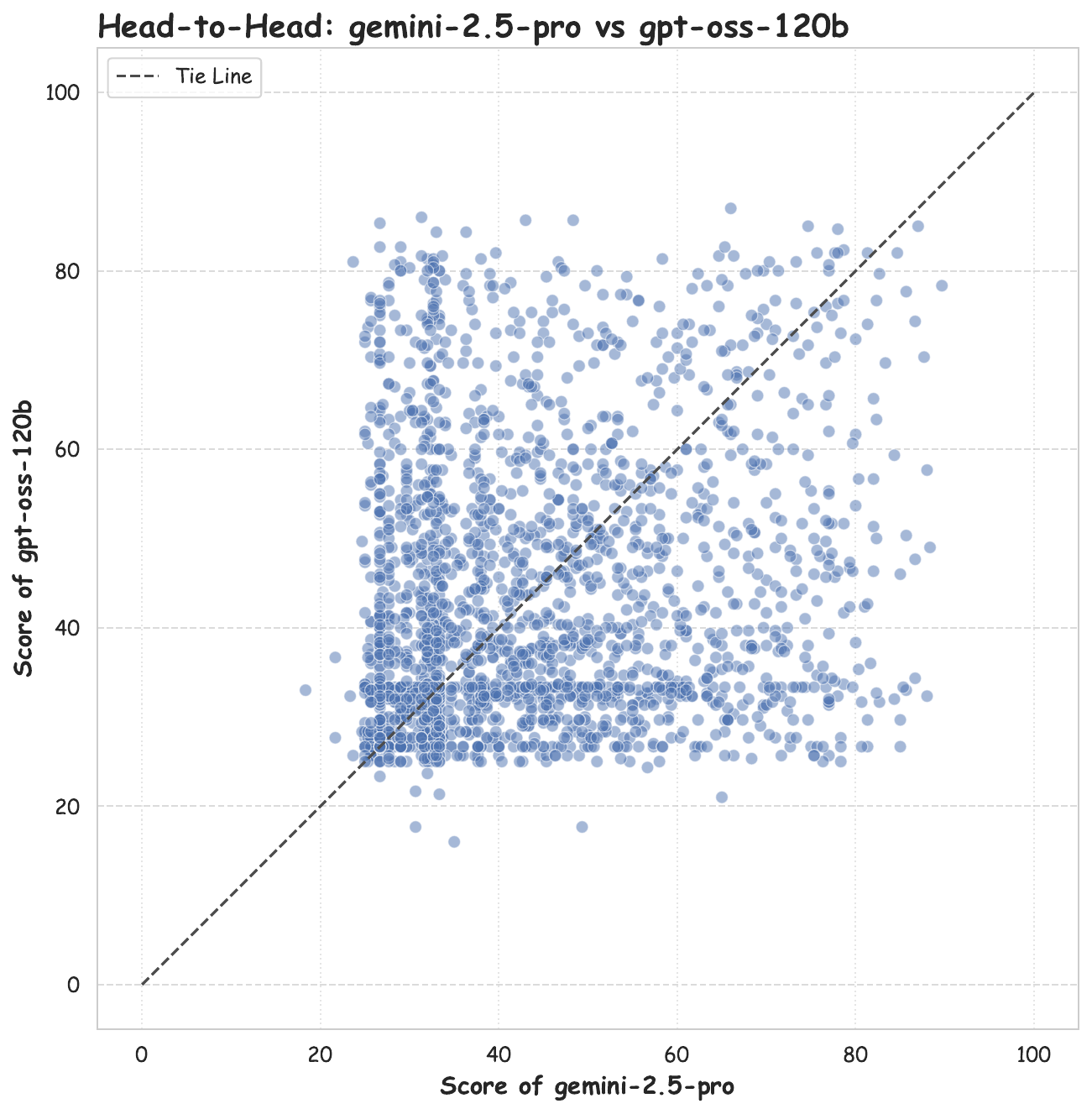}
        \caption{Head-to-head comparison of Gemini-2.5-pro and GPT-OSS-120B.}
    \end{subfigure}
    \begin{subfigure}[b]{0.24\textwidth}
        \centering
        \includegraphics[width=\textwidth]{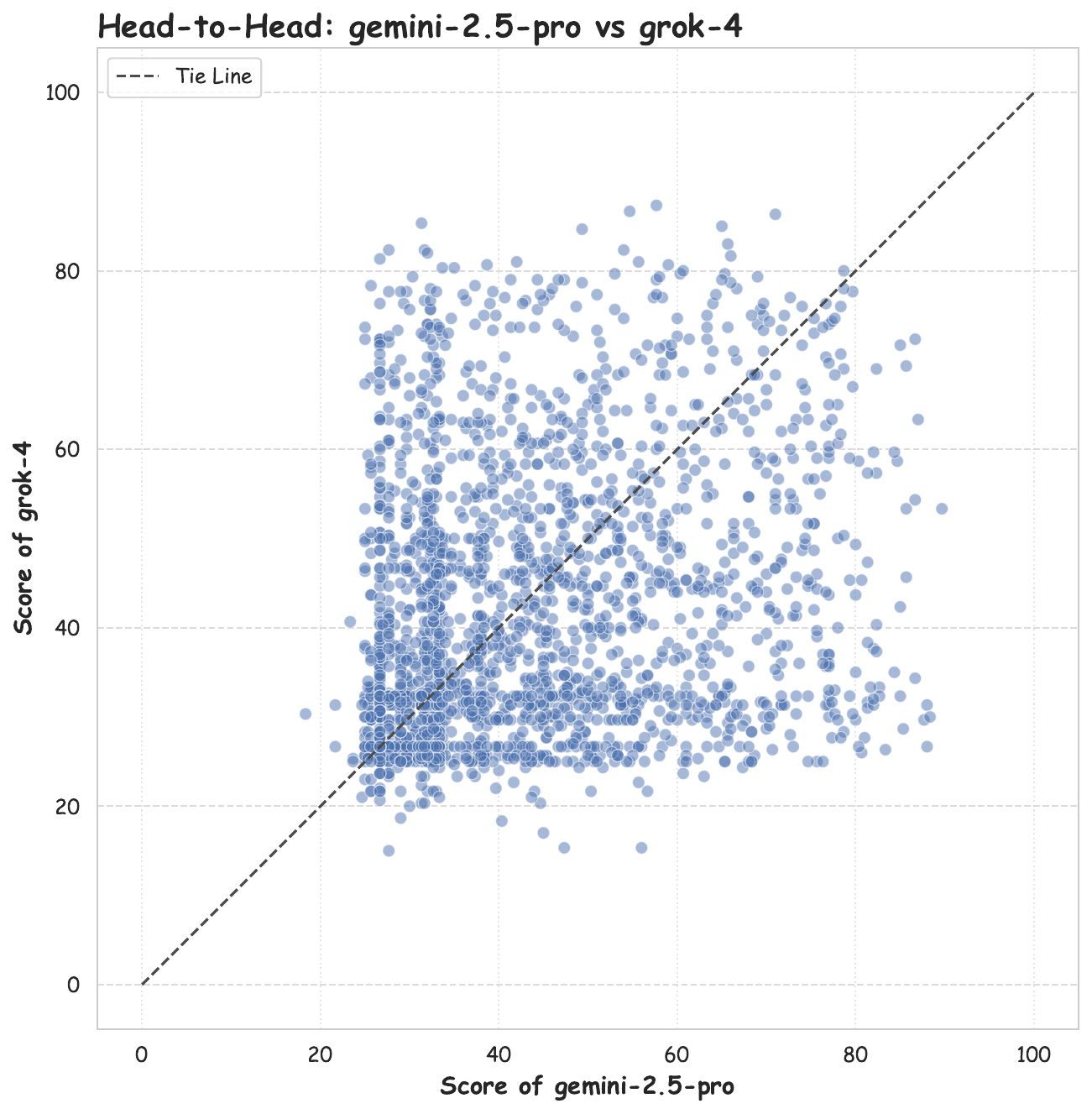}
        \caption{Head-to-head comparison of Gemini-2.5-pro and Grok-4.}
    \end{subfigure}
    \begin{subfigure}[b]{0.24\textwidth}
        \centering
        \includegraphics[width=\textwidth]{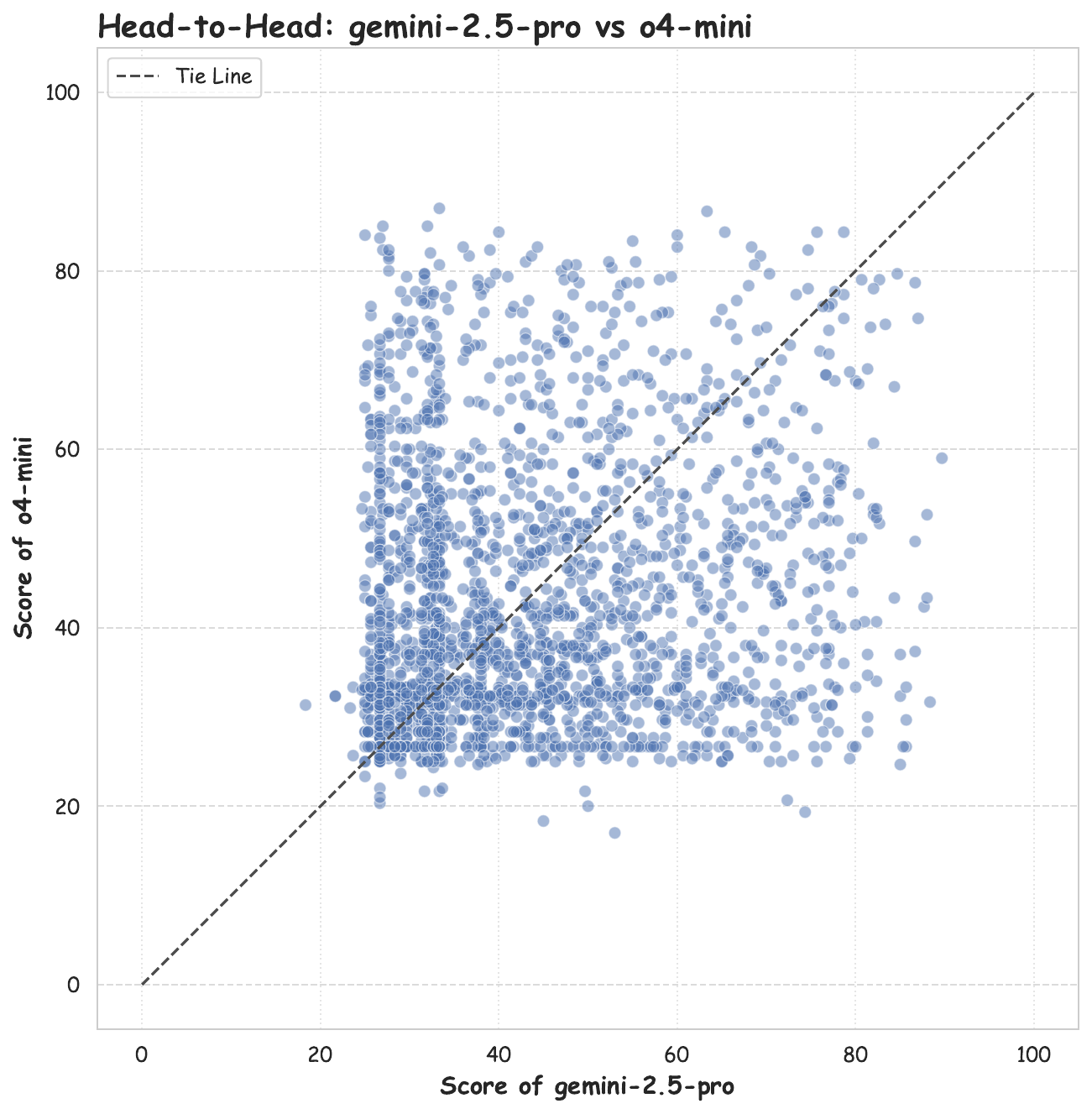}
        \caption{Head-to-head comparison of Gemini-2.5-pro and o4-mini.}
    \end{subfigure}
    \begin{subfigure}[b]{0.24\textwidth}
        \centering
        \includegraphics[width=\textwidth]{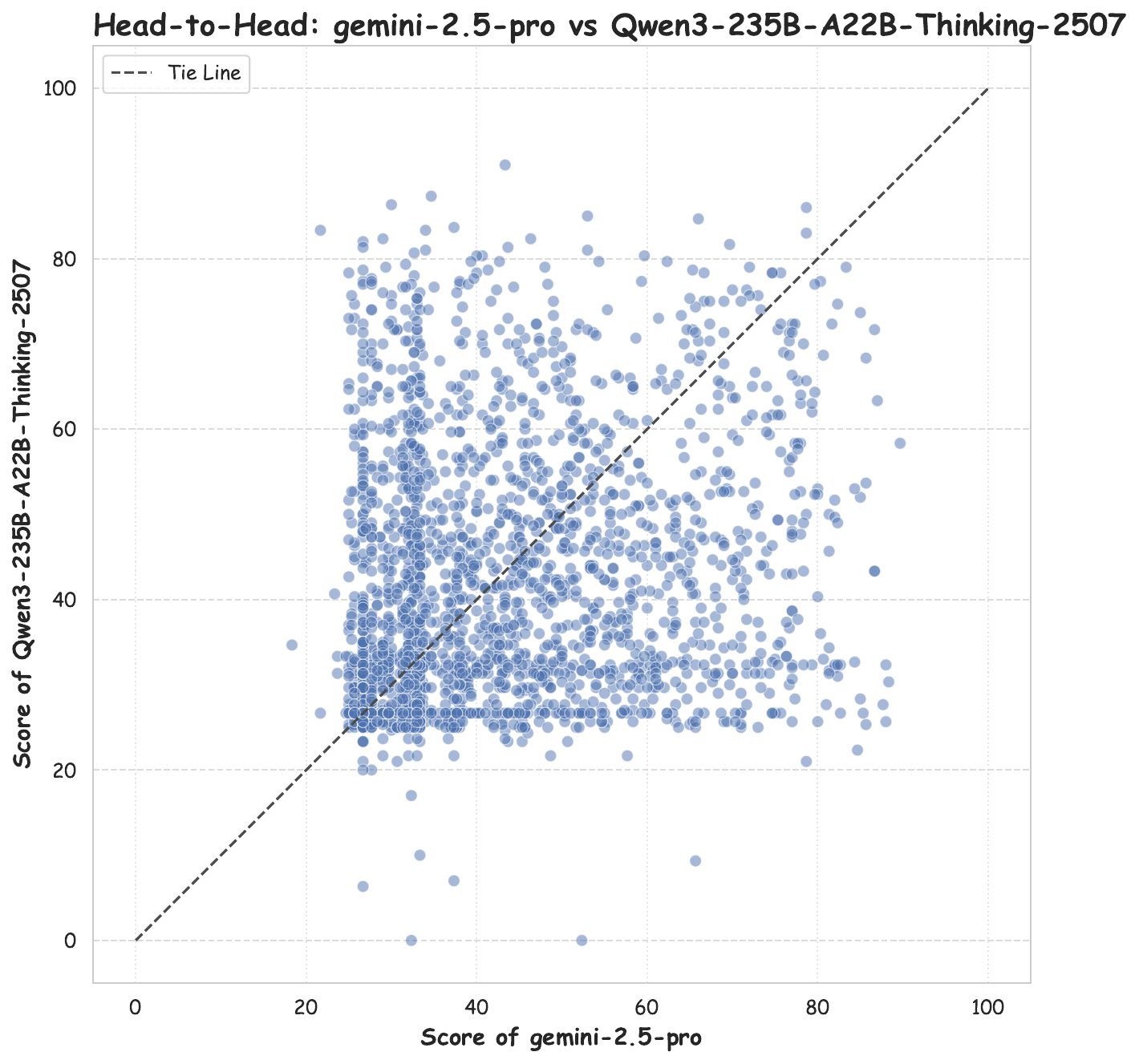}
        \caption{Head-to-head comparison of Gemini-2.5-pro and Qwen3-235B-A22B-Thinking-2507.}
    \end{subfigure}
    \begin{subfigure}[b]{0.24\textwidth}
        \centering
        \includegraphics[width=\textwidth]{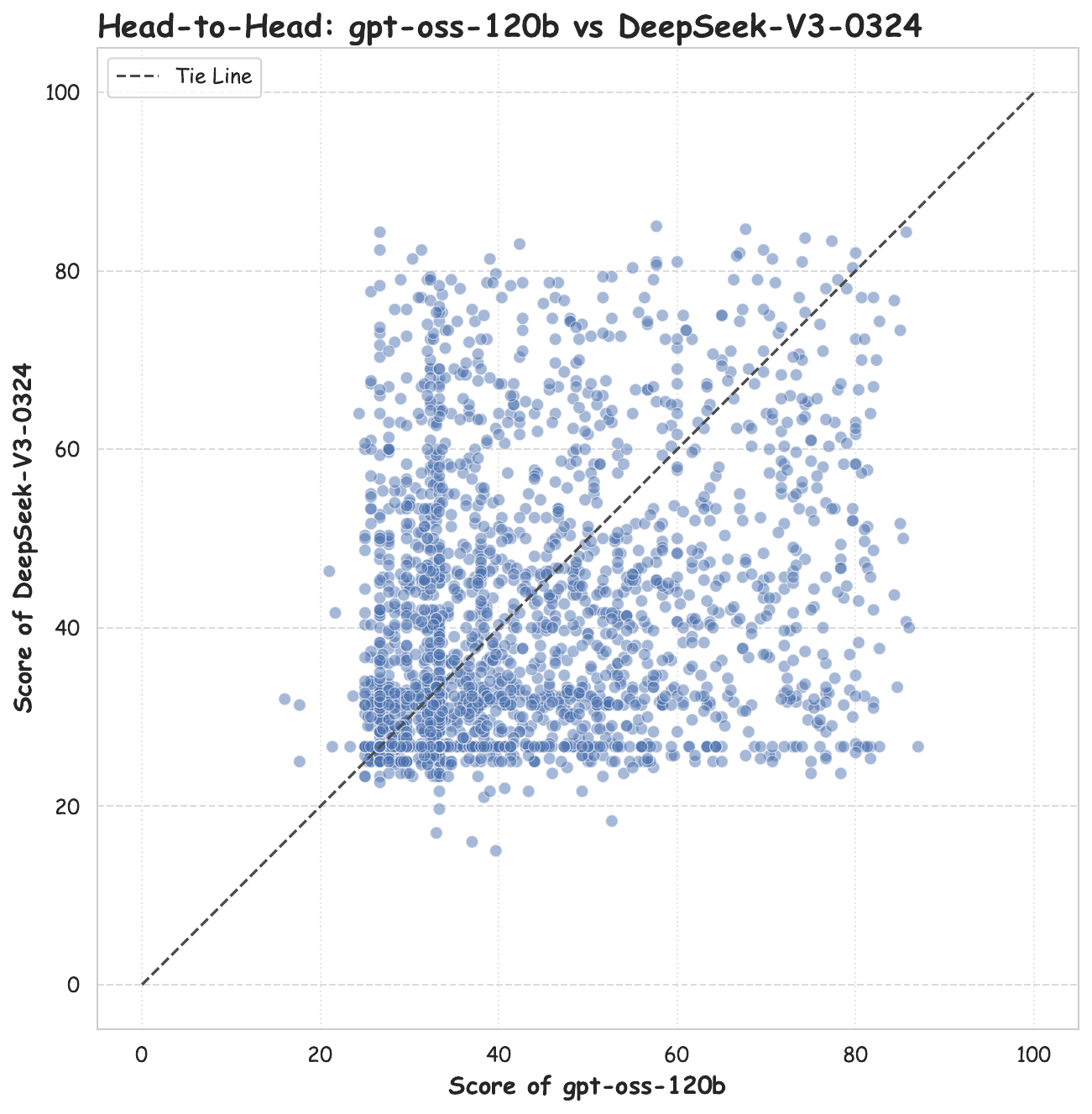}
        \caption{Head-to-head comparison of GPT-OSS-120B and DeepSeek-V3-0324.}
    \end{subfigure}
    \begin{subfigure}[b]{0.24\textwidth}
        \centering
        \includegraphics[width=\textwidth]{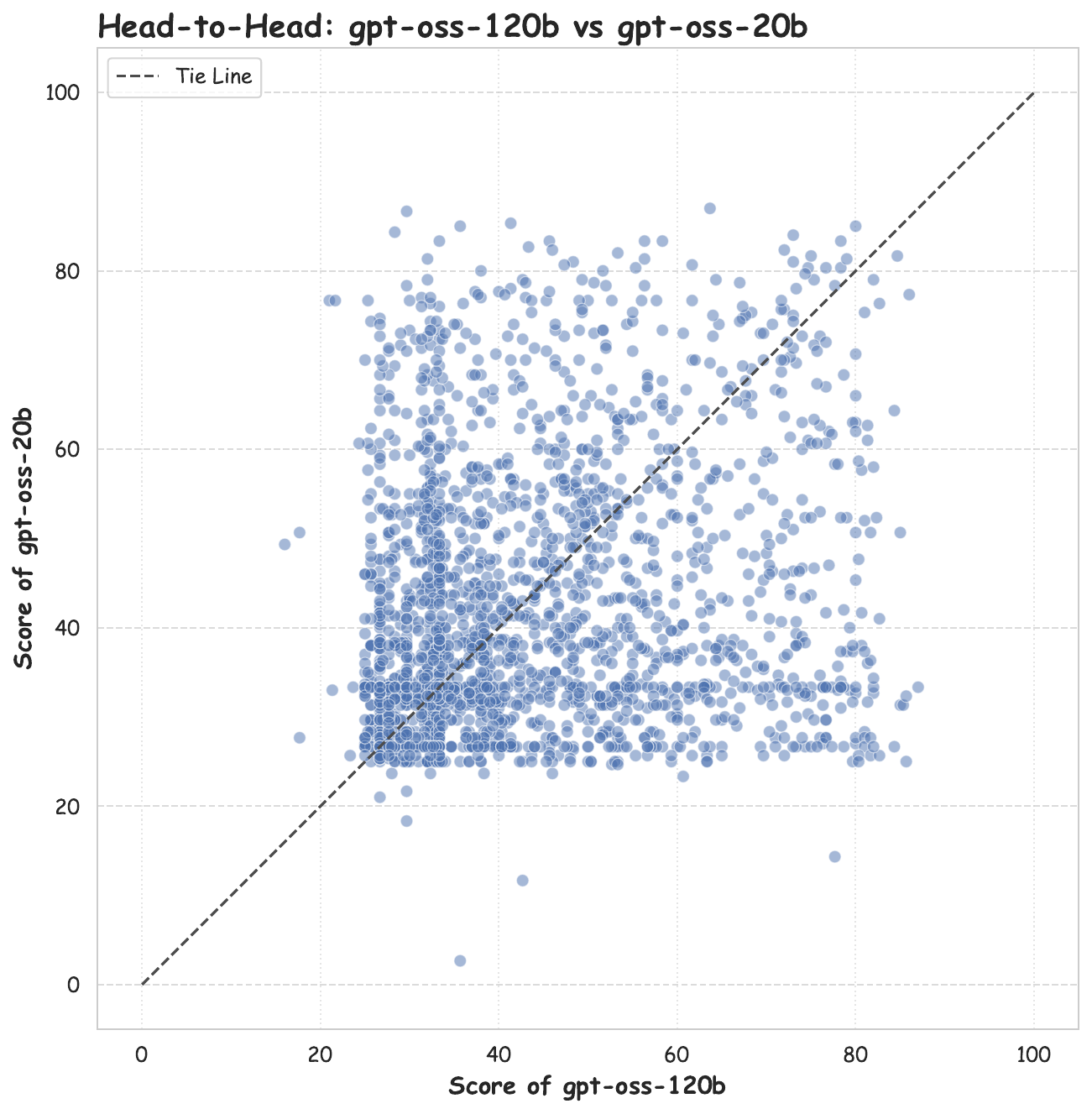}
        \caption{Head-to-head comparison of GPT-OSS-120B and GPT-OSS-20B.}
    \end{subfigure}
    \begin{subfigure}[b]{0.24\textwidth}
        \centering
        \includegraphics[width=\textwidth]{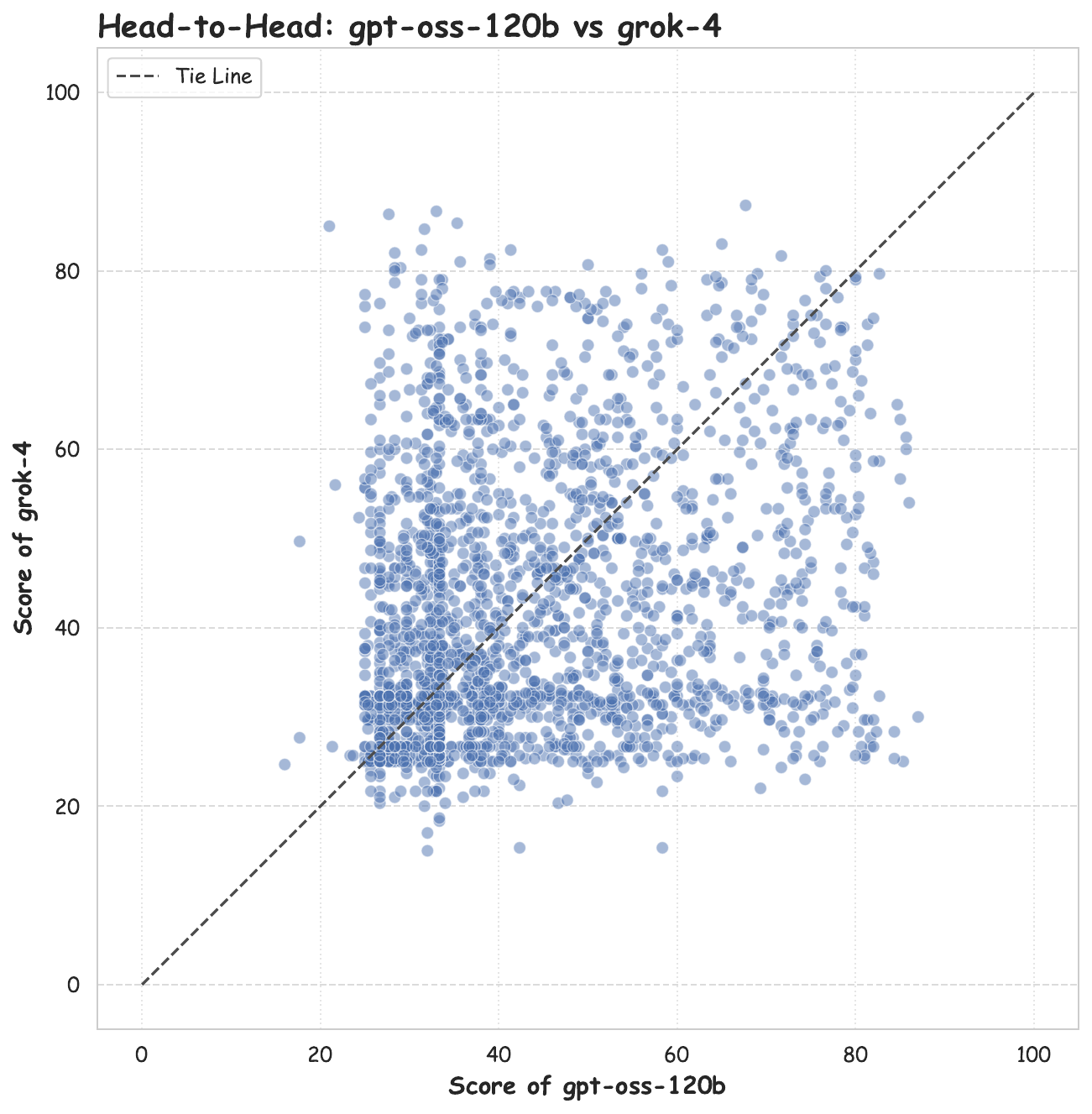}
        \caption{Head-to-head comparison of GPT-OSS-120B and Grok-4.}
    \end{subfigure}
    \begin{subfigure}[b]{0.24\textwidth}
        \centering
        \includegraphics[width=\textwidth]{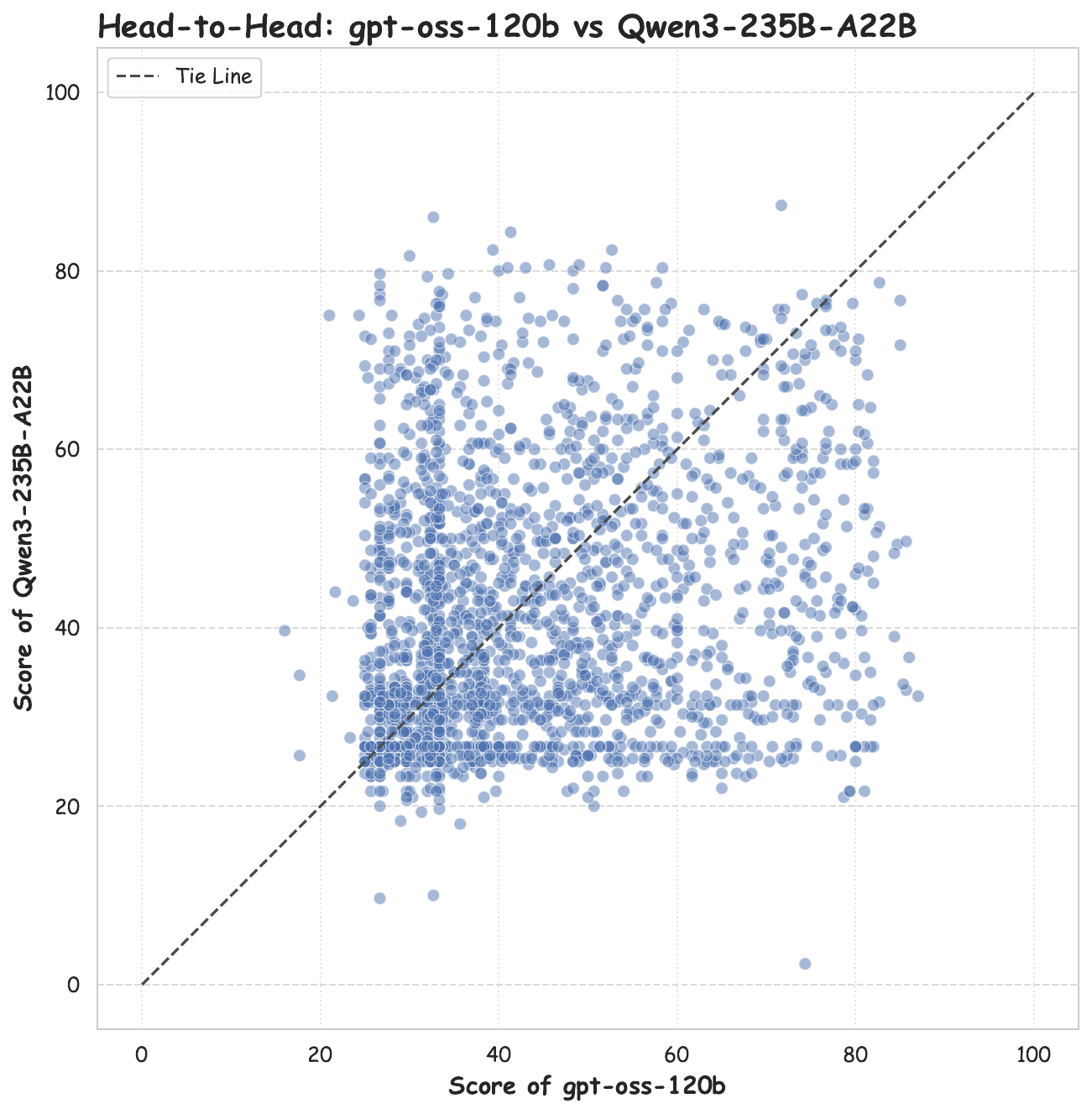}
        \caption{Head-to-head comparison of GPT-OSS-120B and Qwen3-235B-A22B.}
    \end{subfigure}
    \begin{subfigure}[b]{0.24\textwidth}
        \centering
        \includegraphics[width=\textwidth]{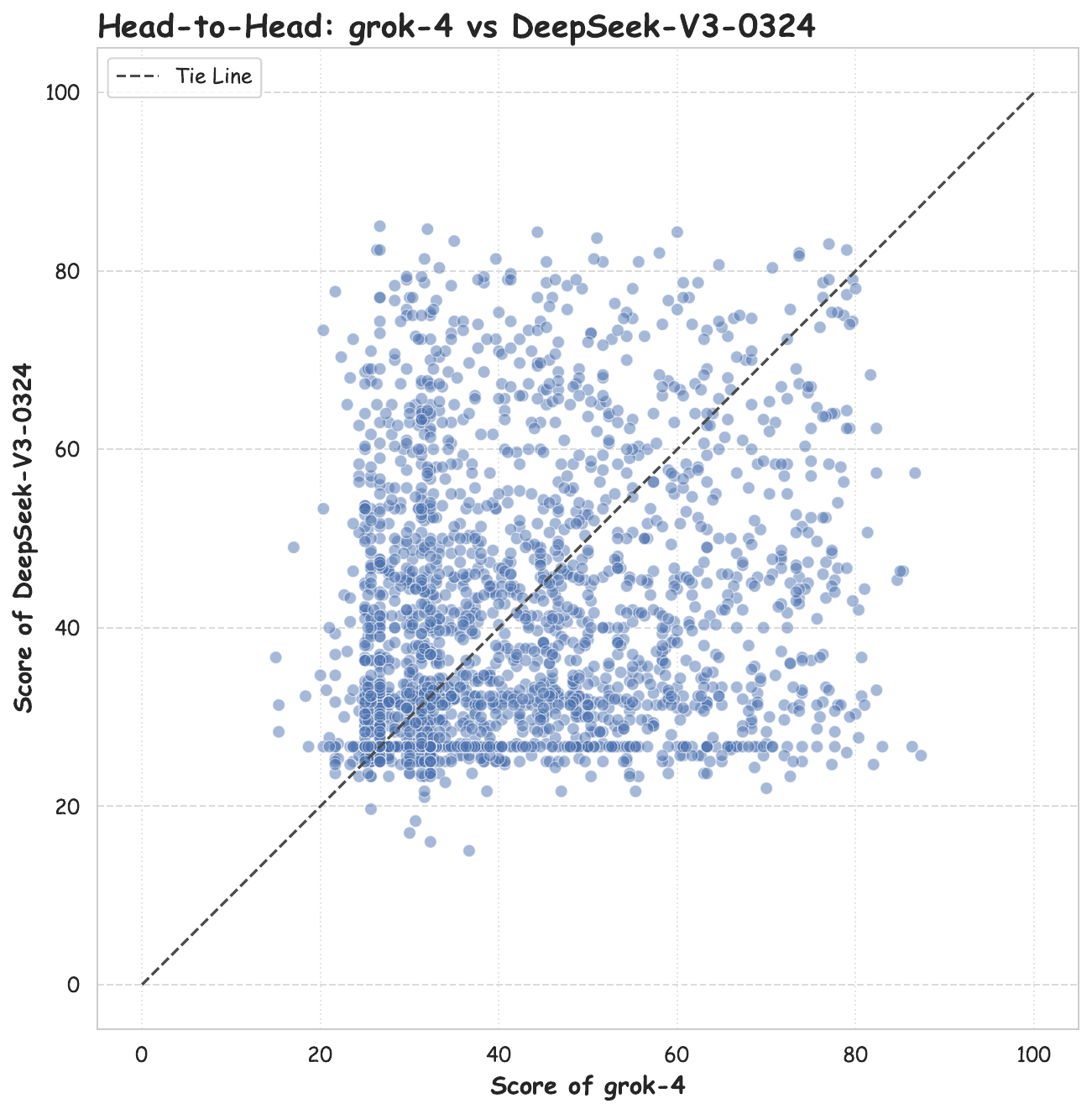}
        \caption{Head-to-head comparison of Grok-4 and DeepSeek-V3-0324.}
    \end{subfigure}
    \begin{subfigure}[b]{0.24\textwidth}
        \centering
        \includegraphics[width=\textwidth]{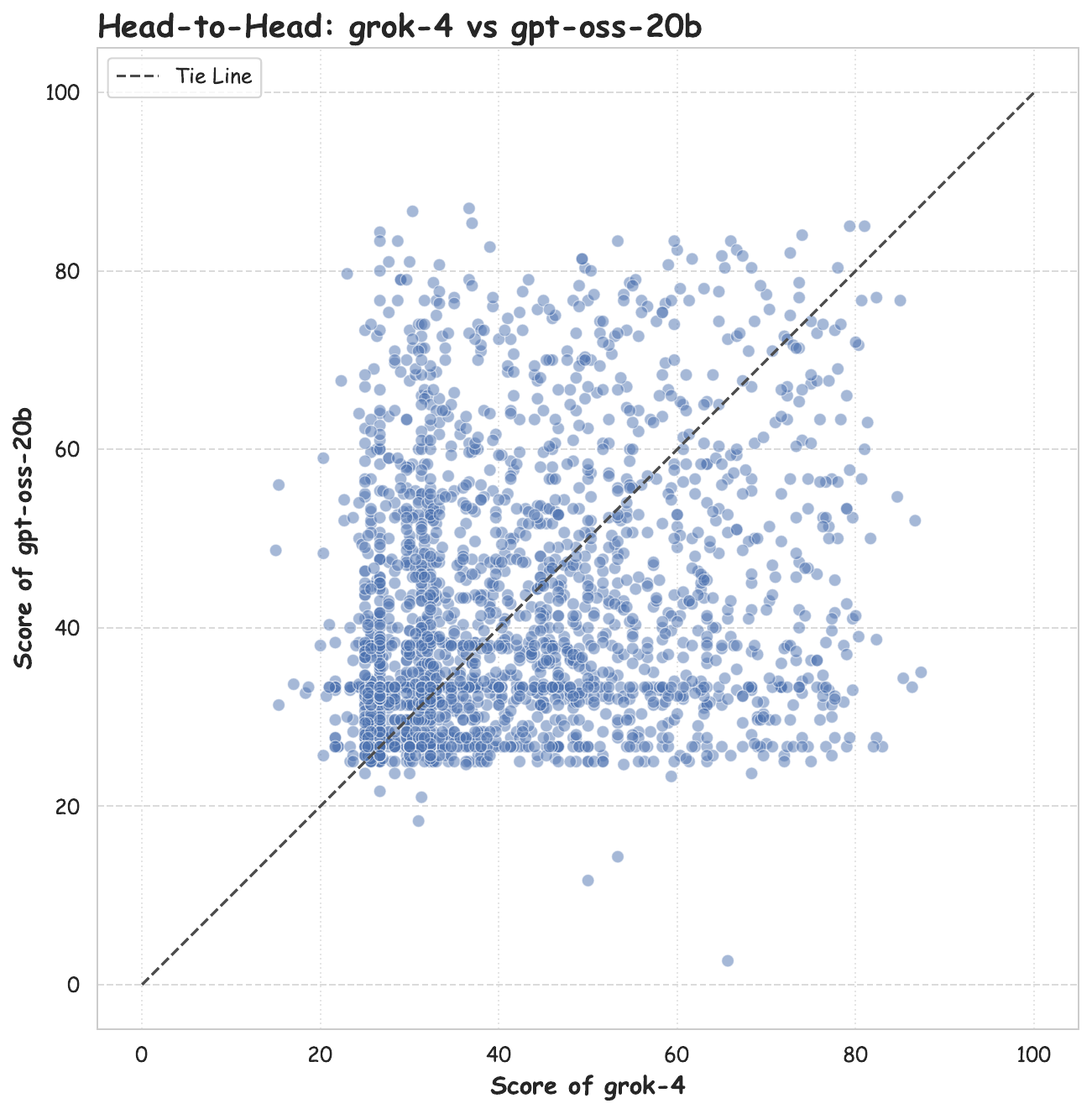}
        \caption{Head-to-head comparison of Grok-4 and GPT-OSS-20B.}
    \end{subfigure}
    \begin{subfigure}[b]{0.24\textwidth}
        \centering
        \includegraphics[width=\textwidth]{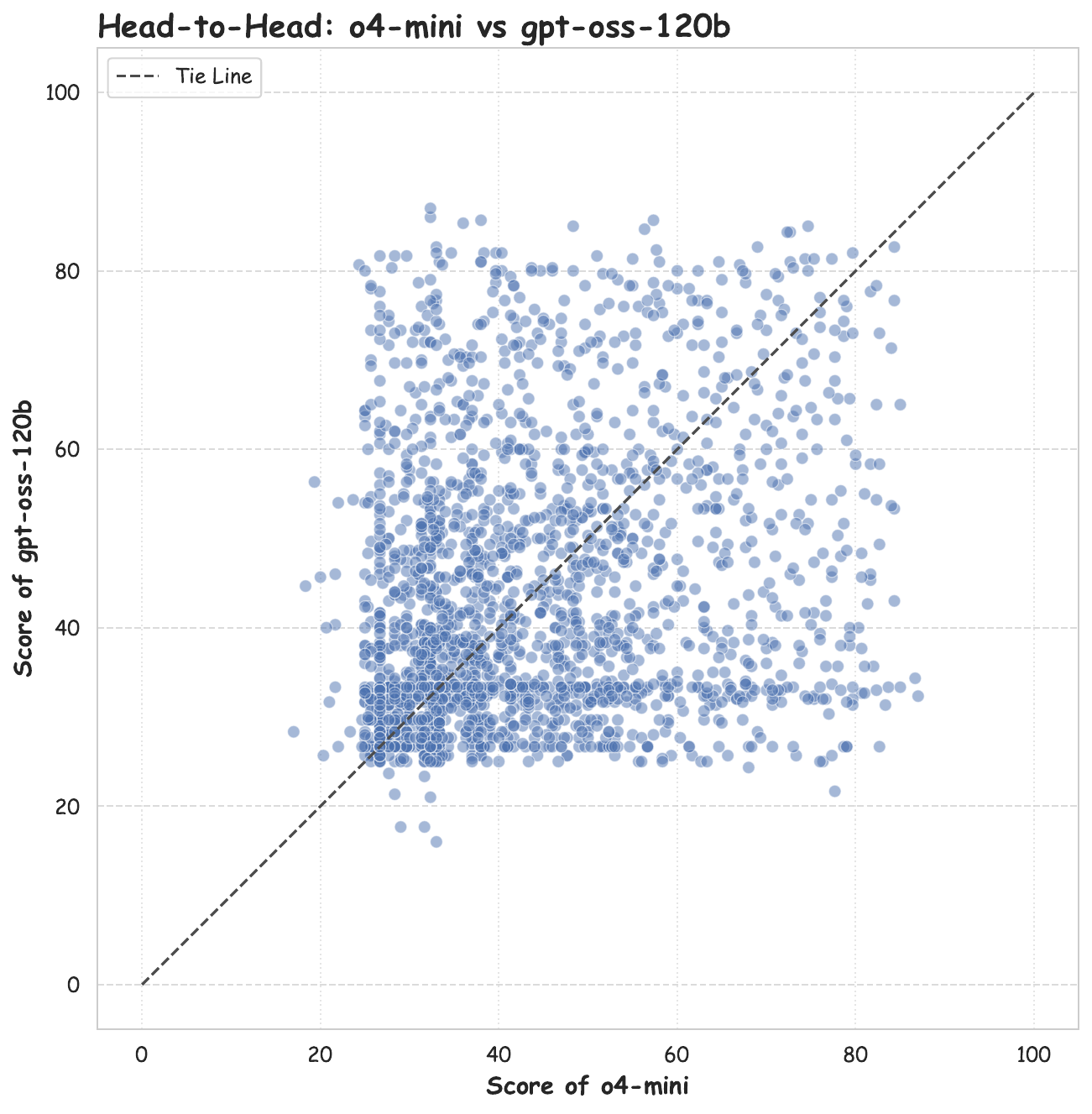}
        \caption{Head-to-head comparison of o4-mini and GPT-OSS-120B.}
    \end{subfigure}
    \begin{subfigure}[b]{0.24\textwidth}
        \centering
        \includegraphics[width=\textwidth]{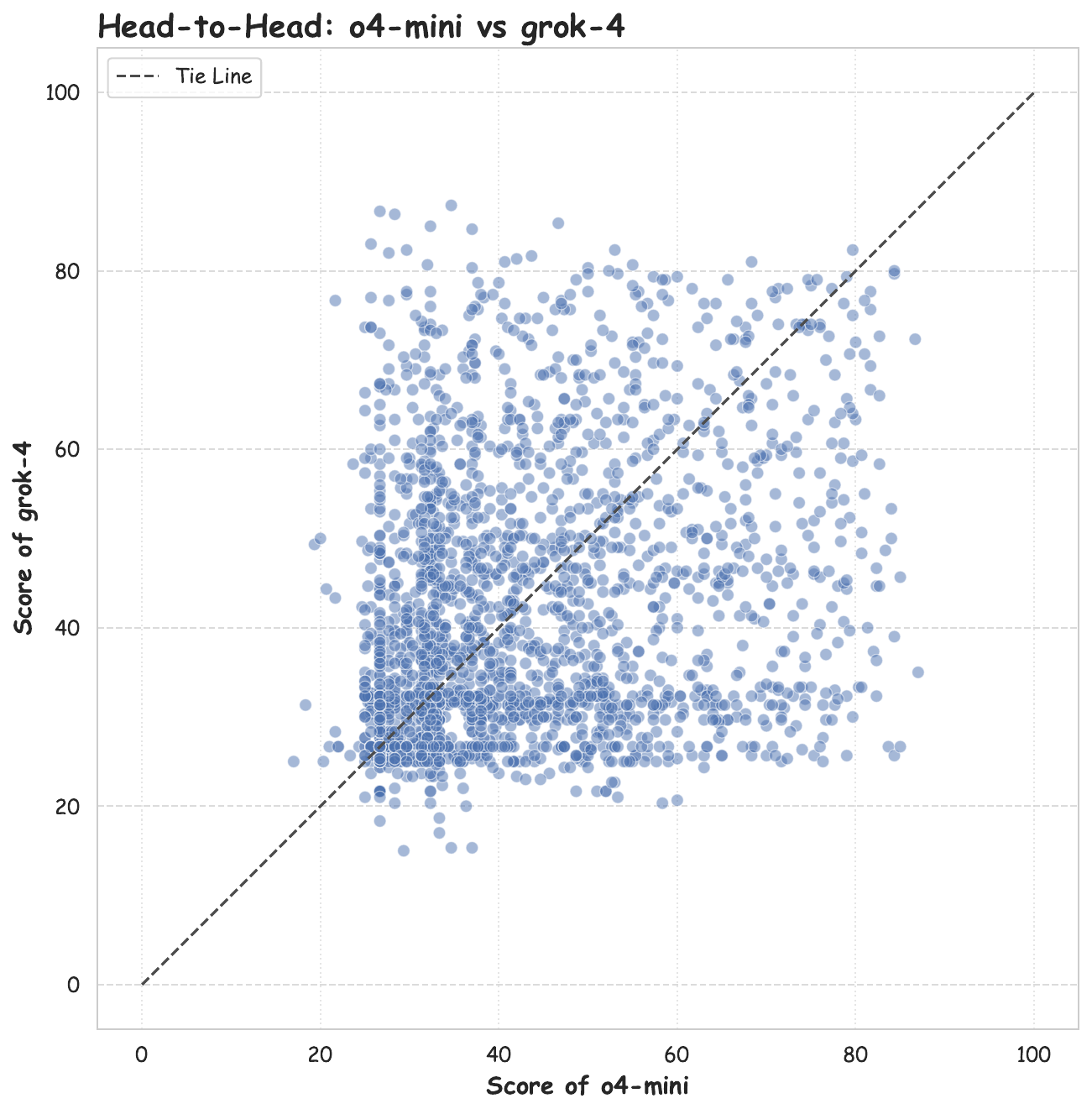}
        \caption{Head-to-head comparison of o4-mini and Grok-4.}
    \end{subfigure}
    \begin{subfigure}[b]{0.24\textwidth}
        \centering
        \includegraphics[width=\textwidth]{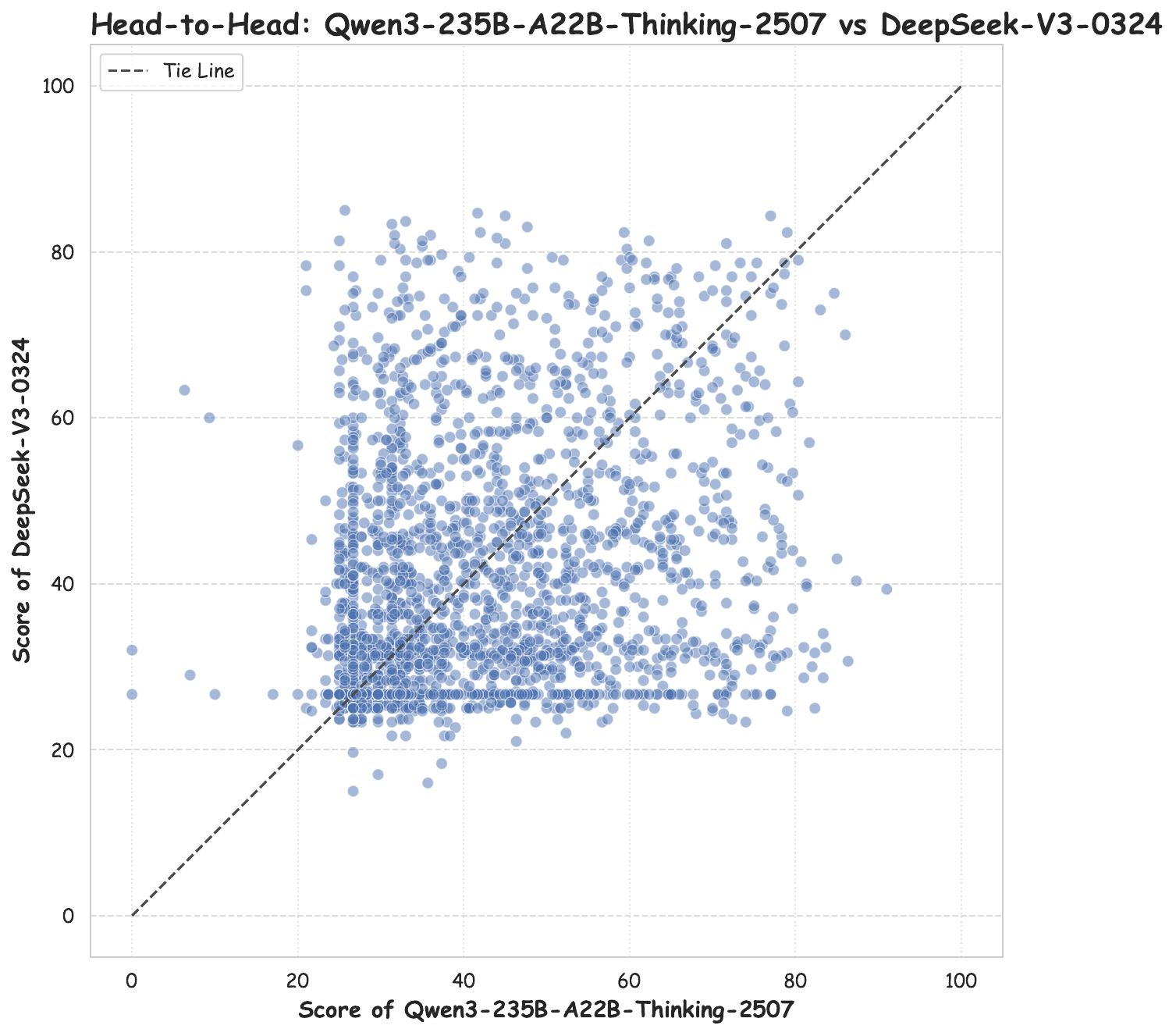}
        \caption{Head-to-head comparison of Qwen3-235B-A22B-Thinking-2507 and DeepSeek-V3-0324.}
    \end{subfigure}
    \begin{subfigure}[b]{0.24\textwidth}
        \centering
        \includegraphics[width=\textwidth]{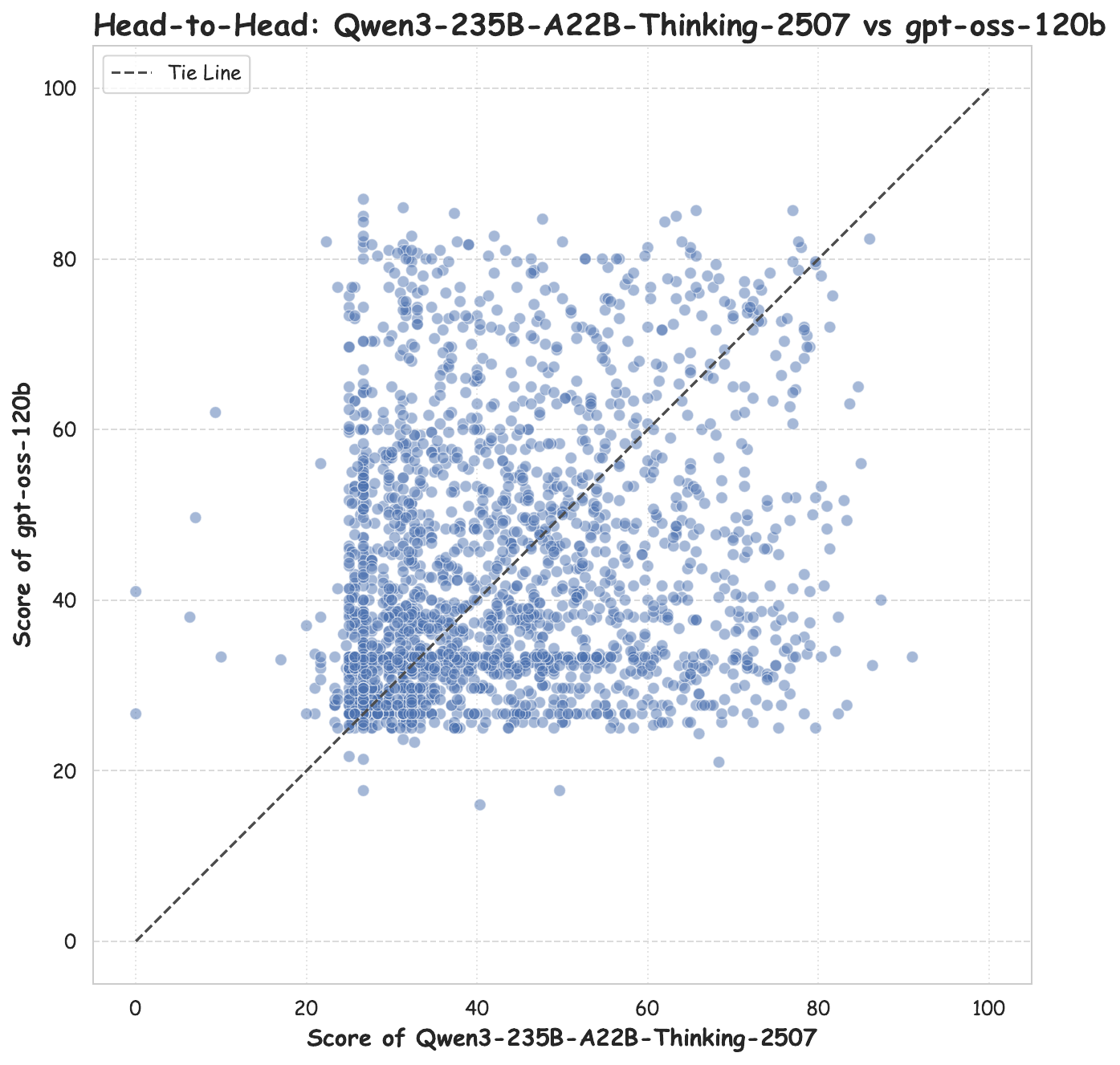}
        \caption{Head-to-head comparison of Qwen3-235B-A22B-Thinking-2507 and GPT-OSS-120B.}
    \end{subfigure}
    \begin{subfigure}[b]{0.24\textwidth}
        \centering
        \includegraphics[width=\textwidth]{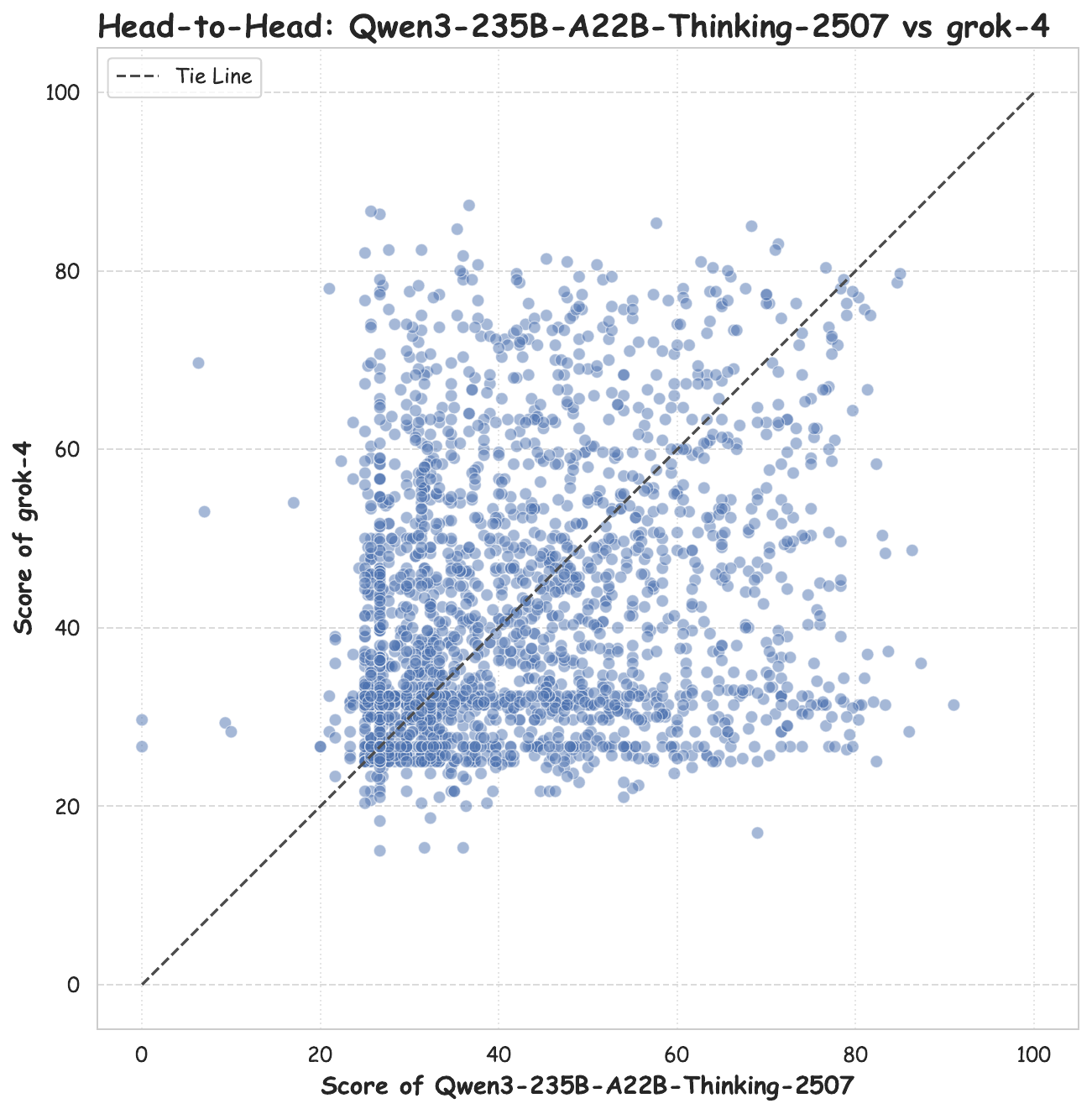}
        \caption{Head-to-head comparison of Qwen3-235B-A22B-Thinking-2507 and Grok-4.}
    \end{subfigure}
    \begin{subfigure}[b]{0.24\textwidth}
        \centering
        \includegraphics[width=\textwidth]{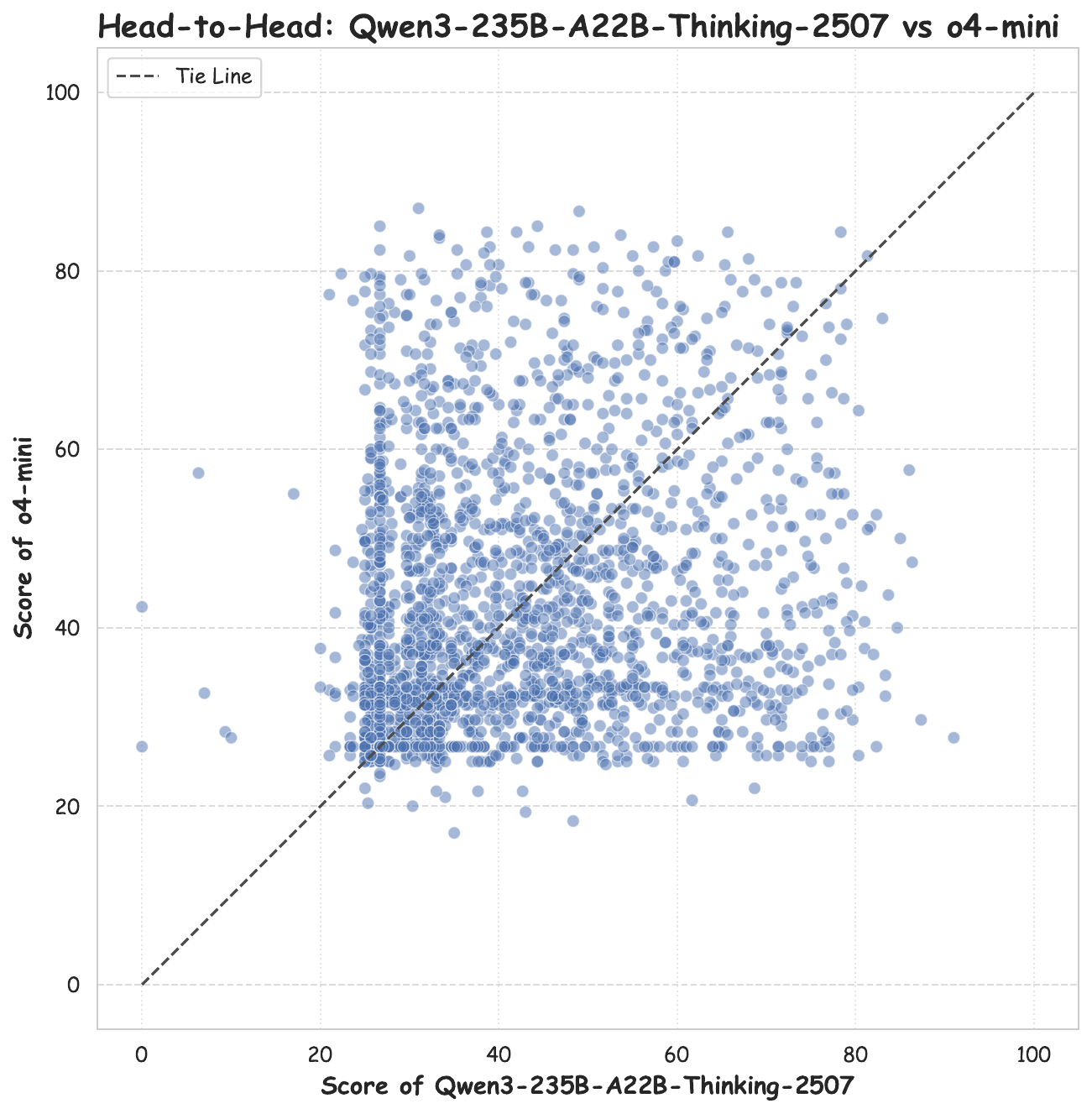}
        \caption{Head-to-head comparison of Qwen3-235B-A22B-Thinking-2507 and o4-mini.}
    \end{subfigure}
    \caption{Direct performance comparisons between selected model pairs showing competitive advantages across individual games.}
    \label{fig:figures/figures_evals/10_head_to_head}
  \end{figure*}

In Figure~\ref{fig:figures/figures_evals/10_head_to_head}, the head-to-head comparisons reveal nuanced competitive dynamics between leading models. Points above the diagonal line indicate games where the y-axis model outperforms the x-axis model, and vice versa. The scatter patterns show that even among top-tier models, performance advantages are game-specific rather than universal. Some model pairs exhibit complementary strengths, suggesting potential ensemble benefits. The analysis also reveals that certain games consistently favor particular model architectures, indicating systematic biases in problem-solving approaches.

\section{Comprehensive Performance Heatmap}

\begin{figure*}[ht]
  \centering
  \includegraphics[width=1.0\textwidth]{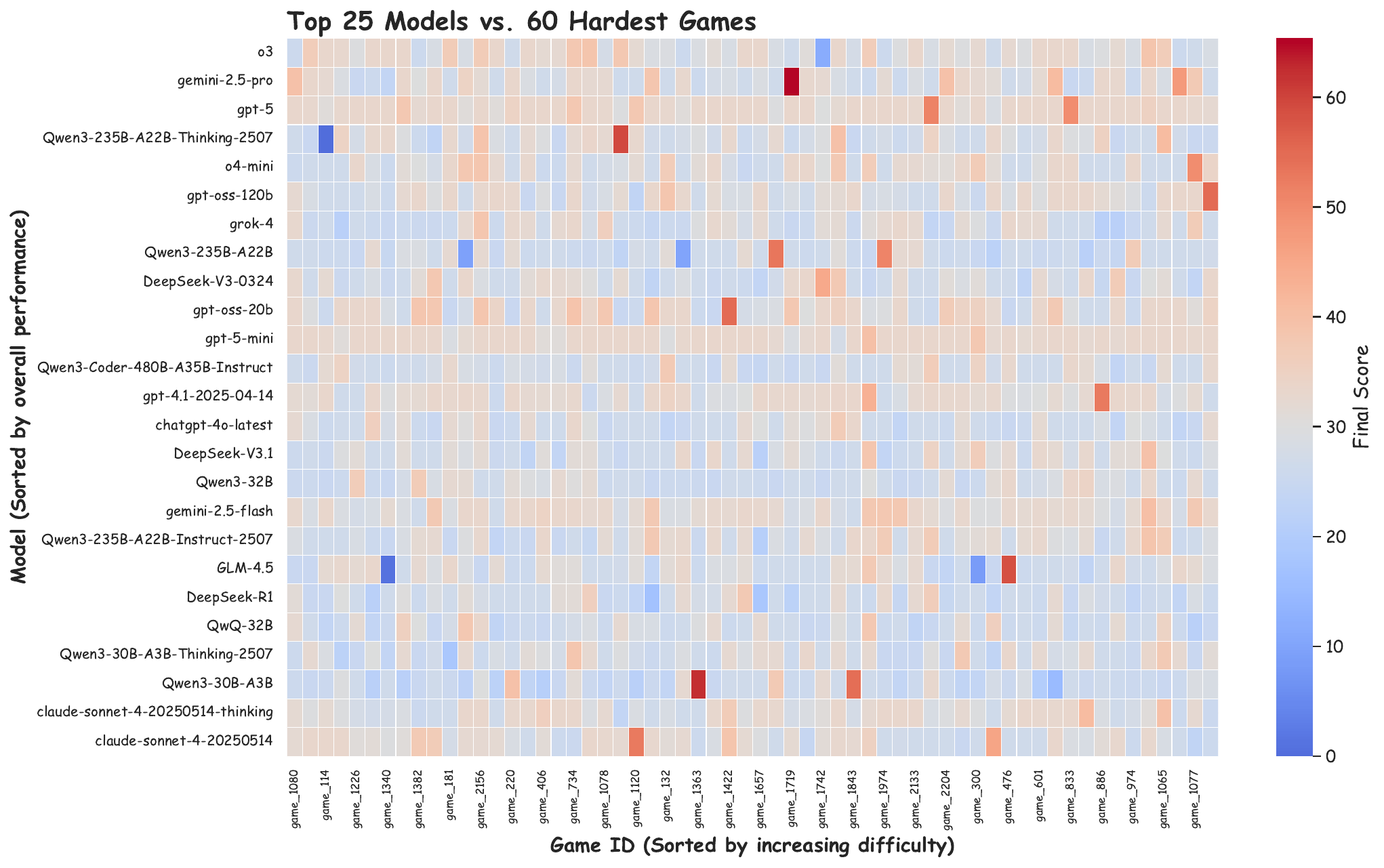}
  \caption{Performance matrix of top 25 models on the 60 most challenging games, ordered by increasing difficulty and overall model performance.}
  \label{fig:figures/figures_evals/14_model_performance_heatmap.pdf}
\end{figure*}

In Figure~\ref{fig:figures/figures_evals/14_model_performance_heatmap.pdf}, the performance heatmap provides a granular view of model capabilities across the most challenging benchmark subset. The clear progression from lighter (better performance) to darker (poorer performance) colors as difficulty increases confirms the validity of our difficulty ordering. Notably, even the highest-performing models struggle with the rightmost games, indicating these represent genuine frontier challenges. The heatmap also reveals interesting patterns where certain models show unexpected strength on specific difficult games, suggesting specialized capabilities that average performance metrics might obscure. The clustering of similar performance patterns across model families reinforces the architectural influence on problem-solving approaches.

\clearpage

\section{Seed Code Dataset Quality Analysis}

To provide comprehensive insights into the characteristics and quality of our seed dataset, we conducted extensive statistical analysis across multiple dimensions. The analysis encompasses cluster distribution, quality metrics, file characteristics, module usage patterns, and structural complexity.

\begin{figure*}[ht]
\centering
\includegraphics[width=0.7\textwidth]{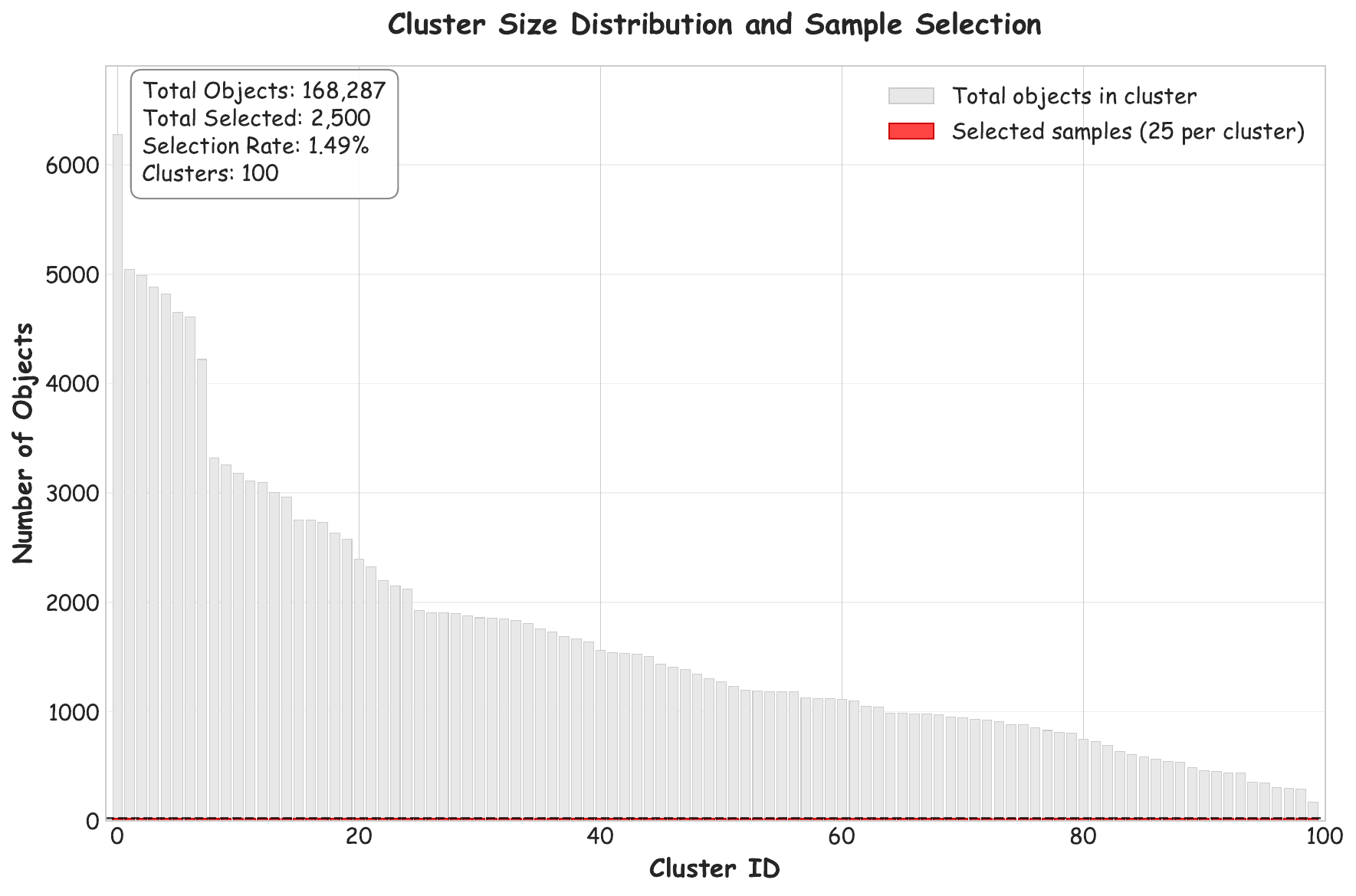}
\caption{Cluster size distribution and sample selection strategy, showing uniform selection of 25 samples from each of the 100 clusters across 168,287 total objects.}
\label{fig:figures/figures_seed/01_cluster_coverage.pdf}
\vspace{-20pt}
\end{figure*}

\begin{figure*}[ht]
\centering
\includegraphics[width=0.75\textwidth]{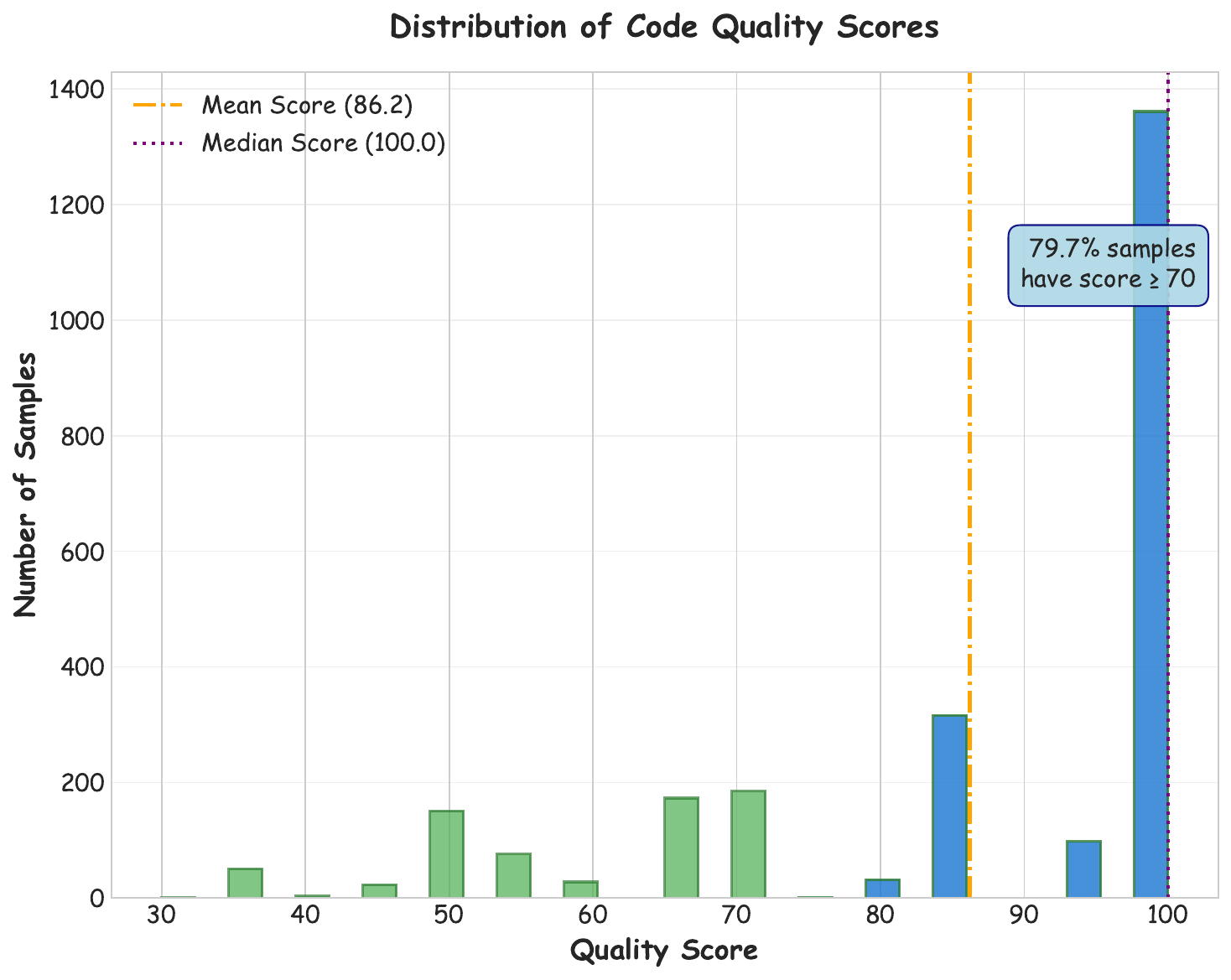}
\caption{Distribution of code quality scores across 2,500 selected samples. And 2500 is the seed set selected after clustering, and 2219 is the final test set after LLM pipeline and manual verification.}
\label{fig:figures/figures_seed/02_quality_distribution.pdf}
\end{figure*}

\paragraph{Cluster Coverage and Sample Selection}
Figure~\ref{fig:figures/figures_seed/01_cluster_coverage.pdf} illustrates the distribution of objects across our 100 clusters and the uniform selection strategy employed. The analysis reveals significant variation in cluster sizes, ranging from hundreds to thousands of objects per cluster, with a total of 168,287 objects processed. Our systematic selection of 25 samples per cluster ensures balanced representation across all functional categories, achieving a 1.49\% overall selection rate. This uniform sampling strategy effectively mitigates bias toward popular game types while maintaining diversity across the entire functional spectrum.

\begin{figure*}[t!]
\centering
\includegraphics[width=0.75\textwidth]{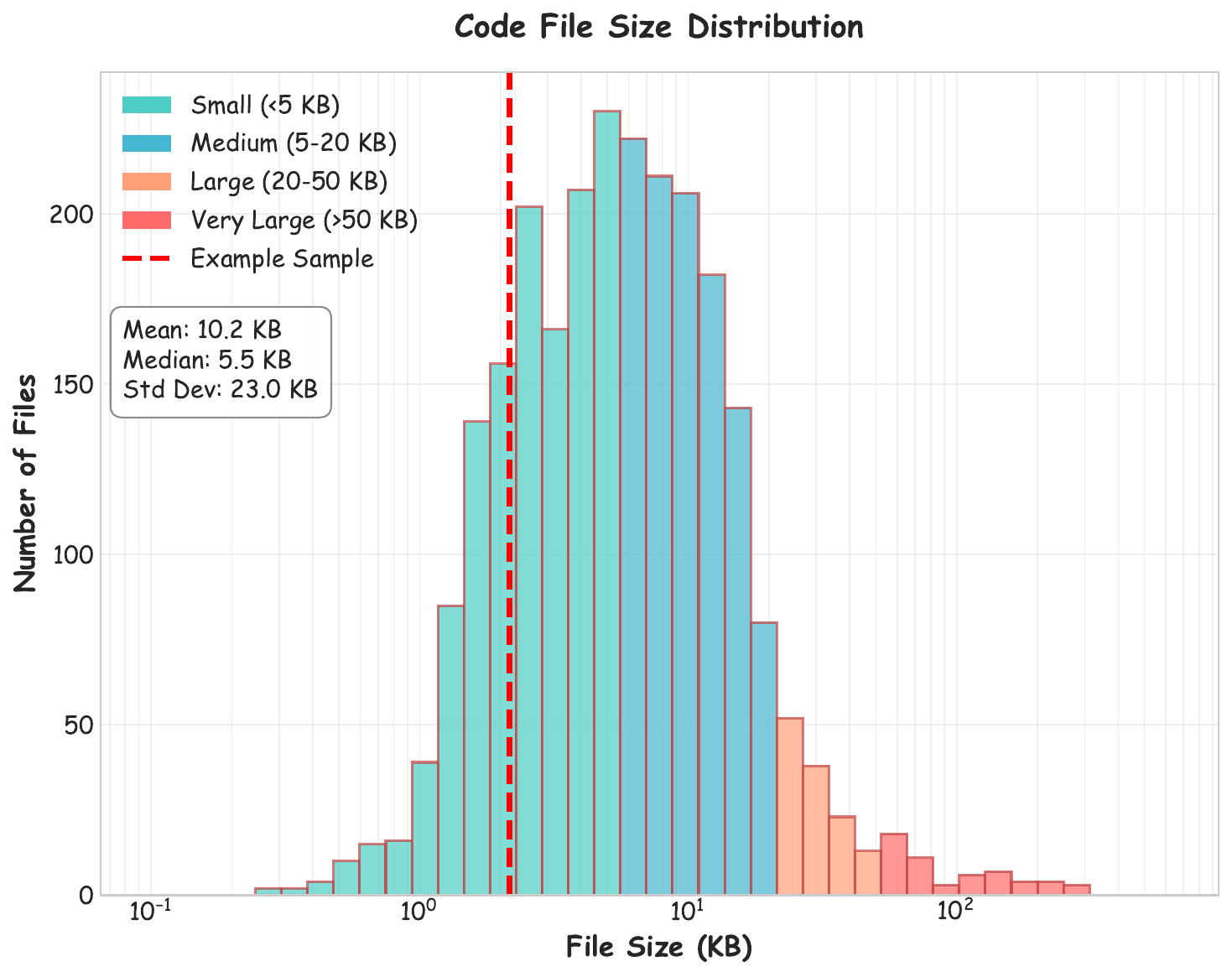}
\caption{File size distribution showing log-normal characteristics with mean 10.2 KB and median 5.5 KB, indicating predominantly compact but complete implementations.}
\label{fig:figures/figures_seed/03_file_size_distribution.pdf}
\vspace{-20pt}
\end{figure*}

\begin{figure*}[t!]
\centering
\includegraphics[width=0.75\textwidth]{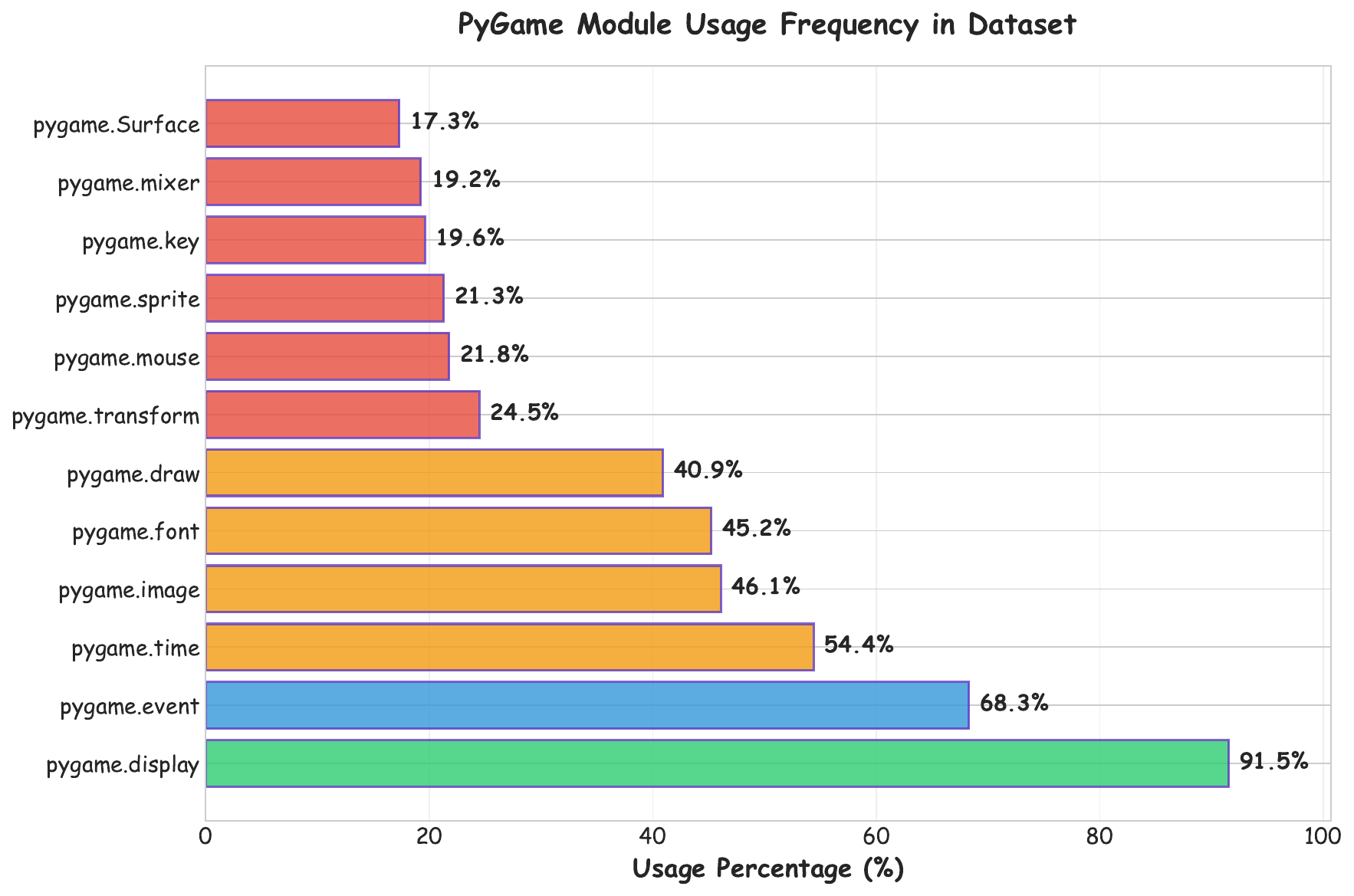}
\caption{Frequency analysis of Pygame module usage, with core modules like display (91.5\%) and event handling (68.3\%) showing high adoption rates.}
\label{fig:figures/figures_seed/04_pygame_module_usage.pdf}
\end{figure*}

\paragraph{Quality Score Distribution Analysis}
Figure~\ref{fig:figures/figures_seed/02_quality_distribution.pdf} demonstrates the high quality of our curated dataset, with 79.7\% of samples achieving quality scores above 70. The distribution exhibits a strong right skew with a mean score of 86.2 and a median of 100.0, indicating that our clustering-based selection successfully identified structurally complete and well-implemented code samples. The concentration of samples in the high-quality range validates our selection methodology and ensures that the benchmark provides reliable reference implementations for evaluation purposes.

\begin{figure*}[t!]
\centering
\includegraphics[width=0.8\textwidth]{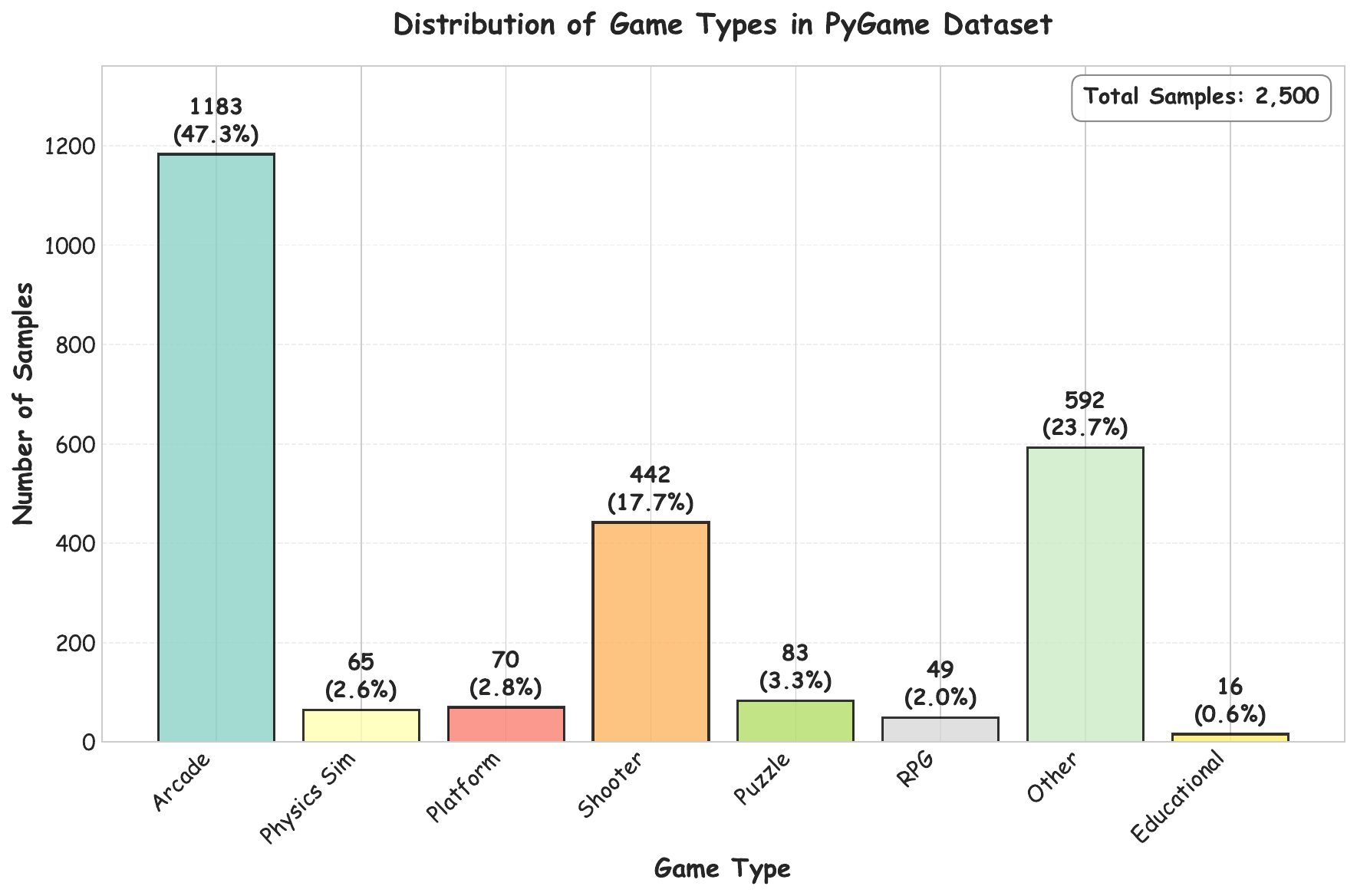}
\caption{Distribution of game types in the dataset, representation across eight distinct genres.}
\label{fig:figures/figures_seed/05_game_type_distribution.pdf}
\vspace{-20pt}
\end{figure*}

\begin{figure*}[t!]
\centering
\includegraphics[width=0.8\textwidth]{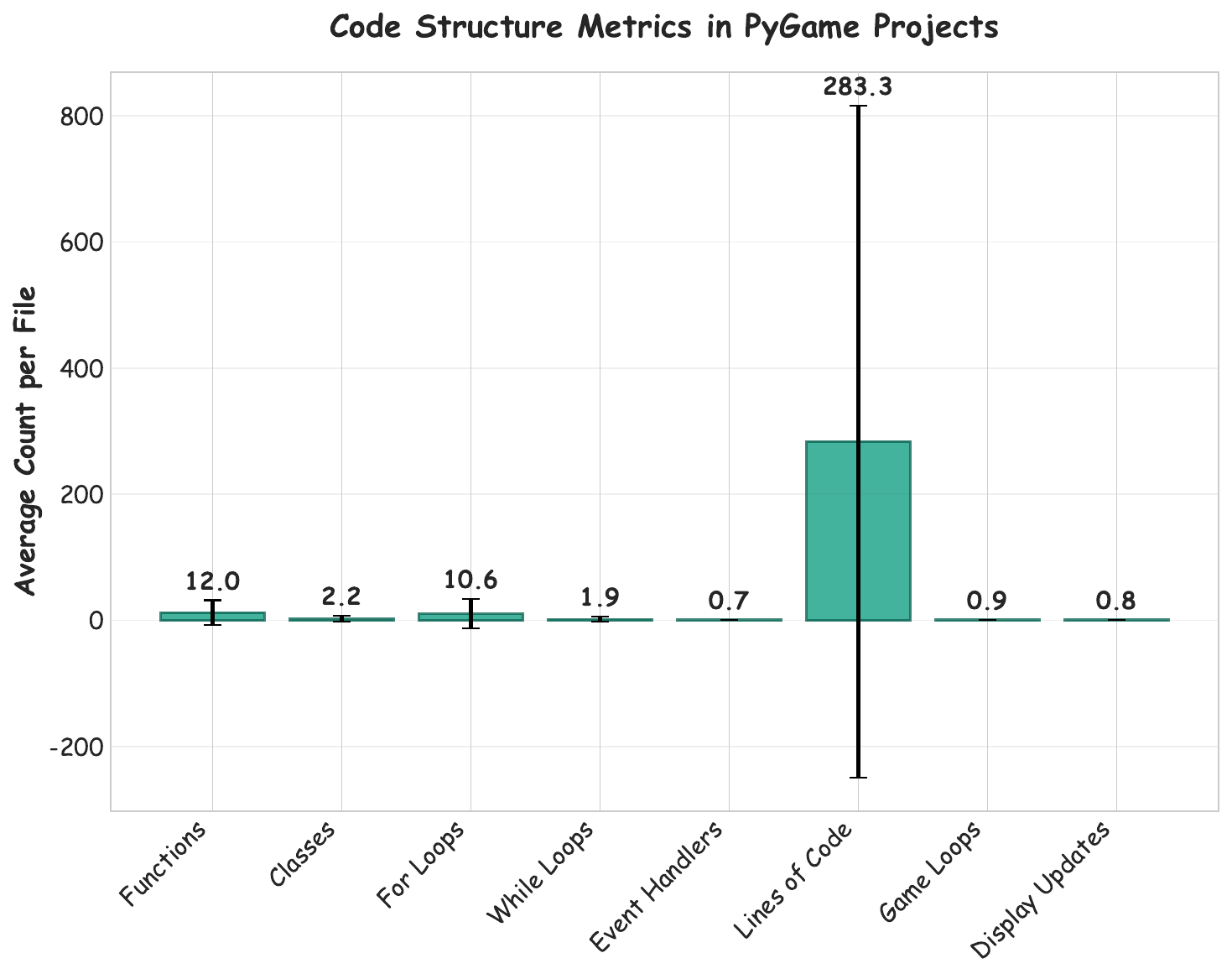}
\caption{Structural metrics of code samples, showing average counts of functions (12.0), classes (2.2), and other programming constructs per file.}
\label{fig:figures/figures_seed/06_code_structure_features.pdf}

\end{figure*}

\paragraph{File Size Characteristics}
The file size analysis in Figure~\ref{fig:figures/figures_seed/03_file_size_distribution.pdf} reveals a log-normal distribution with a mean of 10.2 KB and a median of 5.5 KB. This size distribution indicates that most games in our dataset are compact, self-contained implementations suitable for educational and prototyping purposes, while still including complex examples exceeding 50 KB. The predominance of smaller files (under 20 KB) aligns with typical Pygame project patterns and ensures computational efficiency during evaluation while maintaining functional completeness.

\paragraph{Pygame Module Usage Patterns}
Figure~\ref{fig:figures/figures_seed/04_pygame_module_usage.pdf} quantifies the frequency of core Pygame API usage across our dataset. The analysis shows that fundamental modules like \texttt{pygame.display} (91.5\%) and \texttt{pygame.event} (68.3\%) are nearly universal, confirming adherence to standard Pygame development patterns. The moderate usage of advanced features like \texttt{pygame.sprite} (21.3\%) and \texttt{pygame.mixer} (19.2\%) indicates a balanced representation of both basic and sophisticated game development techniques within our corpus.

\paragraph{Game Type Distribution}
The game type analysis in Figure~\ref{fig:figures/figures_seed/05_game_type_distribution.pdf} demonstrates substantial diversity in our dataset, with arcade games comprising the largest category (47.3\%) followed by shooter games (17.7\%) and other miscellaneous types (23.7\%). This distribution reflects the natural prevalence of different game genres in the Pygame community while ensuring adequate representation of specialized categories like physics simulations, RPGs, and educational games. The balanced representation across game types enhances the benchmark's ability to evaluate diverse programming patterns and game mechanics.

\begin{figure}[t!]
\centering
\includegraphics[width=0.8\textwidth]{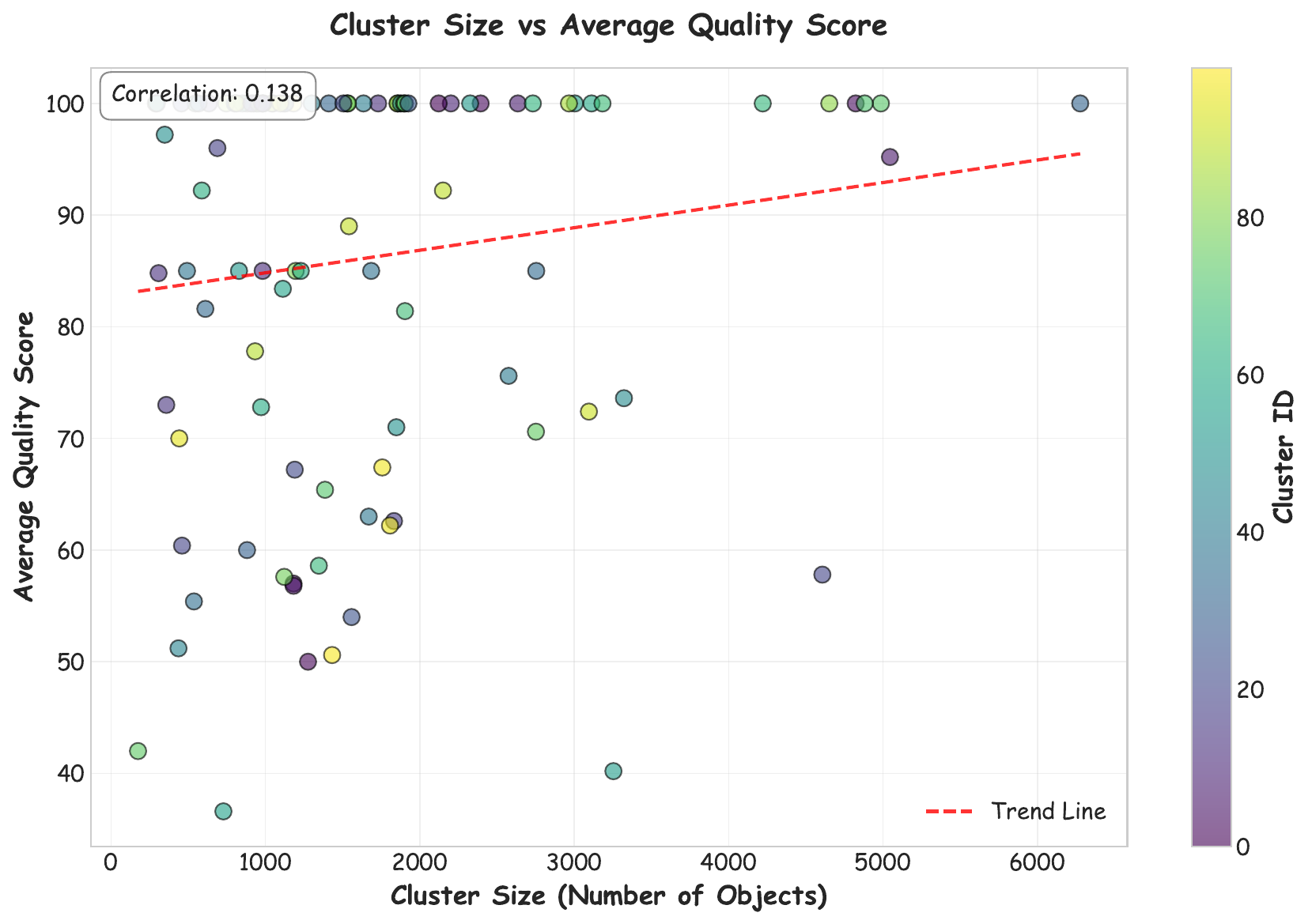}
\caption{Correlation analysis between cluster size and average quality scores, showing weak correlation (0.138) that validates quality-based selection within clusters.}
\label{fig:figures/figures_seed/07_cluster_quality_analysis.pdf}
\end{figure}

\begin{figure*}[t!]
\centering
\includegraphics[width=0.8\textwidth]{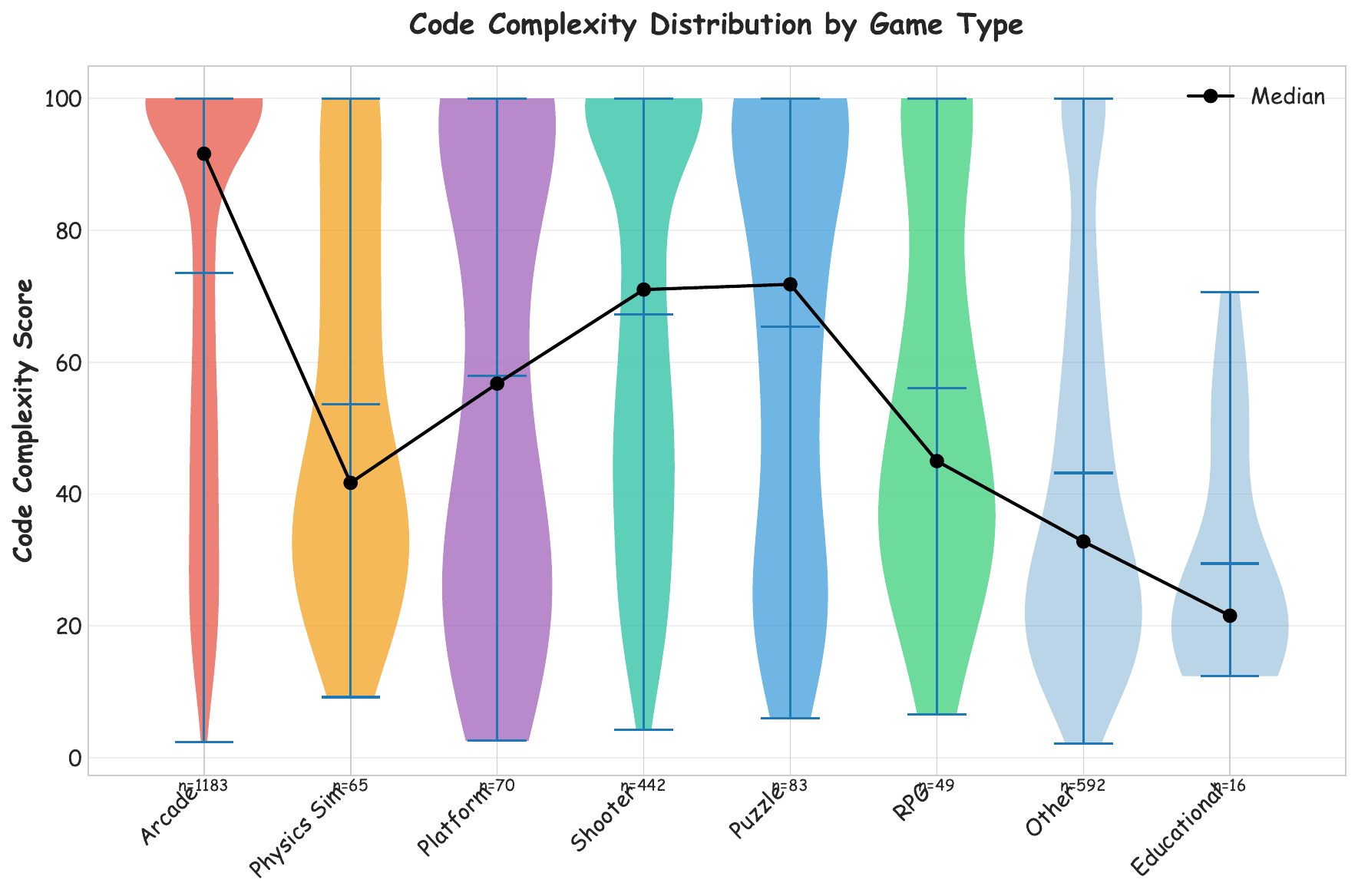}
\caption{Box plot analysis of code complexity scores across game types, with physics simulations and RPGs showing highest complexity variance.}
\label{fig:figures/figures_seed/08_complexity_by_game_type.pdf}
\end{figure*}

\paragraph{Code Structure Analysis}
Figure~\ref{fig:figures/figures_seed/06_code_structure_features.pdf} provides quantitative insights into the structural characteristics of our dataset. The average code file contains 12.0 functions, 2.2 classes, and 283.3 lines of code, indicating well-structured implementations that follow object-oriented programming principles. The prevalence of for loops (10.6 per file) and event handlers reflects the iterative and interactive nature of game programming, while the consistent presence of game loops and display updates confirms adherence to standard Pygame architectural patterns.

\begin{figure*}[t!]
\centering
\includegraphics[width=0.8\textwidth]{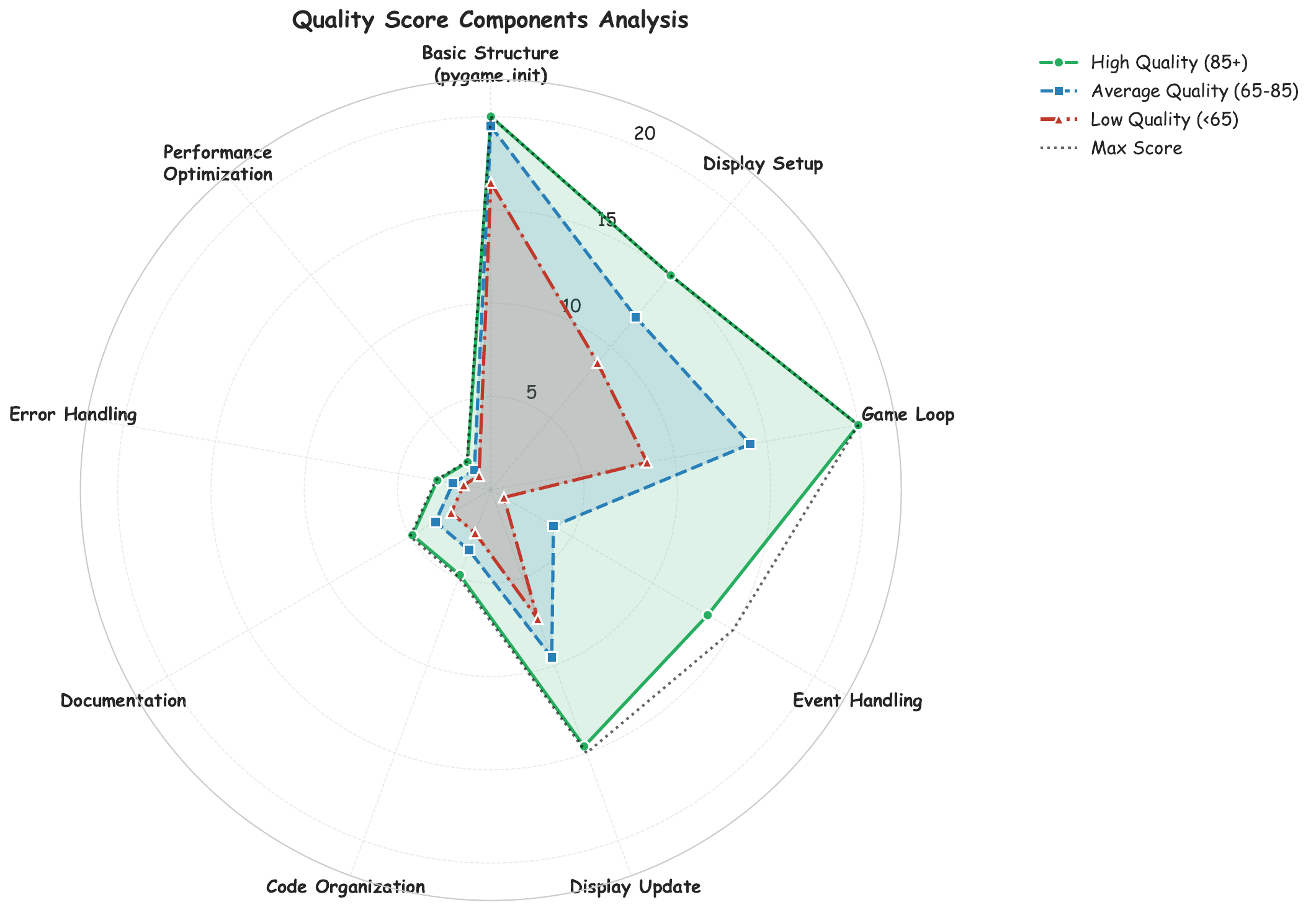}
\caption{Radar chart analysis of quality components across different score tiers, highlighting strengths in structure and organization for high-quality samples.}
\label{fig:figures/figures_seed/09_quality_components_radar.pdf}
\end{figure*}

\paragraph{Cluster Quality Correlation}
The scatter plot in Figure~\ref{fig:figures/figures_seed/07_cluster_quality_analysis.pdf} examines the relationship between cluster size and average quality scores. The weak positive correlation (0.138) suggests that larger clusters do not necessarily contain higher-quality code, validating our quality-based selection approach within each cluster. This analysis confirms that our methodology successfully identifies the best exemplars from each functional group regardless of the cluster's overall size, ensuring consistent quality across diverse game categories.

\paragraph{Complexity by Game Type}
Figure~\ref{fig:figures/figures_seed/08_complexity_by_game_type.pdf} reveals significant variation in code complexity across different game genres. Physics simulation and RPG games exhibit the highest complexity scores, reflecting their sophisticated mechanics and state management requirements. In contrast, educational and puzzle games show lower complexity, aligning with their focus on simplicity and clarity. This complexity distribution ensures that our benchmark captures the full spectrum of programming challenges inherent in different game development domains.

\paragraph{Quality Components Radar Analysis}
The radar chart in Figure~\ref{fig:figures/figures_seed/09_quality_components_radar.pdf} provides a multi-dimensional view of quality factors across different performance tiers. High-quality samples (85+) consistently excel across all dimensions, particularly in basic structure, display setup, and code organization. The analysis reveals that documentation and error handling are key differentiators between quality tiers, while basic functionality components like pygame initialization and game loops are well implemented across all levels. This comprehensive quality assessment ensures that our dataset maintains high standards while capturing diverse implementation approaches.

\section{System Architecture and Performance}

\begin{center}
\begin{tcolorbox}[title=Complete Pipeline Performance, showcase] 
    \begin{tcolorbox}[title=End-to-End Workflow (Generation → Recording → Evaluation), context]
        \begin{tcolorbox}[query]
            Pipeline Components:\\
            1. \textbf{Code Generation}: OpenAI API with parallel processing\\
            2. \textbf{Game Recording}: Optimized pygame execution with media capture\\
            3. \textbf{Multi-modal Evaluation}: Code + Screenshot + Video analysis
        \end{tcolorbox}
        \begin{tcolorbox}[query] 
            Performance Optimizations:\\
            - Async I/O for file operations\\
            - Batch processing for efficiency\\
            - Configurable worker pools\\
            - Resume capability for interrupted runs\\
            - Streaming API responses
        \end{tcolorbox}
        \begin{tcolorbox}[query]
            Quality Assurance:\\
            - Automatic retry mechanisms\\
            - JSON validation with error handling\\
            - Progress tracking and recovery\\
            - Comprehensive logging and statistics
        \end{tcolorbox}
    \end{tcolorbox}
    \vspace{1pt}
    \begin{center}
    \begin{tcbraster}[raster columns=3, raster column skip=0.2em, raster valign=top, 
    raster force size=false, raster equal height]
    \begin{tcolorbox}[title=Overall Success\phantom{p}, goldanswer]
    >80\% end-to-end success rate from requirement to final evaluation score.
    \end{tcolorbox}
    \begin{tcolorbox}[title=Scalability\phantom{p}, abab]
    Handles 1000+ games with configurable parallelization and resource management.
    \end{tcolorbox}
    \begin{tcolorbox}[title=Reliability\phantom{p}, others]
    Robust error handling with automatic recovery and detailed failure analysis.
    \end{tcolorbox}
    \end{tcbraster}
    \end{center}
\end{tcolorbox} 
\end{center}

\section{Game Code Generation Pipeline}

\begin{center}
\begin{tcolorbox}[title=Game Code Generation Case, showcase] 
    \begin{tcolorbox}[title=User Request (Game Requirement \ding{247} + System Prompt \ding{46} + Code Template \ding{49}), context]
        \begin{tcolorbox}[query]
            \ding{247} Generate a complete pygame code based on the following game requirement: \\
            "Create a simple Snake game where the player controls a snake to eat food and grow longer. The game should have collision detection and score display."
        \end{tcolorbox}
        \begin{tcolorbox}[query] 
            \ding{46} System Prompt: "You are a pygame game development expert, good at quickly developing small games based on requirements."\\
            
            Requirements:\\
            1. Generate a complete and runnable pygame code\\
            2. The game should automatically run for 10 seconds and then exit\\
            3. Include all necessary import statements, especially `import time`\\
            4. Add time-based automatic exit mechanism\\
            5. Add a visual timer showing elapsed time\\
            6. Set reasonable FPS
        \end{tcolorbox}
        \begin{tcolorbox}[query]
            \ding{49} Code Template Structure:\\
\begin{verbatim}```python
import pygame
import time
start_time = time.time()
# In main loop:
current_time = time.time()
if current_time - start_time >= 10:
    running = False
```\end{verbatim}
        \end{tcolorbox}
    \end{tcolorbox}
    \vspace{1pt}
    \begin{center}
    \begin{tcbraster}[raster columns=3, raster column skip=0.2em, raster valign=top, 
    raster force size=false, raster equal height]
    \begin{tcolorbox}[title=Generated Code\phantom{p}, goldanswer]
    Complete pygame Snake game with automatic exit, timer display, and proper game loop implementation.
    \end{tcolorbox}
    \begin{tcolorbox}[title=Success Rate\phantom{p}, abab]
    85.3\% of generated codes compile and run successfully.
    \end{tcolorbox}
    \begin{tcolorbox}[title=Avg Generation Time\phantom{p}, others]
    2.3 seconds per game with parallel processing.
    \end{tcolorbox}
    \end{tcbraster}
    \end{center}
\end{tcolorbox} 
\end{center}

\section{Game Recording and Media Capture}

\begin{center}
\begin{tcolorbox}[title=Game Recording Pipeline, showcase] 
    \begin{tcolorbox}[title=Recording Process (Code Execution \ding{247} + Screenshot Capture \ding{46} + Video Recording \ding{49}), context]
        \begin{tcolorbox}[query]
            \ding{247} Execute generated pygame code with optimized performance:\\
            - Record duration: 10-30 seconds\\
            - Video FPS: 3-30 (configurable)\\
            - Screenshot format: JPG (faster) or PNG\\
            - Async I/O for better performance
        \end{tcolorbox}
        \begin{tcolorbox}[query] 
            \ding{46} Screenshot Capture at specific timestamps:
\begin{verbatim}```python
_screenshot_times = [0, 1, 2, 3, 4, 5, 6, 7, 8, 9]
# Capture at each second for consistent evaluation
```\end{verbatim}\\

Optimized surface conversion:
\begin{verbatim}```python
def _pygame_surface_to_cv2_optimized(surface):
    raw = pygame.image.tobytes(surface, 'RGB')
    img = np.frombuffer(raw, dtype=np.uint8)
    return cv2.cvtColor(img, cv2.COLOR_RGB2BGR)
```\end{verbatim}
        \end{tcolorbox}
        \begin{tcolorbox}[query]
            \ding{49} Video Generation with optimized settings:\\
            - Use mp4v codec for fast encoding\\
            - Batch write frames for efficiency\\
            - Configurable frame interval based on target FPS\\
            - Background thread for async I/O operations
        \end{tcolorbox}
    \end{tcolorbox}
    \vspace{1pt}
    \begin{center}
    \begin{tcbraster}[raster columns=3, raster column skip=0.2em, raster valign=top, 
    raster force size=false, raster equal height]
    \begin{tcolorbox}[title=Output Media\phantom{p}, goldanswer]
    10 screenshots + 1 gameplay video per game with consistent quality.
    \end{tcolorbox}
    \begin{tcolorbox}[title=Processing Speed\phantom{p}, abab]
    Average 1.2s per game with 20 parallel workers.
    \end{tcolorbox}
    \begin{tcolorbox}[title=Success Rate\phantom{p}, others]
    92.7\% games successfully recorded with complete media files.
    \end{tcolorbox}
    \end{tcbraster}
    \end{center}
\end{tcolorbox} 
\end{center}

\section{Multi-Modal Game Evaluation System}

\begin{center}
\begin{tcolorbox}[title=Game Evaluation Framework, showcase] 
    \begin{tcolorbox}[title=Evaluation Components (Code Analysis \ding{247} + Screenshot Review \ding{46} + Video Assessment \ding{49}), context]
        \begin{tcolorbox}[query]
            \ding{247} Code Quality Evaluation (0-100 points):\\
            - Functionality (0-25): Implementation completeness\\
            - Code Quality (0-25): Structure and readability\\
            - Game Logic (0-25): Logic correctness\\
            - Technical Implementation (0-25): pygame usage efficiency
        \end{tcolorbox}
        \begin{tcolorbox}[query] 
            \ding{46} Visual Quality Assessment from Screenshots (0-100 points):\\
            - Visual Completeness (0-25): UI elements presence\\
            - UI Design (0-25): Layout and visual effects\\
            - Function Display (0-25): Key features visibility\\
            - Overall Quality (0-25): Visual completion level\\\\
          
            Up to 20 screenshots analyzed per game for comprehensive coverage.
        \end{tcolorbox}
        \begin{tcolorbox}[query]
            \ding{49} Dynamic Behavior Analysis from Video (0-100 points):\\
            - Animation Effect (0-25): Smoothness and naturalness\\
            - Interaction Logic (0-25): User input responsiveness\\
            - Game Flow (0-25): Gameplay continuity\\
            - Dynamic Quality (0-25): Overall playability
        \end{tcolorbox}
    \end{tcolorbox}
    \vspace{1pt}
    \begin{center}
    \begin{tcbraster}[raster columns=3, raster column skip=0.2em, raster valign=top, 
    raster force size=false, raster equal height]
    \begin{tcolorbox}[title=Final Score\phantom{p}, goldanswer]
    Average of three evaluation components with automatic retry mechanism for robust scoring.
    \end{tcolorbox}
    \begin{tcolorbox}[title=Retry Mechanism\phantom{p}, abab]
    Up to 10 attempts with JSON parsing validation for reliable results.
    \end{tcolorbox}
    \begin{tcolorbox}[title=Evaluation Speed\phantom{p}, others]
    4 parallel processes with streaming responses for efficient processing.
    \end{tcolorbox}
    \end{tcbraster}
    \end{center}
\end{tcolorbox} 
\end{center}

\end{document}